%% file: master.tex
\documentclass[11pt]{book} 
\usepackage{amsmath} 
\usepackage{amssymb} 
\usepackage{makeidx} 
\usepackage{epsfig}
\usepackage{graphicx} 

\newtheorem{definition}{Definition}
\newtheorem{example}{Example}
\newtheorem{theorem}{Theorem}
\newtheorem{property}{Property}
\newenvironment{proof}{Proof}

\newtheorem{proposition}{Proposition}

\begin{document} 

\thispagestyle{empty}
\begin{center}
{\bf \Large Haibin Wang \\ Florentin Smarandache \\ Yan-Qing Zhang \\ Rajshekhar Sunderraman \\}
\vspace{0.6in}
{\bf \LARGE Interval Neutrosophic Sets and Logic:\\ Theory and Applications in Computing} 

\vspace{1.5in}

\begin{figure}[htb!]
\includegraphics{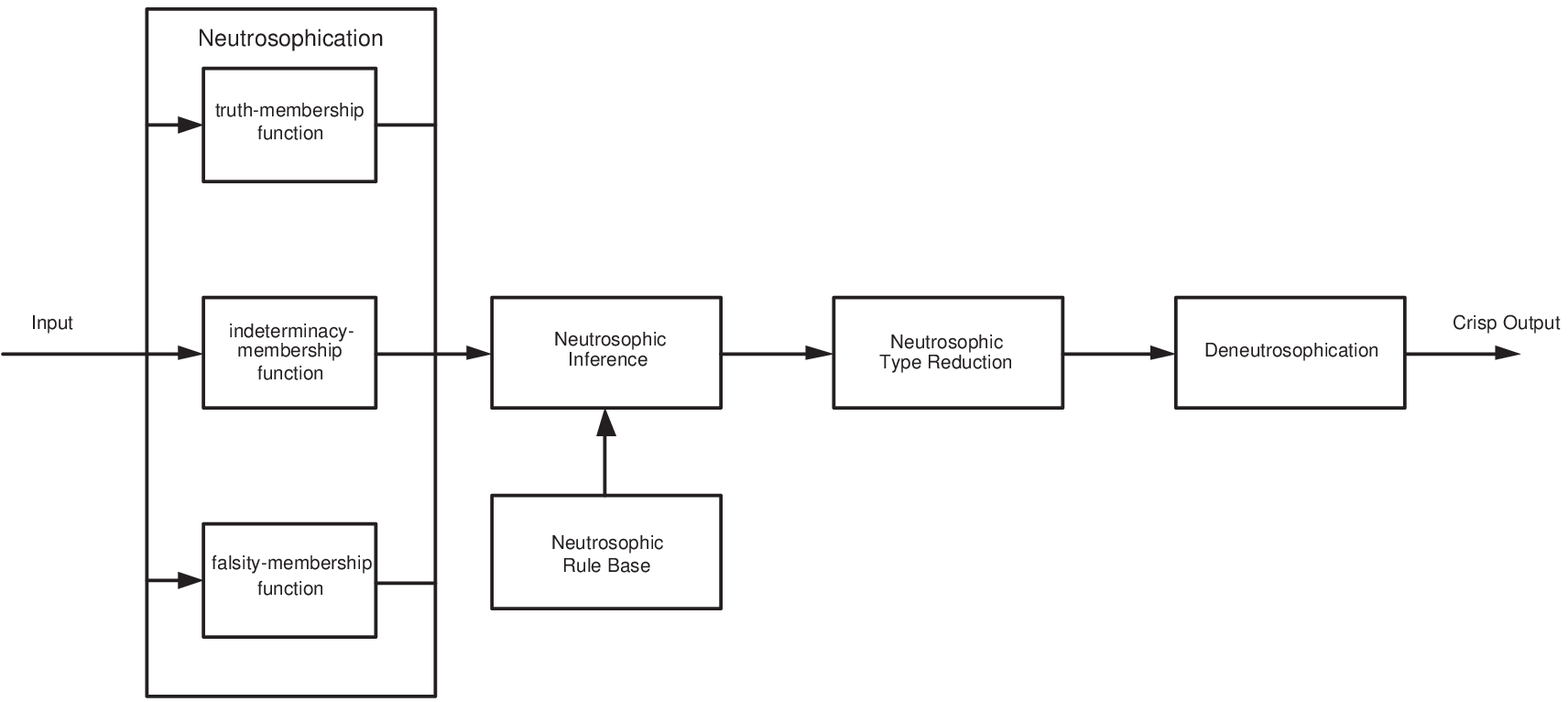}
\end{figure}
\end{center}


\hfill{ }{\bf 2005}

\newpage
\thispagestyle{empty}
\begin{center}
{\bf \LARGE Interval Neutrosophic Sets and Logic:} \\
{\bf \LARGE Theory and Applications in Computing} \\

\vspace{0.6in}
{\bf \Large Haibin Wang$^1$ \\ Florentin Smarandache$^2$ \\ Yan-Qing Zhang$^1$ \\ Rajshekhar Sunderraman$^1$ \\} 

\vspace{0.4in}
{\bf $^1$Department of Computer Science \\
Georgia State University \\
Atlanta, GA 30302, USA \\
}
\vspace{0.4in}
{\bf 
$^2$Department of Mathematics and Science \\
University of New Mexico \\
Gallup, NM 87301, USA \\
}

\vspace{3in}
\end{center}

\newpage
\thispagestyle{empty}

\vspace{0.2in}



\vspace{0.6in}
{\bf Peer Reviewers}: This book has been peer reviewed and recommended for publication by: 

Dr. Albena Tchamova, Bulgarian Academy of Sciences, Sofia, Bulgaria. 

Dr. W. B. Vasantha Kandasamy, Indian Institute of Technology Madras, 
Chennay, India. 

Dr. Feijun Song, Florida Atlantic University, Dania, USA. 

Dr. Xiaohong Yuan, North Carolina A $\&$ T State University, Greensboro, USA

\vspace{0.3in}


\vspace{0.3in}


\vspace{0.3in}

Many books can be downloaded from the following E-Library of Science: 

http://www.gallup.unm.edu/$\sim$smarandache/eBooks-otherformats.htm   

\vspace{1.2in}


\vspace{0.2in}


\frontmatter 
\tableofcontents 
\include{preface} 

\mainmatter 

\include{chaptr1}

\include{chaptr2}

\include{chaptr3}

\include{chaptr4} 


\bibliographystyle{amsalpha} 
\bibliography{ref} 
\end{document}

%% file: chaptr1.tex
\chapter{Interval Neutrosophic Sets} 
\label{INS} 

{\small
A neutrosophic set is a part of neutrosophy that studies the origin, nature,
and scope of neutralities, as well as their interactions with different
ideational spectra. The neutrosophic set is a powerful general formal
framework that has been recently proposed. However, the neutrosophic set needs to
be specified from a technical point of view. Now 
we define the set-theoretic operators on an instance
of a neutrosophic set, and call it an Interval Neutrosophic Set (INS). We prove 
various properties of INS, which are connected to operations
and relations over INS. Finally, we introduce and prove the convexity of 
interval
neutrosophic sets.
}

\section{Introduction}
The concept of fuzzy sets was introduced by Zadeh in 1965~\cite{ZAD65}.
Since then fuzzy sets and fuzzy logic have been applied in many real 
applications to handle uncertainty. The traditional fuzzy set uses one real
number $\mu_A(x) \in [0, 1]$ to represent the grade of membership of fuzzy
set $A$ defined on universe $X$. Sometimes $\mu_A(x)$ itself is uncertain
and hard to be defined by a crisp value. So the concept of interval valued
fuzzy sets was proposed~\cite{TUR86} to capture the uncertainty of grade of
membership. Interval valued fuzzy set uses an interval value $[\mu_{A}^L(x),
\mu_{A}^U(x)]$ with $0 \leq \mu_{A}^L(x) \leq \mu_{A}^U(x) \leq 1$ to represent
the grade of membership of fuzzy set $A$. In some applications such as 
expert system, belief system and information fusion, we should consider not
only the truth-membership supported by the evidence but also the 
falsity-membership against by the evidence. That is beyond the scope of fuzzy
sets and interval valued fuzzy sets. In 1986, Atanassov introduced the
intuitionistic fuzzy sets~\cite{ATA86} that is a generalization of fuzzy sets
and provably equivalent to interval valued fuzzy sets. 
The intuitionistic fuzzy sets consider 
both truth-membership and falsity-membership. Later on, intuitionistic fuzzy
sets were extended to the interval valued intuitionistic fuzzy sets~\cite{ATA89}. The interval valued intuitionistic fuzzy set uses a pair of intervals
$[t^-, t^+], \mbox{  } 0 \leq t^-  \leq t^+ \leq 1$ and 
$[f^-, f^+], \mbox{  } 0 \leq f^- \leq f^+ \leq 1$ with $t^+ + f^+ \leq 1$ to 
describe the degree of true belief
and false belief. Because of the restriction that $t^+ + f^+ \leq 1$, 
intuitionistic fuzzy sets and interval valued intuitionistic fuzzy sets can 
only handle incomplete information not the indeterminate information and 
inconsistent information which exists commonly in belief systems. For example,
when we ask the opinion of an expert about certain statement, he or she may
say that the possibility that the statement is true is between $0.5$ and $0.7$
and the statement is false is between $0.2$ and $0.4$ and the degree that
he or she is not sure is between $0.1$ and $0.3$. Here is another example, 
suppose there are 10 voters during a voting process. In time $t_1$, three vote
``yes", two vote ``no" and five are undecided, using neutrosophic notation, it
can be expressed as $x(0.3,0.5,0.2)$. In time $t_2$, three vote ``yes", two 
vote ``no", two give up and three are undecided, it then can be expressed as
$x(0.3,0.3,02)$. That is beyond the scope of the intuitionistic fuzzy set. 
So, the notion of neutrosophic set is more general and overcomes the 
aforementioned issues.

In neutrosophic set, indeterminacy is quantified explicitly and \\ 
truth-membership, 
indeterminacy-membership and falsity-membership are independent. This assumption 
is very important in many applications such as information fusion in which we try to
combine the data from different sensors. 
Neutrosophy was introduced by Florentin Smarandache in 1980. ``It is a branch
of philosophy which studies the origin, nature and scope of neutralities, as
well as their interactions with different ideational spectra"~\cite{SMA99}.  
Neutrosophic set is a powerful general formal framework which generalizes the
concept of the classic set, fuzzy set~\cite{ZAD65}, 
interval valued fuzzy set~\cite{TUR86}, 
intuitionistic fuzzy set~\cite{ATA86}, 
interval valued intuitionistic fuzzy set~\cite{ATA89}, 
paraconsistent set~\cite{SMA99}, dialetheist set~\cite{SMA99}, 
paradoxist set~\cite{SMA99}, tautological set~\cite{SMA99}.
A neutrosophic set $A$ defined on universe $U$. $x = x(T, I, F) \in A$ with
$T, I$ and $F$ being the real standard or non-standard subsets of $]0^-, 1^+[$.
$T$ is the degree of truth-membership function in the set $A$, $I$ is the 
degree of indeterminacy-membership function in the set $A$ and $F$ is the
degree of falsity-membership function in the set $A$. 

The neutrosophic set 
generalizes the above mentioned sets from philosophical point of view. 
From scientific or
engineering point of view, the neutrosophic set and set-theoretic operators
need to  be specified. Otherwise, it will be difficult to apply in the real 
applications. In this chapter, we define the set-theoretic operators on an
instance of neutrosophic set called Interval Neutrosophic Set (INS). We
call it as ``interval" because it is subclass of neutrosophic set, that is
we only consider the subunitary interval of $[0, 1]$.

An interval neutrosophic set $A$ defined on universe $X$, $x = x(T, I, F) 
\in A$
with $T$, $I$ and $F$ being the subinterval of $[0, 1]$.
The interval neutrosophic set can represent uncertain, imprecise, incomplete 
and inconsistent information which exist in real world. The interval 
neutrosophic set generalizes the following sets:
\begin{enumerate}
\item the \emph{classical set}, $I = \emptyset$, $\inf T = \sup T = 0$ or $1$, $\inf F = \sup F
= 0$ or $1$ and $\sup T + \sup F = 1$. 
\item the \emph{fuzzy set}, $I = \emptyset$, $\inf T = \sup T \in [0, 1]$, $\inf F = \sup F \in
[0, 1]$ and $\sup T + \sup F = 1$.
\item the \emph{interval valued fuzzy set}, $I = \emptyset$, $\inf T, \sup T, \inf F, \sup F \in
[0, 1]$, $\sup T + \inf F = 1$ and $\inf T + \sup F = 1$.
\item the \emph{intuitionistic fuzzy set}, $I = \emptyset$, $\inf T = \sup T \in [0, 1]$,
$\inf F = \sup F \in [0, 1]$ and $\sup T + \sup F \leq 1$.
\item the \emph{interval valued intuitionistic fuzzy set}, $I = \emptyset$, $\inf T, \sup T,
\inf F, \sup F \in [0, 1]$ and $\sup T + \sup F \leq 1$.
\item the \emph{paraconsistent set}, $I = \emptyset$, $\inf T = \sup T \in [0, 1]$,
$\inf F = \sup F \in [0, 1]$ and $\sup T + \sup F > 1$.
\item the \emph{interval valued paraconsistent set}, $I = \emptyset$, $\inf T, \sup T, \inf F,
\sup F \in [0, 1]$ and $\inf T + \inf F > 1$.
\end{enumerate}
  
\begin{figure}[hbtp]
\centering
\includegraphics{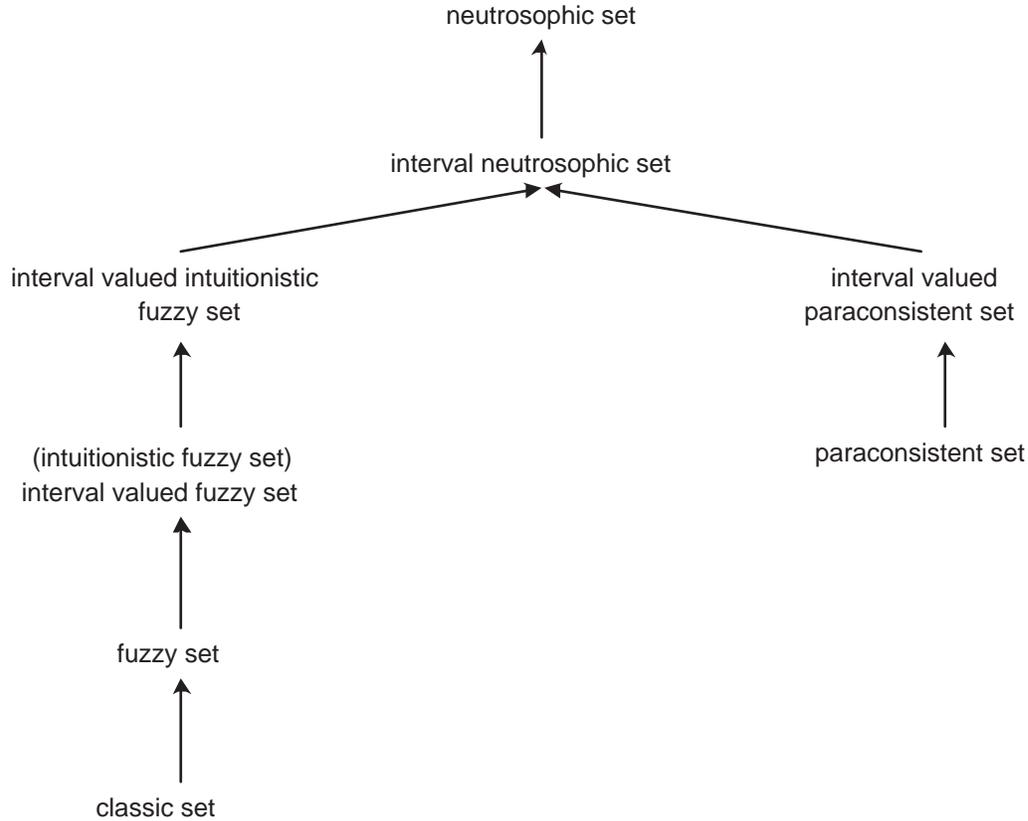}
\caption{Relationship among interval neutrosophic set and other sets}
\label{fig0}
\end{figure}

The relationship among interval neutrosophic set and other sets is 
illustrated in Fig~\ref{fig0}.

Note that $\rightarrow$ in Fig.~\ref{fig0} such as $a \rightarrow b$ means
that b is a generalization of a.

We define the set-theoretic operators on the interval neutrosophic set.
Various properties of INS are proved, which are connected to the operations
and relations over INS.                           

The rest of chapter is organized as follows. Section~\ref{chapter1:section2} gives a brief 
overview of the neutrosophic set. Section~\ref{chapter1:section3} gives the definition of 
the interval neutrosophic set and set-theoretic operations.
Section~\ref{chapter1:section4} presents some
properties of set-theoretic operations. Section~\ref{chapter1:section5} defines 
the convexity of the interval neutrosophic sets and proves some 
properties of convexity. Section~\ref{chapter1:section6} concludes the
chapter. To maintain a smooth flow throughout the chapter, we present the proofs
to all theorems in Appendix.  

\section{Neutrosophic Set}
\label{chapter1:section2}
This section gives a brief overview of concepts of neutrosophic set defined 
in~\cite{SMA99}. Here, we use different notations to express the same meaning.
Let $S_1$ and $S_2$ be two real standard or non-standard 
subsets, then $S_1 \oplus S_2 = \{x | x=s_1 + s_2, s_1 \in S_1 \mbox{ and } s_2 \in S_2\}$, $\{1^+\} \oplus S_2 = \{x | x=1^+ + s_2, s_2 \in S_2\}$. $S_1 \ominus S_2 = \{x | x = s_1 - s_2, s_1 \in S_1 \mbox{ and } s_2 \in S_2\}$, $\{1^+\} \ominus S_2 = \{x | x = 1^+ - s_2, s_2 \in S_2\}$. $S_1 \odot S_2 = \{x | x = s_1 \cdot s_2, s_1 \in S_1 \mbox{ and } s_2 \in S_2\}$. 

\begin{definition}[Neutrosophic Set]
Let $X$ be a space of points (objects), with a generic element in $X$ denoted
by $x$. \\
A neutrosophic set $A$ in $X$ is characterized by a \emph{truth-membership function}
$T_A$, a \emph{indeterminacy-membership function} $I_A$ and a 
\emph{falsity-membership} function $F_A$. $T_A(x), I_A(x)$ and $F_A(x)$ are
real standard or non-standard subsets of $]0^-, 1^+[$. That is 

\begin{eqnarray}
   T_A:X & \rightarrow & ]0^-, 1^+[, \\
   I_A:X & \rightarrow & ]0^-, 1^+[, \\
   F_A:X & \rightarrow & ]0^-, 1^+[.
\end{eqnarray}

There is no restriction on the sum of $T_A(x)$, $I_A(x)$ and $F_A(x)$, so 
$0^- \leq \sup T_A(x) + \sup I_A(x) + \sup F_A(x) \leq 3^+$.
\end{definition}

\begin{definition}
The \emph{complement} of a neutrosophic set $A$ is denoted by $\bar{A}$ and 
is defined by

\begin{eqnarray}
T_{\bar{A}}(x) & = & \{1^+\} \ominus T_A(x), \\
I_{\bar{A}}(x) & = & \{1^+\} \ominus I_A(x), \\
F_{\bar{A}}(x) & = & \{1^+\} \ominus F_A(x), 
\end{eqnarray}
for all $x$ in $X$.
\end{definition}

\begin{definition}[Containment]
A neutrosophic set $A$ is \emph{contained} in the other neutrosophic set $B$, 
$A \subseteq B$,
if and only if 

\begin{eqnarray}
\inf T_A(x) \leq \inf T_B(x) & , & \sup T_A(x) \leq \sup T_B(x), \\
\inf I_A(x) \geq \inf I_B(x) &,  & \sup I_A(x) \geq \sup I_B(x), \\
\inf F_A(x) \geq \inf F_B(x) & , & \sup F_A(x) \geq \sup F_B(x),
\end{eqnarray}
for all $x$ in $X$. 

\end{definition}

\begin{definition}[Union]
The \emph{union} of two neutrosophic sets $A$ and $B$ 
is a neutrosophic set
$C$, written as $C = A \cup B$, whose truth-membership, 
indeterminacy-membership and falsity-membership
functions are related to those of $A$ and $B$ by

\begin{eqnarray}
T_C(x) & = & T_A(x) \oplus T_B(x) \ominus T_A(x) \odot T_B(x), \\
I_C(x) & = & I_A(x) \oplus I_B(x) \ominus I_A(x) \odot I_B(x), \\
F_C(x) & = & F_A(x) \oplus F_B(x) \ominus F_A(x) \odot F_B(x),
\end{eqnarray}
for all $x$ in $X$.
\end{definition}

\begin{definition}[Intersection]
The \emph{intersection} of two neutrosophic sets $A$ and $B$ 
is a neutrosophic set $C$, written as $C = A \cap B$, whose truth-membership,
indeterminacy-membership and
falsity-membership functions are related to those of $A$ and $B$ by

\begin{eqnarray}
T_C(x) & = & T_A(x) \odot T_B(x), \\
I_C(x) & = & I_A(x) \odot I_B(x), \\
F_C(x) & = & F_A(x) \odot F_B(x),
\end{eqnarray}
for all $x$ in $X$.

\end{definition}

\begin{definition}[Difference]
The \emph{difference} of two neutrosophic sets $A$ and $B$ is a neutrosophic
set $C$, written as $C = A \setminus B$, whose truth-membership,
indeterminacy-membership and falsity-membership functions are related to 
those of $A$ and $B$ by

\begin{eqnarray} 
T_C(x) & = & T_A(x) \ominus T_A(x) \odot T_B(x), \\
I_C(x) & = & I_A(x) \ominus I_A(x) \odot I_B(x), \\
F_C(x) & = & F_A(x) \ominus F_A(x) \odot F_B(x),
\end{eqnarray}
for all $x$ in $X$.

\end{definition}

\begin{definition}[Cartesian Product]
Let $A$ be the neutrosophic set defined on universe $E_1$ and $B$ be the 
neutrosophic set defined on universe $E_2$. If $x(T_{A}^1,I_{A}^1,F_{A}^1) \in A$ and $y(T_{A}^2,I_{A}^2,F_{A}^2) \in B$, then the \emph{cartesian product}
of two neutrosophic sets $A$ and $B$ is defined by
\begin{equation}
(x(T_{A}^1,I_{A}^1,F_{A}^1),y(T_{A}^2,I_{A}^2,F_{A}^2)) \in A \times B
\end{equation}  
\end{definition}

\section{Interval Neutrosophic Set}
\label{chapter1:section3}
In this section, we present the notion of the \emph{interval neutrosophic set} 
(INS).
The interval neutrosophic set (INS) is an instance of neutrosophic set which
can be used in real scientific and engineering applications. 

\begin{definition}[Interval Neutrosophic Set]
Let $X$ be a space of points (objects), with a generic element in $X$ denoted
by $x$. \\
An interval neutrosophic set (INS) $A$ in $X$ is characterized by 
truth-membership function $T_A$, indeterminacy-membership function $I_A$ and 
falsity-membership function $F_A$.
For each point $x$ in $X$,
$T_A(x), I_A(x), F_A(x) \subseteq [0, 1]$.
\end{definition}

\begin{figure}[hbt]
\centering
\includegraphics[width=2.5in]{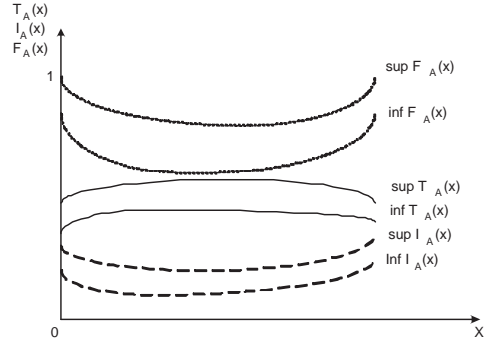}
\caption{Illustration of interval neutrosophic set in $R^1$}
\label{fig1}
\end{figure}

An interval neutrosophic set (INS) in $R^1$ is illustrated in 
Fig.~\ref{fig1}.

When $X$ is continuous, an INS $A$ can be written as

\begin{equation}
A = \int_{X} \langle T(x), I(x), F(x) \rangle / x, \mbox{ } x \in X
\end{equation}

When $X$ is discrete, an INS $A$ can be written as

\begin{equation}
A = \sum_{i=1}^{n} \langle T(x_i), I(x_i), F(x_i) \rangle / x_i, \mbox{ } x_i \in X  
\end{equation}

Consider parameters such as capability, trustworthiness and price of semantic
Web services. These parameters are commonly used to define quality of service
of semantic Web services.
In this section, we will use the evaluation of quality of service of semantic
Web services~\cite{WZR04} as running example to illustrate every set-theoretic
operation on interval neutrosophic set.

\begin{example}
\label{chapter1:example1}
Assume that $X = [x_1, x_2, x_3]$. $x_1$ is capability, $x_2$ is trustworthiness and $x_3$ is price. The values of $x_1, x_2$ and $x_3$ are in $[0,1]$. They
are obtained from the questionnaire of some domain experts, their option could
be degree of good, degree of indeterminacy and degree of poor.
$A$ is an interval  neutrosophic set 
of $X$ defined by \\
$A = \langle [0.2,0.4],[0.3,0.5],[0.3,0.5] \rangle/x_1 + \langle [0.5,0.7],[0,0.2],[0.2,0.3] \rangle/x_2 + \\ 
\langle [0.6,0.8],[0.2,0.3],[0.2,0.3] \rangle/x_3$. \\ 
$B$ is an interval neutrosophic set of $X$ defined by \\
$B = \langle [0.5,0.7],[0.1,0.3],[0.1,0.3] \rangle/x_1 + \langle [0.2,0.3],[0.2,0.4],[0.5,0.8] \rangle/x_2 + \\
\langle [0.4,0.6],[0,0.1],[0.3,0.4] \rangle/x_3$. \\
\end{example}

\begin{definition}
An interval neutrosophic set $A$ is \emph{empty} if and only if its 
$\inf T_A(x) = \sup T_A(x) = 0$, $\inf I_A(x) = \sup I_A(x) = 1$ and $\inf F_A(x) = \sup T_A(x) = 0$,
for all $x$ in $X$.
\end{definition}

We now present the set-theoretic operators on
interval neutrosophic set. 

\begin{definition}[Containment]
An interval neutrosophic set $A$ is \emph{contained} in the other interval 
neutrosophic set $B$, $A \subseteq B$, if and only if
\begin{eqnarray}
\inf T_A(x) \leq \inf T_B(x) & , & \sup T_A(x) \leq \sup T_B(x), \\
\inf I_A(x) \geq \inf I_B(x) & , & \sup I_A(x) \geq \sup I_B(x), \\
\inf F_A(x) \geq \inf F_B(x) & , & \sup F_A(x) \geq \sup F_B(x),
\end{eqnarray}
for all $x$ in $X$. 
\end{definition}

\begin{definition}
Two interval neutrosophic sets $A$ and $B$ are \emph{equal}, written as 
$A = B$, if and only if $A \subseteq B$ and $B \subseteq A$
\end{definition}

Let $\underline 0$ = $\langle 0, 1, 1 \rangle$ and 
$\underline 1$ = $\langle 1, 0, 0 \rangle$.

\begin{definition}[Complement]
Let $C_N$ denote a neutrosophic
\emph{complement} of $A$. Then $C_N$ is a function
\[
  C_N : N \rightarrow N 
\]
and $C_N$ must satisfy at least the following three axiomatic requirements:
\begin{enumerate}
\item $C_N(\underline 0)$ = $\underline 1$ and $C_N(\underline 1)$ = $\underline 0$ (boundary conditions).
\item Let $A$ and $B$ be two interval neutrosophic sets defined on $X$,  if
$A(x) \leq B(x)$, then $C_N(A(x)) \geq C_N(B(x))$, for all $x$ in $X$. (monotonicity).
\item Let $A$ be an interval neutrosophic set defined on $X$, then
$C_N(C_N(A(x))) = A(x)$, for all $x$ in $X$. (involutivity).
\end{enumerate}
\hfill{\space} $\Box$
\end{definition}
 
There are many functions which satisfy the requirement to be the complement 
operator
of interval neutrosophic sets. Here we give one example.
\begin{definition}[Complement $C_{N_1}$]
The complement of an interval neutrosophic set $A$ is denoted by $\bar{A}$
and is defined by
\begin{eqnarray}
T_{\bar A}(x) &=& F_A(x), \\
\inf I_{\bar A}(x) &=& 1 - \sup I_{A}(x), \\
\sup I_{\bar A}(x) &=& 1 - \inf I_{A}(x), \\
F_{\bar A}(x) &=& T_A(x),
\end{eqnarray}
for all $x$ in $X$.
\hfill{\space}  $\Box$
\end{definition}

\begin{example}
\label{chapter1:example2}
Let $A$ be the interval neutrosophic set defined in Example~\ref{chapter1:example1}. 
Then, \\
${\bar{A}} = \langle [0.3,0.5],[0.5,0.7],[0.2,0.4] \rangle/x_1 + \langle [0.2,0.3],[0.8,1.0],
[0.5,0.7] \rangle/x_2 + 
\langle [0.2,0.3],[0.7,0.8],[0.6,0.8] \rangle/x_3$. \\
 
\end{example}

\begin{definition}[$N$-norm]
Let $I_N$ denote a neutrosophic
\emph{intersection} of two interval neutrosophic sets $A$ and $B$. 
Then $I_N$ is a function
\[
  I_N : N \times N \rightarrow N
\]
and $I_N$ must satisfy at least the following four axiomatic requirements:
\begin{enumerate}
\item $I_N(A(x), \underline 1) = A(x)$, for all $x$ in $X$. (boundary condition).
\item $B(x) \leq C(x)$ implies $I_N(A(x), B(x)) \leq I_N(A(x), C(x))$, for all $x$ in $X$. (monotonicity).
\item $I_N(A(x), B(x)) = I_N(B(x), A(x))$, for all $x$ in $X$. (commutativity).
\item $I_N(A(x), I_N(B(x), C(x))) = I_N(I_N(A(x), B(x)), C(x))$, for all $x$ in $X$. (associativity).
\end{enumerate}
\hfill{\space} $\Box$
\end{definition}

Here we give one example of intersection of two interval neutrosophic sets
which satisfies above $N$-norm axiomatic requirements. Other different
definitions can be given for different applications.

\begin{definition}[Intersection $I_{N_1}$]
The intersection of two interval neutrosophic sets $A$ and $B$ is an interval
neutrosophic set $C$, written as $C = A \cap B$, whose truth-membership,
indeterminacy-membership, and false-membership are related to those of $A$
and $B$ by
\begin{eqnarray}
\inf T_C(x) &=& \min(\inf T_A(x),\inf T_B(x)), \\
\sup T_C(x) &=& \min(\sup T_A(x),\sup T_B(x)), \\
\inf I_C(x) &=& \max(\inf I_A(x),\inf I_B(x)), \\
\sup I_C(x) &=& \max(\sup I_A(x),\sup I_B(x)), \\
\inf F_C(x) &=& \max(\inf F_A(x),\inf F_B(x)), \\
\sup F_C(x) &=& \max(\sup F_A(x),\sup F_B(x)),
\end{eqnarray}
for all $x$ in $X$.
\hfill{\space}  $\Box$
\end{definition}

\begin{example}
\label{chapter1:example3}
Let $A$ and $B$ be the interval neutrosophic sets defined in
Example~\ref{chapter1:example1}.
Then,
$A \cap B = \langle [0.2,0.4],[0.3,0.5],[0.3,0.5] \rangle/x_1 + \\ 
\langle [0.2,0.3],
[0.2,0.4],
[0.5,0.8] \rangle/x_2 + 
\langle [0.4,0.6],[0.2,0.3],[0.3,0.4] \rangle/x_3$. \\
\end{example}

\begin{theorem}
$A \cap B$ is the largest interval neutrosophic set contained in 
both $A$ and $B$.
\end{theorem}

\begin{definition}[$N$-conorm]
Let $U_N$ denote a neutrosophic
\emph{union} of two interval neutrosophic sets $A$ and $B$.
Then $U_N$ is a function
\[
  U_N : N \times N \rightarrow N
\]
and $U_N$ must satisfy at least the following four axiomatic requirements:
\begin{enumerate}
\item $U_N(A(x), \underline 0) = A(x)$, for all $x$ in $X$. (boundary condition).
\item $B(x) \leq C(x)$ implies $U_N(A(x), B(x)) \leq U_N(A(x), C(x))$, for all $x$ in $X$. (monotonicity).

\item $U_N(A(x), B(x)) = U_N(B(x), A(x))$, for all $x$ in $X$. (commutativity).
\item $U_N(A(x), U_N(B(x), C(x))) = U_N(U_N(A(x), B(x)), C(x))$, for all $x$ in $X$. (associativity).
\end{enumerate}
\hfill{\space} $\Box$
\end{definition}

Here we give one example of union of two interval neutrosophic sets
which satisfies above $N$-conorm axiomatic requirements. Other different
definitions can be given for different applications.

\begin{definition}[Union $U_{N_1}$]
The union of two interval neutrosophic sets $A$ and $B$ is an interval
neutrosophic set $C$, written as $C = A \cup B$, whose truth-membership,
indeterminacy-membership, and false-membership are related to those of $A$
and $B$ by
\begin{eqnarray}
\inf T_C(x) &=& \max(\inf T_A(x),\inf T_B(x)), \\
\sup T_C(x) &=& \max(\sup T_A(x),\sup T_B(x)), \\
\inf I_C(x) &=& \min(\inf I_A(x),\inf I_B(x)), \\
\sup I_C(x) &=& \min(\sup I_A(x),\sup I_B(x)), \\
\inf F_C(x) &=& \min(\inf F_A(x),\inf F_B(x)), \\
\sup F_C(x) &=& \min(\sup F_A(x),\sup F_B(x)),
\end{eqnarray}
for all $x$ in $X$.
\hfill{\space}  $\Box$
\end{definition}

\begin{example}
\label{chapter1:example4}
Let $A$ and $B$ be the interval neutrosophic sets defined in 
Example~\ref{chapter1:example1}.
Then,
$A \cup B = \langle [0.5,0.7],[0.1,0.3],[0.1,0.3] \rangle/x_1 + \\ 
\langle [0.5,0.7],
[0,0.2],
[0.2,0.3] \rangle/x_2 + 
\langle [0.6,0.8],[0,0.1],[0.2,0.3] \rangle/x_3$. \\
\end{example}

The intuition behind the union operator is that if one of elements in $A$
and $B$ is true then it is true in $A \cup B$, only both are
indeterminate and false in $A$ and $B$ then it is indeterminate and false
in $A \cup B$. The other operators should be understood similarly.

\begin{theorem}
$A \cup B$ is the smallest interval neutrosophic set containing 
both $A$ and $B$.
\end{theorem}

\begin{theorem}

Let $P$ be the power set of all interval neutrosophic sets defined in
the universe X. Then $\langle P;I_{N_1}, U_{N_1} \rangle$ is a
distributive lattice. 
\end{theorem}
\begin{proof}
Let $A, B, C$ be the arbitrary interval neutrosophic sets defined on $X$.
It is easy to verify that $A \cap A = A, A \cup A = A$ (idempotency), 
$A \cap B = B \cap A, A \cup B = B \cup A$ (commutativity), $(A \cap B) \cap C = A \cap (B \cap C), (A \cup B) \cup C = A \cup (B \cup C)$ (associativity), and
$A \cap (B \cup C) = (A \cap B) \cup (A \cap C), A \cup (B \cap C) = (A \cup B) \cap (A \cup C)$ (distributivity). 
\end{proof}

\begin{definition}[Interval neutrosophic relation]
Let $X$ and $Y$ be two non-empty crisp sets. An interval neutrosophic relation
$R(X, Y)$ is a subset of product space $X \times Y$, and is characterized by
the truth membership function $T_{R}(x,y)$, the indeterminacy membership function
$I_{R}(x,y)$, and the falsity membership function $F_{R}(x,y)$, 
where $x \in X$ and $y \in Y$ and $T_{R}(x,y), I_{R}(x,y), F_{R}(x,y) \subseteq [0,1]$.
\end{definition}

\begin{definition}[Interval Neutrosophic Composition Functions]
The membership functions
for the composition of interval neutrosophic relations
$R(X,Y)$ and $S(Y,Z)$ are given by the \emph{interval neutrosophic sup-star 
composition} of $R$ and $S$ 
\begin{eqnarray}
T_{R \circ S} (x,z) &=& \sup_{y \in Y} \min(T_{R}(x,y), T_{S}(y,z)), \\
I_{R \circ S} (x,z) &=& \sup_{y \in Y} \min(I_{R}(x,y), I_{S}(y,z)), \\
F_{R \circ S} (x,z) &=& \inf_{y \in Y} \max(F_{R}(x,y), F_{S}(y,z)).
\end{eqnarray}
\end{definition}

If $R$ is an interval neutrosophic set rather than an interval neutrosophic
relation, then $Y = X$ and  \\
$\sup_{y \in Y} \min(T_{R}(x,y), T_{S}(y,z))$ becomes
$\sup_{y \in Y} \min(T_{R}(x), T_{S}(y,z))$, which is only a function of the output
variable $z$. 
It is similar for $\sup_{y \in Y} \min(I_{R}(x,y), I_{S}(y,z))$ and $\inf_{y \in Y} \max(F_{R}(x,y), F_{S}(y,z))$. Hence, the notation of 
$T_{R \circ S} (x,z)$ can be simplified to
$T_{R \circ S} (z)$, so that in the case of $R$ being just an interval neutrosophic set,
\begin{eqnarray}
T_{R \circ S} (z) &=& \sup_{x \in X} \min(T_{R}(x), T_{S}(x,z)), \\
I_{R \circ S} (z) &=& \sup_{x \in X} \min(I_{R}(x), I_{S}(x,z)), \\
F_{R \circ S} (z) &=& \inf_{x \in X} \max(F_{R}(x), F_{S}(x,z)).
\end{eqnarray}

\begin{definition}[Difference]
The \emph{difference} of two interval neutrosophic sets $A$ and $B$ 
is an interval neutrosophic 
set $C$, written as $C = A \setminus B$, whose truth-membership,
indeterminacy-membership and falsity-membership functions are related to
those of $A$ and $B$ by         
\begin{eqnarray}
\inf T_C(x) & = & \min(\inf T_A(x), \inf F_B(x)), \\
\sup T_C(x) & = & \min(\sup T_A(x), \sup F_B(x)), \\
\inf I_C(x) & = & \max(\inf I_A(x), 1 - \sup I_B(x)), \\
\sup I_C(x) & = & \max(\sup I_A(x), 1 - \inf I_B(x)), \\
\inf F_C(x) & = & \max(\inf F_A(x), \inf T_B(x)), \\
\sup F_C(x) & = & \max(\sup F_A(x), \sup T_B(x)),
\end{eqnarray}
for all $x$ in $X$.  
\end{definition}

\begin{example}
\label{chapter1:example5}
Let $A$ and $B$ be the interval neutrosophic sets defined in
Example~\ref{chapter1:example1}.
Then,
$A \setminus B = \langle [0.1,0.3],[0.7,0.9],[0.5,0.7] \rangle/x_1 + \\ 
\langle [0.5,0.7],
[0.6,0.8],
[0.2,0.3] \rangle/x_2 + 
\langle [0.3,0.4],[0.9,1.0],[0.4,0.6] \rangle/x_3$. \\
\end{example}

\begin{theorem}
$A \subseteq B \leftrightarrow \bar{B} \subseteq \bar{A}$
\end{theorem}

\begin{definition}[Addition]
The \emph{addition} of two interval neutrosophic sets $A$ and $B$
is an interval neutrosophic
set $C$, written as $C = A + B$, whose truth-membership,
indeterminacy-membership and falsity-membership functions are related to
those of $A$ and $B$ by
\begin{eqnarray}
\inf T_C(x) & = & \min(\inf T_A(x) + \inf T_B(x), 1), \\
\sup T_C(x) & = & \min(\sup T_A(x) + \sup T_B(x), 1), \\
\inf I_C(x) & = & \min(\inf I_A(x) + \inf I_B(x), 1), \\
\sup I_C(x) & = & \min(\sup I_A(x) + \sup I_B(x), 1), \\
\inf F_C(x) & = & \min(\inf F_A(x) + \inf F_B(x), 1), \\
\sup F_C(x) & = & \min(\sup F_A(x) + \sup F_B(x), 1),
\end{eqnarray}
for all $x$ in $X$.
\end{definition}  

\begin{example}
\label{chapter1:example6}
Let $A$ and $B$ be the interval neutrosophic sets defined in
Example~\ref{chapter1:example1}.
Then,
$A + B = \langle [0.7,1.0],[0.4,0.8],[0.4,0.8] \rangle/x_1 + \\
\langle [0.7,1.0],
[0.2,0.6],
[0.7,1.0] \rangle/x_2 + 
\langle [1.0,1.0],[0.2,0.4],[0.5,0.7] \rangle/x_3$. \\
\end{example}

\begin{definition}[Cartesian product]
The \emph{cartesian product} of two interval neutrosophic sets $A$ defined 
on universe $X_1$ and $B$ defined on universe $X_2$
is an interval neutrosophic
set $C$, written as $C = A \times B$, whose truth-membership,
indeterminacy-membership and falsity-membership functions are related to
those of $A$ and $B$ by
\begin{eqnarray}
\inf T_C(x,y) & = &\inf T_A(x) + \inf T_B(y) - \inf T_A(x) \cdot \inf T_B(y),\\
\sup T_C(x,y) & = &\sup T_A(x) + \sup T_B(y) - \sup T_A(x) \cdot \sup T_B(y),\\
\inf I_C(x,y) & = & \inf I_A(x) \cdot \sup I_B(y), \\
\sup I_C(x,y) & = & \sup I_A(x) \cdot \sup I_B(y), \\
\inf F_C(x,y) & = & \inf F_A(x) \cdot \inf F_B(y), \\
\sup F_C(x,y) & = & \sup F_A(x) \cdot \sup F_B(y),
\end{eqnarray}
for all $x$ in $X_1$, $y$ in $X_2$.
\end{definition}  

\begin{example}
\label{chapter1:example7}
Let $A$ and $B$ be the interval neutrosophic sets defined in
Example~\ref{chapter1:example1}.
Then,
$A \times B = \langle [0.6,0.82],[0.03,0.15],[0.03,0.15] \rangle/x_1 + \\
\langle [0.6,0.79],
[0,0.08],
[0.1,0.24] \rangle/x_2 + 
\langle [0.76,0.92],[0,0.03],[0.03,0.12] \rangle/x_3$. \\
\end{example}

\begin{definition}[Scalar multiplication]
The \emph{scalar multiplication} of interval neutrosophic set $A$ 
is an interval neutrosophic
set $B$, written as $B = a \cdot A $, whose truth-membership,
indeterminacy-membership and falsity-membership functions are related to
those of $A$ by
\begin{eqnarray}
\inf T_B(x) & = & \min(\inf T_A(x) \cdot a, 1), \\
\sup T_B(x) & = & \min(\sup T_A(x) \cdot a, 1), \\
\inf I_B(x) & = & \min(\inf I_A(x) \cdot a, 1), \\
\sup I_B(x) & = & \min(\sup I_A(x) \cdot a, 1), \\
\inf F_B(x) & = & \min(\inf F_A(x) \cdot a, 1), \\
\sup F_B(x) & = & \min(\sup F_A(x) \cdot a, 1),
\end{eqnarray}
for all $x$ in $X$, $a \in R^+$.
\end{definition}  

\begin{definition}[Scalar division]
The \emph{scalar division} of interval neutrosophic set $A$
is an interval neutrosophic
set $B$, written as $B = a \cdot A $, whose truth-membership,
indeterminacy-membership and falsity-membership functions are related to
those of $A$ by
\begin{eqnarray}
\inf T_B(x) & = & \min(\inf T_A(x) / a, 1), \\
\sup T_B(x) & = & \min(\sup T_A(x) / a, 1), \\
\inf I_B(x) & = & \min(\inf I_A(x) / a, 1), \\
\sup I_B(x) & = & \min(\sup I_A(x) / a, 1), \\
\inf F_B(x) & = & \min(\inf F_A(x) / a, 1), \\
\sup F_B(x) & = & \min(\sup F_A(x) / a, 1),
\end{eqnarray}
for all $x$ in $X$, $a \in R^+$.
\end{definition}             

Now we will define two operators: truth-favorite ($\triangle$) and 
false-favorite ($\nabla$) to remove the indeterminacy in the interval 
neutrosophic sets and transform it into interval valued intuitionistic
fuzzy sets or interval valued paraconsistent sets. These two operators are
unique on interval neutrosophic sets.

\begin{definition}[Truth-favorite]
The \emph{truth-favorite} of interval neutrosophic set $A$ is an interval 
neutrosophic set $B$, written as $B = \triangle A$, whose truth-membership
and falsity-membership functions are related to those of $A$ by
\begin{eqnarray}
\inf T_B(x) & = & \min(\inf T_A(x) + \inf I_A(x), 1), \\
\sup T_B(x) & = & \min(\sup T_A(x) + \sup I_A(x), 1), \\
\inf I_B(x) & = & 0, \\
\sup I_B(x) & = & 0, \\
\inf F_B(x) & = & \inf F_A(x), \\
\sup F_B(x) & = & \sup F_A(x),
\end{eqnarray}
for all $x$ in $X$.
\end{definition}

\begin{example}
\label{chapter1:example8}
Let $A$ be the interval neutrosophic set defined in
Example~\ref{chapter1:example1}.
Then,
$\triangle A  = \langle [0.5,0.9],[0,0],[0.3,0.5] \rangle/x_1 +
\langle [0.5,0.9],
[0,0],
[0.2,0.3] \rangle/x_2 + \\
\langle [0.8,1.0],[0,0],[0.2,0.3] \rangle/x_3$. \\
\end{example}

The purpose of truth-favorite operator is to evaluate the maximum of
degree of truth-membership of interval neutrosophic set.

\begin{definition}[False-favorite]
The \emph{false-favorite} of interval neutrosophic set $A$ is an interval 
neutrosophic set $B$, written as $B = \nabla A$, whose truth-membership
and falsity-membership functions are related to those of $A$ by
\begin{eqnarray}
\inf T_B(x) & = & \inf T_A(x), \\
\sup T_B(x) & = & \sup T_A(x), \\
\inf I_B(x) & = & 0, \\
\sup I_B(x) & = & 0, \\
\inf F_B(x) & = & \min(\inf F_A(x) + \inf I_A(x), 1), \\
\sup F_B(x) & = & \min(\sup F_A(x) + \sup I_A(x), 1),
\end{eqnarray}
for all $x$ in $X$.
\end{definition}

\begin{example}
\label{chapter1:example9}
Let $A$ be the interval neutrosophic set defined in
Example~\ref{chapter1:example1}.
Then,
$\nabla A  = \langle [0.2,0.4],[0,0],[0.6,1.0] \rangle/x_1 +
\langle [0.5,0.7],
[0,0],
[0.2,0.5] \rangle/x_2 + \\
\langle [0.6,0.8],[0,0],[0.4,0.6] \rangle/x_3$. \\
\end{example}

The purpose of false-favorite operator is to evaluate the maximum of
degree of false-membership of interval neutrosophic set.

\begin{theorem} 
For every two interval neutrosophic sets $A$ and $B$: \\  
\begin{enumerate}
\item $\triangle (A \cup B) \subseteq \triangle A \cup \triangle B$
\item $\triangle A \cap \triangle B \subseteq \triangle (A \cap B)$
\item $\nabla A \cup \nabla B \subseteq \nabla (A \cup B)$
\item $\nabla (A \cap B) \subseteq \nabla A \cap \nabla B$ 
\end{enumerate}
\end{theorem}

\section{Properties of Set-theoretic Operators}
\label{chapter1:section4}
In this section, we will give some properties of set-theoretic operators 
defined on interval neutrosophic sets as in Section~\ref{chapter1:section3}. The
proof of these properties is left for the readers.

\begin{property}[Commutativity]
$A \cup B = B \cup A$, $A \cap B = B \cap A$, $A + B = B + A$, $A \times B = B \times A$
\end{property}

\begin{property}[Associativity]
$A \cup (B \cup C) = (A \cup B) \cup C$, \\
$A \cap (B \cap C) = (A \cap B) \cap C$, \\
$A + (B + C) = (A + B) + C$, \\
$A \times (B \times C) = (A \times B) \times C$.
\end{property}

\begin{property}[Distributivity]
$A \cup (B \cap C) = (A \cup B) \cap (A \cup C)$,
$A \cap (B \cup C) = (A \cap B) \cup (A \cap C)$.
\end{property}

\begin{property}[Idempotency]
$A \cup A = A$, $A \cap A = A$, $\triangle \triangle A = \triangle A$, $\nabla \nabla A = \nabla A$.
\end{property}

\begin{property}
$A \cap \Phi = \Phi$, $A \cup X = X$, where $\inf T_{\Phi} = \sup T_{\Phi} = 0$,
$\inf I_{\Phi} = \sup I_{\Phi} = \inf F_{\Phi} = \sup F_{\Phi} = 1$ and 
$\inf T_{X} = \sup T_{X} = 1 $,
$\inf I_{X} = \sup I_{X} = \inf F_{X} = \sup F_{X} = 0$. 
\end{property}

\begin{property}
$\triangle (A + B) = \triangle A + \triangle B$, $\nabla (A + B) = \nabla A + \nabla B$.
\end{property}

\begin{property}
$A \cup \Psi = A$, $A \cap X = A$, where $\inf T_{\Phi} = \sup T_{\Phi} = 0 $,
$\inf I_{\Phi} = \sup I_{\Phi} = \inf F_{\Phi} = \sup F_{\Phi} = 1$ and 
$\inf T_{X} = \sup T_{X} = 1$,
$\inf I_{X} = \sup I_{X} = \inf F_{X} = \sup F_{X} = 0$.
\end{property}

\begin{property}[Absorption]
$A \cup (A \cap B) = A$, $A \cap (A \cup B) = A$
\end{property}

\begin{property}[DeMorgan's Laws]
$\overline{A \cup B} = \bar{A} \cap \bar{B}$, 
$\overline{A \cap B} = \bar{A} \cup \bar{B}$.
\end{property}

\begin{property}[Involution]
$\overline{{\overline{A}}} = A$
\end{property}

Here, we notice that by the definitions of complement, union and intersection
of interval neutrosophic set, interval neutrosophic set satisfies the
most properties of class set, fuzzy set and intuitionistic fuzzy set. Same as
fuzzy set and intuitionistic fuzzy set, it does not satisfy the principle of
middle exclude.
 
\section{Convexity of Interval Neutrosophic Set}
\label{chapter1:section5}
We assume that $X$ is a real Euclidean space $E^n$ for correctness.

\begin{definition}[Convexity]
An interval neutrosophic set $A$ is convex if and only if 
\begin{eqnarray}
\inf T_A(\lambda x_1 + (1 - \lambda) x_2) & \geq & \min(\inf T_A(x_1), \inf T_A(x_2)), \\
\sup T_A(\lambda x_1 + (1 - \lambda) x_2) & \geq & \min(\sup T_A(x_1), \sup T_A(
x_2)), \\
\inf I_A(\lambda x_1 + (1 - \lambda) x_2) & \leq & \max(\inf I_A(x_1), \inf I_A(x_2)), \\
\sup I_A(\lambda x_1 + (1 - \lambda) x_2) & \leq & \max(\sup I_A(x_1), \sup I_A(
x_2)), \\
\inf F_A(\lambda x_1 + (1 - \lambda) x_2) & \leq & \max(\inf F_A(x_1), \inf F_A(x_2)), \\
\sup F_A(\lambda x_1 + (1 - \lambda) x_2) & \leq & \max(\sup F_A(x_1), \sup F_A(
x_2)),
\end{eqnarray}
for all $x_1$ and $x_2$ in $X$ and all $\lambda$ in $[0, 1]$. 
\end{definition}
Fig.~\ref{fig1}
is an illustration of convex interval neutrosophic set.

\begin{theorem}
If $A$ and $B$ are convex, so is their intersection. 
\end{theorem}

\begin{definition}[Strongly Convex]
An interval neutrosophic set $A$ is \\ strongly  convex if for any two distinct
points $x_1$ and $x_2$, and any $\lambda$ in the \\ open interval $(0, 1)$, 
\begin{eqnarray}
\inf T_A(\lambda x_1 + (1 - \lambda) x_2) & > & \min(\inf T_A(x_1), \inf T_A(
x_2)), \\
\sup T_A(\lambda x_1 + (1 - \lambda) x_2) & > & \min(\sup T_A(x_1), \sup T_A(
x_2)), \\
\inf I_A(\lambda x_1 + (1 - \lambda) x_2) & < & \max(\inf I_A(x_1), \inf I_A(
x_2)), \\
\sup I_A(\lambda x_1 + (1 - \lambda) x_2) & < & \max(\sup I_A(x_1), \sup I_A(
x_2)), \\
\inf F_A(\lambda x_1 + (1 - \lambda) x_2) & < & \max(\inf F_A(x_1), \inf F_A(
x_2)), \\
\sup F_A(\lambda x_1 + (1 - \lambda) x_2) & < & \max(\sup F_A(x_1), \sup F_A(
x_2)),
\end{eqnarray}
for all $x_1$ and $x_2$ in $X$ and all $\lambda$ in $[0, 1]$. 
\end{definition}

\begin{theorem}
If $A$ and $B$ are strongly convex, so is their intersection.
\end{theorem}

\section{Conclusions}
\label{chapter1:section6}
In this chapter, we have presented an instance of neutrosophic set called
the interval neutrosophic set (INS). The interval neutrosophic set is 
a generalization of classic set, fuzzy set, interval valued fuzzy set,
intuitionistic fuzzy sets, interval valued intuitionistic fuzzy set,
interval type-2 fuzzy 
set~\cite{LM00} and paraconsistent set.
The notions of containment, complement, $N$-norm, $N$-conorm, relation, and 
composition have been defined on interval neutrosophic set. Various 
properties of set-theoretic operators have been proved. 
In the next chapter, we will discuss the interval neutrosophic logic and logic 
inference system based
on interval neutrosophic set.

\section{Appendix}
\setcounter{theorem}{0}

\begin{theorem}
\label{theorem0}
$A \cup B$ is the smallest interval neutrosophic set containing
both $A$ and $B$.
\end{theorem}
\begin{proof}
Let $C = A \cup B$. $\inf T_C = \max(\inf T_{A}, \inf T_{B})$, $\inf T_C \geq
\inf T_A$, $\inf T_{C} \geq \inf T_B$.
$\sup T_{C} = \max(\sup T_A, \sup T_B$,
$\sup T_C \geq \sup T_A$, $\sup T_C \geq \sup T_B$.
$\inf I_C = \min(\inf I_A, \inf I_B)$, $\inf I_C \leq \inf I_A$,
$\inf I_C \leq \inf I_B$, \\
$\sup I_C = \min(\sup I_A, \sup I_B)$, $\sup I_C \leq \sup I_A$,
$\sup I_C \leq \sup I_B$,
$\inf F_C = \min(\inf F_A, \inf F_B)$, $\inf F_{C} \leq \inf F_{A}$,
$\inf F_C \leq \inf F_B$. \\
$\sup F_{C} = \min(\sup F_A, \sup F_B)$, $\sup F_C \leq \sup F_A$,
$\sup F_C \leq  \sup F_B$. 
That means $C$ contains both $A$ and $B$. \\
Furthermore, if $D$ is any extended vague set containing both $A$ and $B$, then
$\inf T_{D} \geq \inf T_{A}$, $\inf T_{D} \geq \inf T_B$, so
$\inf T_{D} \geq \max(\inf T_A, \inf T_B) = \inf T_{C}$.
$\sup T_{D} \geq \sup T_{A}$,
$\sup T_{D} \geq \sup T_{B}$,
so $\sup T_{D} \geq \max(\sup T_{A}, \sup T_{B}) = \sup T_{C}$.
$\inf I_D \leq \inf I_A$, $\inf I_D \leq \inf I_B$, so
$\inf I_{D} \leq \min(\inf I_A, \inf I_B) = \inf I_C$.
$\sup I_D \leq \sup I_A$, $\sup I_D \leq \sup I_B$, so
$\sup I_D \leq \min(\sup I_A, \sup I_B) = \sup I_C$.
$\inf F_{D} \leq \inf F_{A}$,
$\inf F_{D} \leq \inf F_B$, so $\inf F_{D} \leq \min(\inf F_{A}, \inf F_{B})
= \inf F_{C}$.
$\sup F_{D} \leq \sup F_{A}$, $\sup F_{D} \leq \sup F_{B}$, so
$\sup F_{D} \leq \min(\sup F_{A}, \sup F_{B}) = \sup F_{C}$.
That implies $C \subseteq D$.
\end{proof}

\begin{theorem}
$A \cap B$ is the largest interval neutrosophic set contained in
both $A$ and $B$.
\end{theorem}
\begin{proof}
The proof is analogous to the proof of theorem~\ref{theorem0}.
\end{proof}

\begin{theorem}
Let $P$ be the power set of all interval neutrosophic sets defined in
the universe X. Then $\langle P;I_{N_1}, U_{N_1} \rangle$ is a
distributive lattice. 
\end{theorem}
\begin{proof}
Let $A, B, C$ be the arbitrary interval neutrosophic sets defined on $X$.
It is easy to verify that $A \cap A = A, A \cup A = A$ (idempotency), 
$A \cap B = B \cap A, A \cup B = B \cup A$ (commutativity), $(A \cap B) \cap C = A \cap (B \cap C), (A \cup B) \cup C = A \cup (B \cup C)$ (associativity), and
$A \cap (B \cup C) = (A \cap B) \cup (A \cap C), A \cup (B \cap C) = (A \cup B) \cap (A \cup C)$ (distributivity). 
\end{proof}

\begin{theorem}
$A \subseteq B \leftrightarrow \bar{B} \subseteq \bar{A}$
\end{theorem}
\begin{proof}
$A \subseteq B \Leftrightarrow \inf T_A \leq \inf T_B, \sup T_A \leq \sup T_B,
\inf I_A \geq \inf I_B, \sup I_A \geq \sup I_B, \inf F_A \geq \inf F_B,
\sup F_A \geq \sup F_B
 \Leftrightarrow
\inf F_{B} \leq \inf F_{A}, \sup F_{B} \leq \sup F_{A}, \\ 1 - \sup I_B \geq 1 -
 \sup I_A, 1 - \inf I_B \geq 1 - \inf I_A, \inf T_{B} \geq \inf T_{A},
\sup T_{B} \geq \sup T_{A} \Leftrightarrow \bar{B} \subseteq \bar{A}$.
\end{proof}

\begin{theorem}
For every two interval neutrosophic sets $A$ and $B$: \\
\begin{enumerate}
\item $\triangle (A \cup B) \subseteq \triangle A \cup \triangle B$
\item $\triangle A \cap \triangle B \subseteq \triangle (A \cap B)$
\item $\nabla A \cup \nabla B \subseteq \nabla (A \cup B)$
\item $\nabla (A \cap B) \subseteq \nabla A \cap \nabla B$
\end{enumerate}
\end{theorem}
\begin{proof}
We now prove the first identity. Let $C = A \cup B$.\\
$\inf T_C(x) = \max(\inf T_A(x), \inf T_B(x))$, \\
$\sup T_C(x) = \max(\sup T_A(x), \sup T_B(x))$, \\
$\inf I_C(x) = \min(\inf I_A(x), \inf I_B(x))$, \\
$\sup I_C(x) = \min(\sup I_A(x), \sup I_B(x))$, \\
$\inf F_C(x) = \min(\inf F_A(x), \inf F_B(x))$, \\
$\sup F_C(x) = \min(\sup F_A(x), \sup F_B(x))$. \\
$\inf T_{\triangle C}(x) = \min(\inf T_C(x) + \inf I_C(x), 1)$, \\
$\sup T_{\triangle C}(x) = \min(\sup T_C(x) + \sup T_C(x), 1)$, \\
$\inf I_{\triangle C}(x) = \sup I_{\triangle C}(x) = 0$, \\
$\inf F_{\triangle C}(x) = \inf I_C(x)$, \\
$\sup F_{\triangle C}(x) = \sup I_C(x)$. \\
$\inf T_{\triangle A}(x) = \min(\inf T_A(x) + \inf I_A(x), 1)$, \\
$\sup T_{\triangle A}(x) = \min(\sup T_A{x} + \sup I_A(x), 1)$, \\
$\inf I_{\triangle A}(x) = \sup I_{\triangle A}(x) = 0$, \\
$\inf F_{\triangle A}(x) = \inf I_A(x)$, \\
$\sup F_{\triangle A}(x) = \sup I_A(x)$. \\
$\inf T_{\triangle B}(x) = \min(\inf T_B(x) + \inf I_B(x), 1)$, \\
$\sup T_{\triangle B}(x) = \min(\sup T_B{x} + \sup I_B(x), 1)$, \\
$\inf I_{\triangle B}(x) = \sup I_{\triangle B}(x) = 0$, \\
$\inf F_{\triangle B}(x) = \inf I_B(x)$, \\
$\sup F_{\triangle B}(x) = \sup I_B(x)$. \\
$\inf T_{\triangle A \cup \triangle B}(x) = \max(\inf T_{\triangle A}(x), \inf T_{\triangle B}(x))$, \\
$\sup T_{\triangle A \cup \triangle B}(x) = \max(\sup T_{\triangle A}(x), \sup T
_{\triangle B}(x))$, \\
$\inf I_{\triangle A \cup \triangle B}(x) = \sup I_{\triangle A \cup \triangle B}(x) = 0$, \\
$\inf F_{\triangle A \cup \triangle B}(x) = \min(\inf F_{\triangle A}(x), \inf F_{\triangle B}(x))$, \\ 
$\sup F_{\triangle A \cup \triangle B}(x) = \min(\inf F_{\triangle A}(x), \inf F
_{\triangle B}(x))$. \\
Because, \\
$\inf T_{\triangle (A \cup B)} \leq \inf T_{\triangle A \cup \triangle B}$, \\
$\sup T_{\triangle (A \cup B)} \leq \sup T_{\triangle A \cup \triangle B}$, \\
$\inf I_{\triangle (A \cup B)} =  \inf T_{\triangle A \cup \triangle B} = 0$,\\
$\sup I_{\triangle (A \cup B)} =  \sup T_{\triangle A \cup \triangle B} = 0$,\\
$\inf F_{\triangle (A \cup B)} =  \inf F_{\triangle A \cup \triangle B}$,\\
$\sup F_{\triangle (A \cup B)} =  \sup T_{\triangle A \cup \triangle B}$,\\
so, $\triangle (A \cup B) \subseteq \triangle A \cup \triangle B$.
The other identities can be proved in a similar manner.
\end{proof}

\begin{theorem}
\label{theorem4}
If $A$ and $B$ are convex, so is their intersection.
\end{theorem}
\begin{proof}
Let $C = A \cap B$, then \\
$\inf T_C(\lambda x_1 + (1 - \lambda) x_2) \geq \min(\inf T_A(\lambda x_1 + (1 -
 \lambda) x_2), \inf T_B(\lambda x_1 + (1 - \lambda) x_2))$,
$\sup T_C(\lambda x_1 + (1 - \lambda) x_2) \geq \min(\sup T_A(\lambda x_1 + (1 -
 \lambda) x_2), \sup T_B(\lambda x_1 + (1 - \lambda) x_2))$,
$\inf I_C(\lambda x_1 + (1 - \lambda) x_2) \leq \max(\inf I_A(\lambda x_1 + (1 -
 \lambda) x_2), \inf I_B(\lambda x_1 + (1 - \lambda) x_2))$,
$\sup I_C(\lambda x_1 + (1 - \lambda) x_2) \leq \max(\sup I_A(\lambda x_1 + (1 -
 \lambda) x_2), \sup I_B(\lambda x_1 + (1 - \lambda) x_2))$,
$\inf F_C(\lambda x_1 + (1 - \lambda) x_2) \leq \max(\inf F_A(\lambda x_1 + (1 -
 \lambda) x_2), \inf F_B(\lambda x_1 + (1 - \lambda) x_2))$,
$\sup F_C(\lambda x_1 + (1 - \lambda) x_2) \leq \max(\inf F_A(\lambda x_1 + (1 -
 \lambda) x_2), \inf F_B(\lambda x_1 + (1 - \lambda) x_2))$,
Since $A$ and $B$ are convex:
$\inf T_A(\lambda x_1 + (1 - \lambda) x_2) \geq \min(\inf T_A(x1), \inf T_A(x2))
$,
$\sup T_A(\lambda x_1 + (1 - \lambda) x_2) \geq \min(\sup T_A(x1), \sup T_A(x2))
$,
$\inf I_A(\lambda x_1 + (1 - \lambda) x_2) \leq \max(\inf I_A(x1), \inf I_A(x2))
$, \
$\sup I_A(\lambda x_1 + (1 - \lambda) x_2) \leq \\ \max(\sup I_A(x1), \sup I_A(x
2))
$,
$\inf F_A(\lambda x_1 + (1 - \lambda) x_2) \leq \\ \max(\inf F_A(x1), \inf F_A(x
2))
$,
$\sup F_A(\lambda x_1 + (1 - \lambda) x_2) \leq \\ \max(\sup F_A(x1), \sup F_A(x
2))
$, \\
$\inf T_B(\lambda x_1 + (1 - \lambda) x_2) \geq \min(\inf T_B(x1), \inf T_A(x2))
$,
$\sup T_B(\lambda x_1 + (1 - \lambda) x_2) \geq \min(\sup T_B(x1), \sup T_A(x2))
$,
$\inf I_B(\lambda x_1 + (1 - \lambda) x_2) \leq \\ \max(\inf I_B(x1), \inf I_A(x
2))
$,
$\sup I_B(\lambda x_1 + (1 - \lambda) x_2) \leq \\ \max(\sup I_B(x1), \sup I_A(x
2))
$, \\
$\inf F_B(\lambda x_1 + (1 - \lambda) x_2) \leq \max(\inf F_B(x1), \inf F_A(x2))
$, \\
$\sup F_B(\lambda x_1 + (1 - \lambda) x_2) \leq \max(\sup F_B(x1), \sup F_A(x2))
$, \\
Hence, \\
$\inf T_C(\lambda x_1 + (1 - \lambda) x_2) \geq \min(\min(\inf T_A(x_1), \inf T_
A(x_2)) \\
, \min(\inf T_B(x_1), \inf T_B(x_2))) = \min(\min(\inf T_A(x_1), \inf T_B(x_1)),
\\ \min(\inf T_A(x_2), \inf T_B(x_2))) =  \min(\inf T_C(x_1), \inf T_C(x_2))$,
$\sup T_C(\lambda x_1 + (1 - \lambda) x_2) \geq \min(\min(\sup T_A(x_1), \sup T_
A(x_2))
, \\ \min(\sup T_B(x_1), \sup T_B(x_2))) = \min(\min(\sup T_A(x_1), \sup T_B(x_1
)), \\
\min(\sup T_A(x_2), \sup T_B(x_2))) =  \min(\sup T_C(x_1), \sup T_C(x_2))$, \\
$\inf I_C(\lambda x_1 + (1 - \lambda) x_2) \leq \max(\max(\inf I_A(x_1), \\ \inf
 I_
A(x_2))
, \max(\inf I_B(x_1), \inf I_B(x_2))) = \max(\max(\inf I_A(x_1), \\ \inf I_B(x_1
)),
\max(\inf I_A(x_2), \inf I_B(x_2))) = \max(\inf I_C(x_1), \inf I_C(x_2))$, \\
$\sup I_C(\lambda x_1 + (1 - \lambda) x_2) \leq \max(\max(\sup I_A(x_1), \sup I_
A(x_2)) \\
, \max(\sup I_B(x_1), \sup I_B(x_2))) = \max(\max(\sup I_A(x_1), \sup I_B(x_1)),
\\
\max(\sup I_A(x_2), \sup I_B(x_2))) = \max(\sup I_C(x_1), \sup I_C(x_2))$, \\
$\inf F_C(\lambda x_1 + (1 - \lambda) x_2) \leq \max(\max(\inf F_A(x_1), \inf F_
A(x_2))
, \\ \max(\inf F_B(x_1), \inf F_B(x_2))) = \max(\max(\inf F_A(x_1), \inf F_B(x_1
)), \\
\max(\inf F_A(x_2), \inf F_B(x_2))) = \\ \max(\inf F_C(x_1), \inf F_C(x_2))$,
$\sup F_C(\lambda x_1 + (1 - \lambda) x_2) \\ \leq \max(\max(\sup F_A(x_1), \sup
 F_
A(x_2))
, \\ \max(\sup F_B(x_1), \sup F_B(x_2))) = \max(\max(\sup F_A(x_1), \sup F_B(x_1
)), \\
\max(\sup F_A(x_2), \sup F_B(x_2))) = \max(\sup F_C(x_1), \sup F_C(x_2))$.
\end{proof}

\begin{theorem}
If $A$ and $B$ are strongly convex, so is their intersection.
\end{theorem}
\begin{proof}
The proof is analogous to the proof of Theorem~\ref{theorem4}.
\end{proof}

%% file: chaptr2.tex
\chapter{Interval Neutrosophic Logic} 
\label{INL} 

{\small
In this chapter, we present a novel interval neutrosophic logic that generalizes 
the  interval valued fuzzy logic, the intuitionistic 
fuzzy logic and paraconsistent logics which only consider truth-degree
or falsity-degree of a proposition. In the interval neutrosophic logic, we consider not
only truth-degree and falsity-degree but also
indeterminacy-degree which can reliably capture more information under
uncertainty.  
We introduce mathematical definitions
of an interval neutrosophic propositional calculus and an interval neutrosophic
predicate calculus. We propose a general method to design an interval
neutrosophic logic system which 
consists of
neutrosophication, neutrosophic inference, a neutrosophic rule base, 
neutrosophic
type reduction and deneutrosophication. A neutrosophic rule contains input
neutrosophic linguistic variables and output neutrosophic linguistic variables. A
neutrosophic linguistic variable has neutrosophic linguistic values which
defined by interval neutrosophic sets characterized by three membership 
functions: truth-membership, falsity-membership and indeterminacy-membership. 
The interval neutrosophic logic
can be applied to many potential real applications where information is 
imprecise, uncertain, incomplete and inconsistent
such as Web intelligence, medical informatics, bioinformatics, decision making,
etc. 
}

\section{Introduction}

The concept of fuzzy sets was introduced by Zadeh in 1965~\cite{ZAD65}. Since
then fuzzy sets and fuzzy logic have been applied to many real applications
to handle uncertainty. The traditional fuzzy set uses one real value 
$\mu_{A}(x) \in [0, 1]$ to represent the grade of membership of fuzzy set $A$
defined on universe $X$. The corresponding fuzzy logic associates each 
proposition
$p$ with a real value $\mu(p) \in [0, 1]$ which represents the degree of truth.
Sometimes $\mu_{A}(x)$ itself is uncertain and hard to be defined by a crisp
value. So the concept of interval valued fuzzy sets was proposed~\cite{TUR86}
to capture the uncertainty of grade of membership. The interval valued fuzzy 
set
uses an interval value $[\mu_{A}^L(x),\mu_{A}^U(x)]$ with 
$0 \leq \mu_{A}^L(x) \leq \mu_{A}^U(x) \leq 1$ to represent the grade of 
membership of fuzzy set. The traditional fuzzy logic can be easily extended
to the interval valued fuzzy logic. There are related works such as 
type-2 fuzzy sets and type-2
fuzzy logic~\cite{KM98,LM00,MJ02}. The family of fuzzy sets and fuzzy logic
can only handle ``complete" information that is if a grade of truth-membership
is $\mu_{A}(x)$ then a grade of false-membership is $1 - \mu_{A}(x)$ by default.
In some applications such as expert systems, decision making systems and 
information fusion systems, the information is both uncertain and incomplete. That
is beyond the scope of traditional fuzzy sets and fuzzy logic. 
In 1986, Atanassov
introduced the intuitionistic fuzzy set~\cite{ATA86} which is a generalization
of a fuzzy set and provably equivalent to an interval valued fuzzy set. The
intuitionistic fuzzy sets consider both truth-membership and false-membership.
The corresponding intuitionistic fuzzy logic~\cite{ATA88,ATA90,ATA98} 
associates each proposition $p$ with two real values $\mu(p)$-truth degree
and $\nu(p)$-falsity degree, respectively, where $\mu(p), \nu(p) \in [0, 1], \mu(p) + \nu(p) \leq 1$. So intuitionistic fuzzy sets and intuitionistic fuzzy
logic can handle uncertain and incomplete information.
 
However, inconsistent information exists in a lot of real situations such as those mentioned above. It is obvious that the intuitionistic fuzzy logic
cannot reason with inconsistency because it requires $\mu(p) + \nu(p) \leq 1$. 
Generally, two basic approaches are used to solve the inconsistency problem in knowledge bases: 
the belief revision and paraconsistent logics. The goal of the first approach
is to make an inconsistent theory consistent, either by revising it or by
representing it by a consistent semantics. On the other hand, the 
paraconsistent approach allows reasoning in the presence of inconsistency as contradictory information can be derived or introduced without 
trivialization~\cite{ACM02}. de Costa's ${\cal C}_w$ logic~\cite{COS77}
and Belnap's four-valued logic~\cite{BEL77} are two well-known paraconsistent
logics. 

Neutrosophy was introduced by Smarandache in 1995. 
``Neutrosophy is a branch
of philosophy which studies the origin, nature and scope of neutralities, as
well as their interactions with different ideational spectra"~\cite{SMA03}.
Neutrosophy includes neutrosophic probability, neutrosophic sets and 
neutrosophic logic. 
In a neutrosophic set (neutrosophic logic), indeterminacy is 
quantified explicitly and truth-membership (truth-degree), 
indeterminacy-membership (indeterminacy-degree) and false-membership 
(falsity-degree) are independent. The independence assumption is very important
 in a lot of applications such as information fusion when we try to combine
different data from different sensors. A neutrosophic set (neutrosophic logic) is 
different from an
intuitionistic fuzzy
set (intuitionistic fuzzy logic) where indeterminacy membership (indeterminacy-degree) is
$1 - \mu_{A}(x) - \nu_{A}(x)$ ($1 - \mu(p) - \nu(p)$) by default. 

The neutrosophic set generalizes the above mentioned sets from a philosophical
point of view. From a scientific or engineering point of view, the neutrosophic
set and set-theoretic operators need to be specified meaningfully. Otherwise, 
it will be
difficult to apply to the real applications. In chapter~\ref{INS} we discussed a 
special neutrosophic set called an interval neutrosophic set
and defined a set of set-theoretic operators. 
It is natural to define the interval
neutrosophic logic based on interval neutrosophic sets.
In this chapter, we give mathematical definitions of an interval neutrosophic 
propositional calculus and a first order interval neutrosophic predicate 
calculus.

The rest of this chapter is organized as follows. Section~\ref{INPC1} gives the mathematical 
definition of the interval neutrosophic propositional calculus. 
Section~\ref{INPC2} gives the mathematical definition of the first order 
interval 
neutrosophic
predicate calculus. Section~\ref{AAINL} provides one application example of
interval neutrosophic logic as the foundation for the design of interval
neutrosophic logic system.
In section~\ref{conclusion} we conclude the chapter and discuss the future research directions.

\section{Interval Neutrosophic Propositional Calculus}
\label{INPC1}
In this section, we introduce the elements of an interval neutrosophic
propositional calculus based on the definition of the
interval neutrosophic sets by using the notations from the theory of
classical propositional calculus~\cite{MEN87}.

\subsection{Syntax of Interval Neutrosophic Propositional Calculus}
\label{SINPC11}
Here we give the formalization of syntax of the interval neutrosophic
propositional calculus.

\begin{definition}
An \emph {alphabet} of the interval neutrosophic propositional calculus 
consists of three classes of symbols:
\begin{enumerate}
\item A set of \emph {interval neutrosophic propositional variables}, denoted 
by lower-case letters, sometimes indexed;
\item Five \emph {connectives} $\wedge, \vee, \neg, \rightarrow, \leftrightarrow$ which are called conjunction, disjunction, negation, implication, and
biimplication symbols respectively;
\item The parentheses ( and ).
\end{enumerate}
\hfill{\space}  $\Box$
\end{definition}

The alphabet of the interval neutrosophic propositional calculus has
combinations obtained by assembling connectives and interval neutrosophic
propositional variables in strings. The purpose of the construction rules is
to have the specification of distinguished combinations, called formulas.

\begin{definition}
The set of formulas (well-formed formulas) of interval neutrosophic propositional calculus is defined
as follows.
\begin{enumerate}
\item Every \emph {interval neutrosophic propositional variable} is a formula;
\item If $p$ is a formula, then so is $(\neg p)$;
\item If $p$ and $q$ are formulas, then so are
      \begin{enumerate}
        \item $(p \wedge q)$,
        \item $(p \vee q)$,
        \item $(p \rightarrow q)$, and
        \item $(p \leftrightarrow q)$.
      \end{enumerate}
\item No sequence of symbols is a formula which is not required to be by 1, 2,
      and 3. 
\end{enumerate}
\hfill{\space}   $\Box$
\end{definition}

To avoid having formulas cluttered with parentheses, we adopt the following
precedence hierarchy, with the highest precedence at the top:
\begin{center}
$\neg$, \\
$\wedge, \vee$, \\
$\rightarrow, \leftrightarrow$.
\end{center}

Here is an example of the interval neutrosophic propositional calculus formula:
\[
 \neg p_1 \wedge p_2 \vee (p_1 \rightarrow p_3) \rightarrow p_2 \wedge \neg p_3
\]

\begin{definition}
The \emph{language of interval neutrosophic propositional calculus} given by
an alphabet consists of the set of all formulas constructed from the symbols
of the alphabet.
\hfill{\space}  $\Box$
\end{definition}

\subsection{Semantics of Interval Neutrosophic Propositional Calculus}
\label{SINPC12}
The study of interval neutrosophic propositional calculus comprises, 
among others, a \emph {syntax}, which
has the distinction of well-formed formulas, and a \emph {semantics}, the
purpose of which is the assignment of a meaning to well-formed formulas.

To each interval neutrosophic proposition $p$, we associate it with an ordered
triple components $\langle t(p), i(p), f(p) \rangle$, 
where $t(p), i(p), f(p) \subseteq [0, 1]$. $t(p), i(p), f(p)$ is called
truth-degree, indeterminacy-degree and falsity-degree of $p$, respectively.
Let this assignment be provided by an \emph{interpretation function} or
\emph{interpretation} $INL$ 
defined over
a set of propositions $P$ in such a way that
\[
  INL(p) = \langle t(p), i(p), f(p) \rangle.
\]
Hence, the function $INL: P \rightarrow N$ gives the 
truth, indeterminacy and falsity degrees of all propositions in $P$.
We assume that the interpretation function $INL$ assigns to the logical truth
$T: INL(T) = \langle 1,0,0 \rangle$, and to 
$F: INL(F) = \langle 0,1,1 \rangle$.

An interpretation which makes a formula true is a \emph{model} of the formula.

Let $i, l$ be the subinterval of $[0,1]$. Then 
$i + l = [\inf i + \inf l, \sup i + \sup l]$, 
$i - l = [\inf i - \sup l, \sup i - \inf l]$,
$\max(i, l) = [\max(\inf i, \inf l), \max(\sup i, \sup l)]$,
$\min(i, l) = [\min(\inf i, \inf l), \min(\sup i, \sup l)]$.

The semantics of four interval neutrosophic propositional connectives is given
in Table I. Note that $p \leftrightarrow q$ if and only if $p \rightarrow q$ and $q \rightarrow p$. \\
\begin{table*}[hbtf]
\begin{center}
\label{table1}
\caption{Semantics of Four Connectives in Interval Neutrosophic Propositional Logic}

\begin{tabular}{c|c}
\hline
Connectives & Semantics \\
\hline
\hline
$INL(\neg p)$ & $\langle f(p), 1 - i(p), t(p) \rangle$ \\
\hline 
$INL(p \wedge q)$ &  
$\langle \min(t(p),t(q)), \max(i(p),i(q)), \max(f(p),f(q)) \rangle$ \\
\hline
$INL(p \vee q)$ & 
$\langle \max(t(p),t(q)), \min(i(p),i(q)), \min(f(p),f(q)) \rangle$ \\
\hline
$INL(p \rightarrow q)$ &
$\langle \min(1, 1-t(p)+t(q)), \max(0, i(q)-i(p)), \max(0, f(q)-f(p)) \rangle$ \\
\hline

\end{tabular}
\end{center}
\end{table*}

\begin{example} 
$INL(p) = \langle 0.5, 0.4, 0.7 \rangle$ and $INL(q) = \langle 1, 0.7, 0.2 \rangle$. Then, $INL(\neg p) = \langle 0.7, 0.6, 0.5 \rangle$, $INL(p \wedge \neg p) = \langle 0.5, 0.4, 0.7 \rangle$, $INL(p \vee q) = \langle 1, 0.7, 0.2 \rangle$, $INL(p \rightarrow q) = \langle 1, 1, 0 \rangle$. 
\hfill{\space} $\Box$
\end{example}

A given well-formed interval neutrosophic propositional formula will be called
a tautology (valid) if $INL(A) = \langle 1, 1, 0 \rangle$, for all 
interpretation
functions $INL$. It will be called a contradiction (inconsistent) if 
$INL(A) = \langle 0, 0, 1 \rangle$, for all interpretation functions $INL$.

\begin{definition}
Two formulas $p$ and $q$ are said to be \emph {equivalent}, denoted $p = q$,
if and only if the $INL(p) = INL(q)$ for every interpretation function $INL$.
\hfill{\space} $\Box$
\end{definition}

\begin{theorem}
\label{theorem2}
Let $F$ be the set of formulas and $\wedge$ be the meet and $\vee$ the join, 
then $\langle F; \wedge, \vee \rangle$ is a distributive lattice.
\end{theorem}
\begin{proof}
We leave the proof to the reader.
\end{proof}

\begin{theorem}
\label{theorem3}
If $p$ and $p \rightarrow q$ are tautologies, then $q$ is also a tautology.
\end{theorem}
\begin{proof}
Since $p$ and $p \rightarrow q$ are tautologies then for every $INL$,
$INL(p) = INL(p \rightarrow q) = \langle 1,0,0 \rangle$, that is \\
$t(p) = 1, i(p) = f(p) = 0$, 
$t(p \rightarrow q) = \min(1, 1 - t(p) + t(q)) = 1$,
$i(p \rightarrow q) = \max(0, i(q) - f(p)) = 0$, 
$f(p \rightarrow q) = \max(0, f(q) - f(p)) = 0$. Hence, \\
t(q) = 1, i(q) = f(q) = 0. So $q$ is a tautology. 
\end{proof}

\subsection{Proof Theory of Interval Neutrosophic Propositional Calculus}
\label{PTINPC1}
Here we give the proof theory for interval neutrosophic propositional logic
to complement the semantics part.

\begin{definition}
The interval neutrosophic propositional logic is defined by the following
axiom schema.
\begin{center}
$p \rightarrow (q \rightarrow p)$ \\
$p_1 \wedge \ldots \wedge p_m \rightarrow q_1 \vee \ldots q_n$ provided some $p_i$ is some $q_j$ \\ 
$p \rightarrow (q \rightarrow p \wedge q)$ \\
$(p \rightarrow r) \rightarrow ((q \rightarrow r) \rightarrow (p \vee q \rightarrow r))$ \\
$(p \vee q) \rightarrow r$ iff $p \rightarrow r$ and $q \rightarrow r$ \\
$p \rightarrow q$ iff $\neg q \rightarrow \neg p$ \\
$p \rightarrow q$ and $q \rightarrow r$ implies $p \rightarrow r$ \\
$p \rightarrow q$ iff $p \leftrightarrow (p \wedge q)$ iff $q \rightarrow (p \vee q)$
\end{center}
\hfill{\space} $\Box$
\end{definition}

The concept of (formal) deduction of a formula from a set of formulas, that is,
using the standard notation, $\Gamma \vdash p$, is defined as usual; 
in this case,
we say that $p$ is a syntactical consequence of the formulas in $T$.

\begin{theorem}
For interval neutrosophic propositional logic, we have
\begin{enumerate}
\item $\{p\} \vdash p$,
\item $\Gamma \vdash p$ entails $\Gamma \cup \Delta \vdash p$,
\item if $\Gamma \vdash p$ for any $p \in \Delta$ and $\Delta \vdash q$, then
$\Gamma \vdash q$. 
\end{enumerate}
\end{theorem}
\begin{proof}
It is immediate from the standard definition of the syntactical consequence $(\vdash)$.
\end{proof}

\begin{theorem}
In interval neutrosophic propositional logic, we have:
\begin{enumerate}
\item $\neg \neg p \leftrightarrow p$
\item $\neg (p \wedge q) \leftrightarrow \neg p \vee \neg q$
\item $\neg (p \vee q) \leftrightarrow \neg p \wedge \neg q$ 
\end{enumerate}
\end{theorem}
\begin{proof}
Proof is straight forward by following the semantics of interval neutrosophic
propositional logic.
\end{proof}

\begin{theorem}
In interval neutrosophic propositional logic, the following schema do not hold:\begin{enumerate}
\item $p \vee \neg p$
\item $\neg (p \wedge \neg p)$
\item $p \wedge \neg p \rightarrow q$
\item $p \wedge \neg p \rightarrow \neg q$
\item $\{p, p \rightarrow q\} \vdash q$
\item $\{p \rightarrow q, \neg q \} \vdash \neg p$
\item $\{p \vee q, \neg q \} \vdash p$
\item $\neg p \vee q \leftrightarrow p \rightarrow q$
\end{enumerate}
\end{theorem}
\begin{proof}
Immediate from the semantics of interval neutrosophic propositional logic.
\end{proof}

\begin{example}
To illustrate the use of the interval neutrosophic propositional consequence
relation, let's consider the following example. 
\[
   p \rightarrow (q \wedge r)
\]
\[
   r \rightarrow s
\]
\[
   q \rightarrow \neg s
\]
\[
   a
\]
From $p \rightarrow (q \wedge r)$, we get $p \rightarrow q$ and $p \rightarrow r$. From $p \rightarrow q$ and $q \rightarrow \neg s$, we get $p \rightarrow \neg s$. From $p \rightarrow r$ and $r \rightarrow s$, we get $p \rightarrow s$. 
Hence, $p$ is equivalent to $p \wedge s$ and $p \wedge \neg s$.
However, we cannot detach $s$ from $p$ nor $\neg s$ from $p$. This is in part
due to interval neutrosophic propositional logic incorporating neither modus
ponens nor and elimination.
\hfill{\space} $\Box$
\end{example}

\section{Interval Neutrosophic Predicate Calculus}
\label{INPC2}
In this section, we will extend our consideration to the full language of
first order interval neutrosophic predicate logic. 
First we give the
formalization of syntax of first order interval neutrosophic predicate logic as
in classical first-order predicate logic.

\subsection{Syntax of Interval Neutrosophic Predicate Calculus}
\label{SINPC21}

\begin{definition}
An \emph {alphabet} of the first order interval neutrosophic predicate calculus 
consists of seven classes of symbols:
\begin{enumerate}
\item \emph{variables}, denoted by lower-case letters, sometimes indexed;
\item \emph{constants}, denoted by lower-case letters;
\item \emph{function symbols}, denoted by lower-case letters, sometimes indexed;
\item \emph{predicate symbols}, denoted by lower-case letters, sometimes indexed;
\item Five \emph{connectives} $\wedge, \vee, \neg, \rightarrow, \leftrightarrow$ which are called the conjunction, disjunction, negation, implication, and 
biimplication symbols respectively;
\item Two \emph{quantifiers}, the \emph{universal quantifier} $\forall$ (for all) and the \emph{existential quantifier} $\exists$ (there exists);
\item The parentheses ( and ). 
\end{enumerate}
\hfill{\space} $\Box$
\end{definition}

To avoid having formulas cluttered with brackets, we adopt the following precedence hierarchy, with the highest precedence at the top:
\begin{center}
\[
  \neg, \forall, \exists
\]
\[
  \wedge, \vee
\]
\[
  \rightarrow, \leftrightarrow
\]
\end{center}

Next we turn to the definition of the first order interval neutrosophic 
language given by an alphabet.

\begin{definition}
A \emph{term} is defined as follows:
\begin{enumerate}
\item A variable is a term.
\item A constant is a term.
\item If $f$ is an $n$-ary function symbol and $t_1, \ldots, t_n$ are terms,
then $f(t_1, \ldots, f_n)$ is a term.
\end{enumerate}
\hfill{\space} $\Box$
\end{definition}

\begin{definition}
A \emph{(well-formed )formula} is defined inductively as follows:
\begin{enumerate}
\item If $p$ is an $n$-ary predicate symbol and $t_1, \ldots, t_n$ are terms,
then $p(t_1, \ldots, t_n)$ is a formula (called an \emph{atomic formula} or,
more simply, an \emph{atom}).
\item If $F$ and $G$ are formulas, then so are $(\neg F), (F \wedge G), (F \vee G), (F \rightarrow G)$ and $(F \leftrightarrow G)$.
\item If $F$ is a formula and $x$ is a variable, then $(\forall x F)$ and
$(\exists x F)$ are formulas.
\end{enumerate}
\hfill{\space} $\Box$
\end{definition}

\begin{definition}
The \emph{first order interval neutrosophic language} given by an alphabet
consists of the set of all formulas constructed from the symbols of the alphabet.
\hfill{\space} $\Box$
\end{definition}

\begin{example}
$\forall x \exists y (p(x,y) \rightarrow q(x)), \neg \exists x (p(x,a) \wedge q(x))$ are formulas.
\hfill{\space} $\Box$
\end{example} 

\begin{definition}
The \emph{scope} of $\forall x$ (resp. $\exists x$) in $\forall x F$ (resp. $\exists x F$) is $F$. A \emph{bound occurrence} of a variable in a formula is an
occurrence immediately following a quantifier or an occurrence within the 
scope of a quantifier, which has the same variable immediately after the 
quantifier. Any other occurrence of a variable is $\emph{free}$.
\hfill{\space} $\Box$
\end{definition}

\begin{example}
In the formula $\forall x p(x,y) \vee q(x)$, the first two occurrences of $x$
are bound, while the third occurrence is free, since the scope of $\forall x$
is $p(x,y)$ and $y$ is free.
\hfill{\space} $\Box$
\end{example}

\subsection{Semantics of Interval Neutrosophic Predicate Calculus}
\label{SINPC22}
In this section, we study the semantics of interval neutrosophic predicate
calculus, the purpose of which is the assignment of a meaning to well-formed
formulas.
In the interval neutrosophic propositional logic, an interpretation is an
assignment of truth values (ordered triple component) to propositions. 
In the first order interval neutrosophic predicate logic, since there are
variables involved, we have to do more than that. To define an interpretation
for a well-formed formula in this logic, we have to specify two things, the 
domain and an assignment to constants and predicate symbols occurring in the
formula. The following is the formal definition of an interpretation of a
formula in the first order interval neutrosophic predicate logic.

\begin{definition}
An \emph{interpretation function (or interpretation)} of a formula $F$ in the
first order interval neutrosophic predicate logic consists of a nonempty domain
$D$, and an assignment of ``values" to each constant and predicate symbol
occurring in $F$ as follows:
\begin{enumerate}
\item To each constant, we assign an element in $D$.
\item To each $n$-ary function symbol, we assign a mapping from $D^n$ to $D$.
(Note that $D^n = \{(x_1, \ldots, x_n) | x_1 \in D, \ldots, x_n \in D \}$).
\item Predicate symbols get their meaning through evaluation functions $E$ 
which 
assign to each variable $x$ an element $E(x) \in D$. 
To each $n$-ary predicate symbol $p$, there is a function 
$INP(p): D^n \rightarrow N$. So 
$INP(p(x_1, \ldots, x_n)) = INP(p)(E(x_1), \ldots, E(x_n))$. 
\end{enumerate}
\hfill{\space}  $\Box$
\end{definition}

That is, $INP(p)(a_1, \ldots, a_n) = \langle t(p(a_1, \ldots, a_n)), i(p(a_1, \ldots, a_n)), f(p(a_1, \ldots, a_n))$, \\ 
where $t(p(a_1, \ldots, a_n)), i(p(a_1, \ldots, a_n)), f(p(a_1, \ldots, a_n)) \subseteq [0,1]$. They are called truth-degree, indeterminacy-degree and falsity-degree of $p(a_1, \ldots, a_n)$ respectively. We assume that the interpretation function $INP$ assigns to the logical truth $T: INP(T) = \langle 1, 1, 0 \rangle$, and to $F: INP(F) = \langle 0, 0, 1 \rangle$. 

The semantics of four interval
neutrosophic predicate connectives and two quantifiers is given in Table II. For simplification of notation, we use $p$ to denote $p(a_1, \ldots, a_i)$. Note that $p \leftrightarrow q$ if and only if $p \rightarrow q$ and $q \rightarrow p$.

\begin{table*}[hbtf]
\begin{center}
\label{table2}
\caption{Semantics of Four Connectives and Two Quantifiers in Interval Neutrosophic Predicate Logic}

\begin{tabular}{c|c}
\hline
Connectives & Semantics \\
\hline
\hline
$INP(\neg p)$ & $\langle f(p), 1 - i(p), t(p) \rangle$ \\
\hline
$INP(p \wedge q)$ &
$\langle \min(t(p),t(q)), \max(i(p),i(q)), \max(f(p),f(q)) \rangle$ \\
\hline
$INP(p \vee q)$ &
$\langle \max(t(p),t(q)), \min(i(p),i(q)), \min(f(p),f(q)) \rangle$ \\
\hline
$INP(p \rightarrow q)$ &
$\langle \min(1, 1-t(p)+t(q)), \max(0, i(q)-i(p)), \max(0, f(q)-f(p)) \rangle$ \\
\hline
$INP(\forall x F)$ & $\langle \min t(F(E(x))), \min i(F(E(x))), \max f(F(E(x))) \rangle$, $E(x) \in D$ \\
\hline 
$INP(\exists x F)$ & $\langle \max t(F(E(x))), \max i(F(E(x))), \min f(F(E(x))) \rangle$, $E(x) \in D$ \\
\hline 

\end{tabular}
\end{center}
\end{table*}

\begin{example}
Let $D = {1,2,3}$ and $p(1) = \langle 0.5,1,0.4 \rangle, p(2) = \langle 1,0.2,0 \rangle, p(3) = \langle 0.7,0.4,0.7 \rangle$. 
Then $INP(\forall x p(x)) = \langle 0.5,0.2,0.7 \rangle$, 
and $INP(\exists x p(x)) = \langle 1,1,0 \rangle$. 
\hfill{\space} $\Box$
\end{example}

\begin{definition}
A formula $F$ is \emph{consistent (satisfiable)} if and only if there exists
an interpretation $I$ such that $F$ is evaluated to $\langle 1,1,0 \rangle$ in 
$I$. If a formula $F$ is $T$ in an interpretation $I$, we say that $I$ is a 
\emph{model} of $F$ and $I$ \emph{satisfies} $F$.
\hfill{\space}  $\Box$
\end{definition}

\begin{definition}
A formula $F$ is \emph{inconsistent (unsatisfiable)} if and only if there 
exists no interpretation that satisfies $F$.
\hfill{\space} $\Box$
\end{definition}

\begin{definition}
A formula $F$ is \emph{valid} if and only if every interpretation of $F$ satisfies $F$.
\hfill{\space} $\Box$
\end{definition}

\begin{definition}
A formula $F$ is a \emph{logical consequence} of formulas $F_1, \ldots, F_n$ if
and only if for every interpretation $I$, if $F_1 \wedge \ldots \wedge F_n$ is
true in $I$, $F$ is also true in $I$.
\hfill{\space}  $\Box$
\end{definition}

\begin{example}
$(\forall x)(p(x) \rightarrow (\exists y)p(y))$ is valid, $(\forall x)p(x) \wedge (\exists y) \neg p(y)$ is consistent.
\hfill{\space}  $\Box$
\end{example}

\begin{theorem}
There is no inconsistent formula in the first order interval neutrosophic predicate logic.
\end{theorem}
\begin{proof}
It is direct from the definition of semantics of interval neutrosophic predicate logic.
\end{proof}

Note that the first order interval neutrosophic predicate logic can be considered as an extension of the interval neutrosophic propositional logic. When a
formula in the first order logic contains no variables and quantifiers, it
can be treated just as a formula in the propositional logic.

\subsection{Proof Theory of Interval Neutrosophic Predicate Calculus}
\label{PTINPC2}
In this part, we give the proof theory for first order interval neutrosophic
predicate logic to complement the semantics part.

\begin{definition}
The first order interval neutrosophic predicate logic is defined by the following axiom schema.
\begin{center}
$(p \rightarrow q(x)) \rightarrow (p \rightarrow \forall x q(x))$ \\
$\forall x p(x) \rightarrow p(a)$ \\
$p(x) \rightarrow \exists x p(x)$ \\
$(p(x) \rightarrow q) \rightarrow (\exists x p(x) \rightarrow q)$ 
\end{center}
\hfill{\space} $\Box$
\end{definition}

\begin{theorem}
In the first order interval neutrosophic predicate logic, we have:
\begin{enumerate}
\item $p(x) \vdash \forall x p(x)$
\item $p(a) \vdash \exists x p(x)$
\item $\forall x p(x) \vdash p(y)$
\item $\Gamma \cup \{p(x)\} \vdash q$, then $\Gamma \cup \{\exists x p(x)\} \vdash q$
\end{enumerate}
\end{theorem}
\begin{proof}
Directly from the definition of the semantics of first order interval 
neutrosophic predicate logic. 
\end{proof}

\begin{theorem}
In the first order interval neutrosophic predicate logic, the following 
schemes are valid, where $r$ is a formula in which $x$ does not appear free:
\begin{enumerate}
\item $\forall x r \leftrightarrow r$
\item $\exists x r \leftrightarrow r$
\item $\forall x \forall y p(x,y) \leftrightarrow \forall y \forall x p(x,y)$
\item $\exists x \exists y p(x,y) \leftrightarrow \exists y \exists x p(x,y)$
\item $\forall x \forall y p(x,y) \rightarrow \forall x p(x,x)$
\item $\exists x p(x,x) \rightarrow \exists x \exists y p(x,y)$
\item $\forall x p(x) \rightarrow \exists x p(x)$
\item $\exists x \forall y p(x,y) \rightarrow \forall y \exists x p(x,y)$
\item $\forall x (p(x) \wedge q(x)) \leftrightarrow \forall x p(x) \wedge \forall x q(x)$
\item $\exists x (p(x) \vee q(x)) \leftrightarrow \exists x p(x) \vee \exists x q(x)$
\item $p \wedge \forall x q(x) \leftrightarrow \forall x (p \wedge q(x))$
\item $p \vee \forall x q(x) \leftrightarrow \forall x (p \vee q(x))$
\item $p \wedge \exists x q(x) \leftrightarrow \exists x (p \wedge q(x))$
\item $p \vee \exists x q(x) \leftrightarrow \exists x (p \vee q(x))$

\item $\forall x (p(x) \rightarrow q(x)) \rightarrow (\forall x p(x) \rightarrow \forall x q(x))$
\item $\forall x (p(x) \rightarrow q(x)) \rightarrow (\exists x p(x) \rightarrow \exists x q(x))$ 
\item $\exists x (p(x) \wedge q(x)) \rightarrow \exists x p(x) \wedge \exists x q(x)$
\item $\forall x p(x) \vee \forall x q(x) \rightarrow \forall x (p(x) \vee q(x))$
\item $\neg \exists x \neg p(x) \leftrightarrow \forall x p(x)$
\item $\neg \forall x \neg p(x) \leftrightarrow \exists p(x)$
\item $\neg \exists x p(x) \leftrightarrow \forall x \neg p(x)$
\item $\exists x \neg p(x) \leftrightarrow \neg \forall x p(x)$
\end{enumerate}
\end{theorem}
\begin{proof}
It is straightforward from the definition of the semantics and axiomatic 
schema of first order interval neutrosophic predicate logic.
\end{proof}

\section{An Application of Interval Neutrosophic Logic}
\label{AAINL}
In this section we provide one practical application of the interval neutrosophic
logic -- Interval Neutrosophic Logic System (INLS). 
INLS can handle rule 
uncertainty as same as type-2 FLS~\cite{LM00}, besides, it can handle rule
inconsistency without the danger of trivialization. Like the classical FLS,
INLS is also characterized by IF-THEN rules. INLS consists of 
neutrosophication, neutrosophic inference, a neutrosophic rule base, 
neutrosophic type reduction and deneutrosophication. Given an input vector 
$x = (x_1, \ldots, x_n)$, where $x_1, \ldots, x_n$ can be crisp inputs or 
neutrosophic sets, the INLS
will generate a crisp output $y$. The general scheme of INLS
is shown in Fig. 2.1.

\begin{figure}[htbp]
	\centering
		\includegraphics{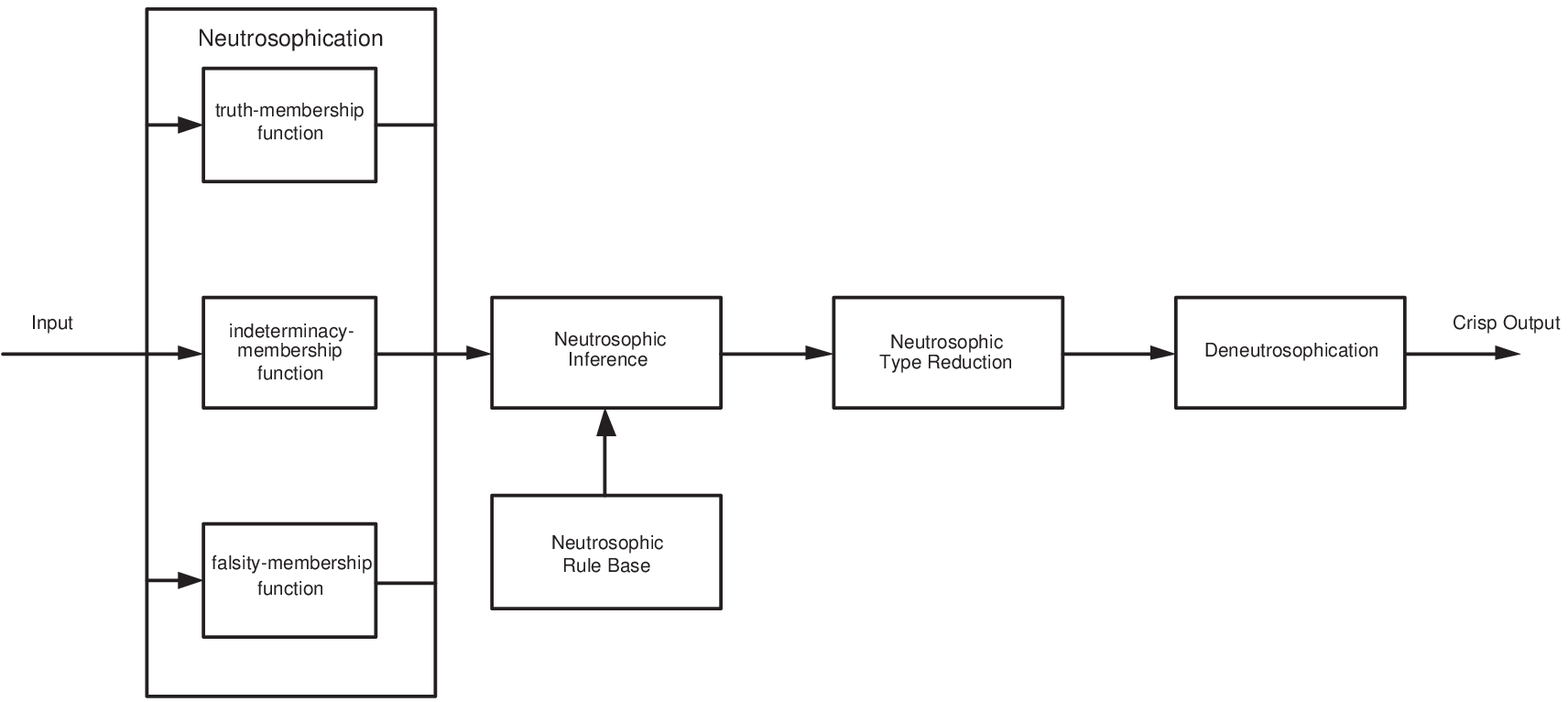}
	\caption{General Scheme of an INLS}
	\label{fig:nls}
\end{figure}

Suppose the neutrosophic rule base consists of $M$ rules in which each rule
has $n$ antecedents and one consequent. Let the $k$th rule be denoted 
by $R^k$ such that IF 
$x_1$ is $A_{1}^k$, $x_2$ is $A_{2}^k$, $\ldots$, and $x_n$ is $A_{n}^k$, THEN $y$
is $B^k$. 
$A_{i}^k$ is an interval neutrosophic set defined on universe $X_i$ with 
truth-membership
function $T_{A_{i}^k}(x_i)$, indeterminacy-membership function 
$I_{A_{i}^k}(x_i)$ and falsity-membership function $F_{A_{i}^k}(x_i)$, where
$T_{A_{i}^k}(x_i), I_{A_{i}^k}(x_i), F_{A_{i}^k}(x_i) \subseteq [0, 1], 1 \leq i \leq n$. 
$B^k$ is
an interval neutrosophic set defined on universe $Y$ with truth-membership
function $T_{B^k}(y)$,
indeterminacy-membership function $I_{B^k}(y)$ and falsity-membership function \\
$F_{B^k}(y)$, where $T_{B^k}(y), I_{B^k}(y), F_{B^k}(y) \subseteq [0, 1]$.
Given fact $x_1$ is $\tilde{A}_{1}^k, x_2$ is $\tilde{A}_{2}^k, \ldots
$,
and $x_n$ is $\tilde{A}_{n}^k$, then consequence $y$ is $\tilde{B}^k$.
$\tilde{A}_{i}^k$ is an interval neutrosophic set defined on universe
$X_i$ with 
truth-membership function $T_{\tilde{A}_{i}^k}(x_i)$, indeterminacy-membership
function $I_{\tilde{A}_{i}^k}(x_i)$ and falsity-membership function 
$F_{\tilde{A}_{i}^k}(x_i)$, where $T_{\tilde{A}_{i}^k}(x_i), I_{\tilde{A}_{i}^k}(x_i), F_{\tilde{A}_{i}^k}(x_i) \subseteq [0, 1], 1 \leq i \leq n$. 
$\tilde{B}^k$ is an interval neutrosophic set defined on universe $Y$ with 
truth-membership function
$T_{\tilde{B}^k}(y)$, indeterminacy-membership function $I_{\tilde{B}^k}(y)$
and falsity-membership function \\ $F_{\tilde{B}^k}(y)$, where $T_{\tilde{B}^k}(y), I_{\tilde{B}^k}(y), F_{\tilde{B}^k}(y) \subseteq [0, 1]$. 
In this chapter, we consider $a_i \leq x_i \leq b_i$ and $\alpha \leq y \leq \beta$. 

An unconditional neutrosophic proposition is expressed by the phrase: 
``$Z$ is $C$", where $Z$ is a variable that receives values $z$ from a universal set $U$, and $C$ is an interval neutrosophic set defined on $U$ that represents a neutrosophic predicate. Each neutrosophic proposition $p$ is associated
with $\langle t(p), i(p), f(p) \rangle$ with $t(p), i(p), f(p) \subseteq [0, 1]$. In general, for any value $z$ of $Z$, $\langle t(p), i(p), f(p) \rangle = \langle T_{C}(z), I_{C}(z), F_{C}(z) \rangle$.
 		
For implication $p \rightarrow q$, we define the semantics as:
\begin{eqnarray}
\sup t_{p \rightarrow q} &=& \min (\sup t(p), \sup t(q)), \\
\inf t_{p \rightarrow q} &=& \min (\inf t(p), \inf t(q)), \\
\sup i_{p \rightarrow q} &=& \max (\sup i(p), \sup i(q)), \\
\inf i_{p \rightarrow q} &=& \max (\inf i(p), \inf i(q)), \\
\sup f_{p \rightarrow q} &=& \max (\sup f(p), \sup f(q)), \\
\inf f_{p \rightarrow q} &=& \max (\inf f(p), \inf f(q)),
\end{eqnarray}
where $t(p),i(p),f(p),t(q),i(q),f(q) \subseteq [0,1]$. 

Let $X = X_1 \times \cdots \times X_n$. 
The truth-membership function, indeterminacy-membership function and 
falsity-membership function $T_{\tilde{B}^k}(y), I_{\tilde{B}^k}(y), F_{\tilde{B}^k}(y)$ of a fired $k$th rule can be represented using the definition of 
interval neutrosophic composition functions (1.44--1.46) and the 
semantics of conjunction and disjunction defined in Table 2.2 and 
equations (2.1--2.6) as:
\begin{eqnarray}
\sup T_{\tilde{B}^k}(y) &=& \sup_{x \in X}\min(\sup T_{\tilde{A}_{1}^k}(x_1), \sup T_{A_{1}^k}(x_1), \ldots, \sup T_{\tilde{A}_{n}^k}(x_n), \sup T_{A_{n}^k}(x_n), \sup T_{B^k}(y)), \\ 
\inf T_{\tilde{B}^k}(y) &=& \sup_{x \in X}\min(\inf T_{\tilde{A}_{1}^k}(x_1), \inf T_{A_{1}^k}(x_1), \ldots, \inf T_{\tilde{A}_{n}^k}(x_n), \inf T_{A_{n}^k}(x_n), \inf T_{B^k}(y)), \\
\sup I_{\tilde{B}^k}(y) &=& \sup_{x \in X}\max(\sup I_{\tilde{A}_{1}^k}(x_1), \sup I_{A_{1}^k}(x_1), \ldots, \sup I_{\tilde{A}_{n}^k}(x_n), \sup I_{A_{n}^k}(x_n), \sup I_{B^k}(y)), \\ 
\inf I_{\tilde{B}^k}(y) &=& \sup_{x \in X}\max(\inf I_{\tilde{A}_{1}^k}(x_1), \inf I_{A_{1}^k}(x_1), \ldots, \inf I_{\tilde{A}_{n}^k}(x_n), \inf I_{A_{n}^k}(x_n), \inf I_{B^k}(y)), \\
\sup F_{\tilde{B}^k}(y) &=& \inf_{x \in X}\max(\sup F_{\tilde{A}_{1}^k}(x_1), \sup F_{A_{1}^k}(x_1), \ldots, \sup F_{\tilde{A}_{n}^k}(x_n), \sup F_{A_{n}^k}(x_n), \sup F_{B^k}(y)), \\
\inf F_{\tilde{B}^k}(y) &=& \inf_{x \in X}\max(\inf F_{\tilde{A}_{1}^k}(x_1), \inf F_{A_{1}^k}(x_1), \ldots, \inf F_{\tilde{A}_{n}^k}(x_n), \inf F_{A_{n}^k}(x_n), \inf F_{B^k}(y)),
\end{eqnarray}
where $y \in Y$.

Now, we give the algorithmic description of INLS. 

BEGIN

Step 1: Neutrosophication

The purpose of neutrosophication is to map inputs into interval 
neutrosophic input sets. 
Let $G_{i}{^k}$ be an interval neutrosophic input set to represent the result of
neutrosophication of $i$th input variable of $k$th rule, then
\begin{eqnarray}
\sup T_{G_{i}{^k}}(x_i) &=& \sup_{x_i \in X_i}\min(\sup T_{\tilde{A}_{i}^k}(x_i), \sup T_{A_{i}^k}(x_i)), \\ 
\inf T_{G_{i}^k}(x_i) &=& \sup_{x_i \in X_i}\min(\inf T_{\tilde{A}_{i}^k}(x_i), \inf T_{A_{i}^k}(x_i)), \\ 
\sup I_{G_{i}^k}(x_i) &=& \sup_{x_i \in X_i}\max(\sup I_{\tilde{A}_{i}^k}(x_i), \sup I_{A_{i}^k}(x_i)), \\
\inf I_{G_{i}^k}(x_i) &=& \sup_{x_i \in X_i}\max(\inf I_{\tilde{A}_{i}^k}(x_i), \inf I_{A_{i}^k}(x_i)), \\
\sup F_{G_{i}^k}(x_i) &=& \inf_{x_i \in X_i}\max(\sup F_{\tilde{A}_{i}^k}(x_i), \sup F_{A_{i}^k}(x_i)), \\
\inf F_{G_{i}^k}(x_i) &=& \inf_{x_i \in X_i}\max(\inf F_{\tilde{A}_{i}^k}(x_i), \inf F_{A_{i}^k}(x_i)),
\end{eqnarray}
where $x_i \in X_i$.

If $x_i$ are crisp inputs, then equations (50--55) are simplified 
to
\begin{eqnarray}
\sup T_{G_{i}^k}(x_i) &=& 
\sup T_{A_{i}^k}(x_i)
, \\
\inf T_{G_{i}^k}(x_i) &=&  
\inf T_{A_{i}^k}(x_i)
, \\
\sup I_{G_{i}^k}(x_i) &=& 
\sup I_{A_{i}^k}(x_i)
, \\
\inf I_{G_{i}^k}(x_i) &=& 
\inf I_{A_{i}^k}(x_i)
, \\
\sup F_{G_{i}^k}(x_i) &=& 
\sup F_{A_{i}^k}(x_i)
, \\
\inf F_{G_{i}^k}(x_i) &=& 
\inf F_{A_{i}^k}(x_i)
,
\end{eqnarray}
where $x_i \in X_i$.

Fig. 2 shows the conceptual diagram for neutrosophication of a crisp input $x_1$. 

\begin{figure}[htb!]
\begin{center}
\includegraphics[scale=0.6]{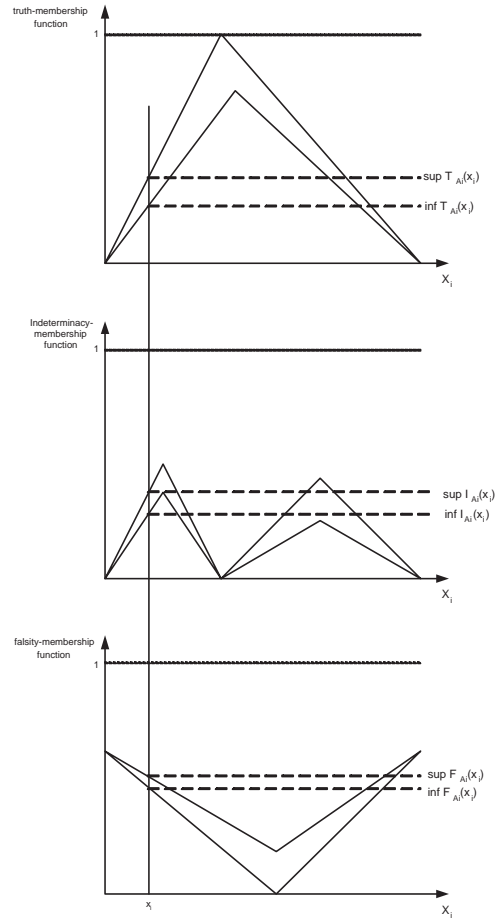}
\end{center}
\caption{Conceptual Diagram for Neutrosophication of Crisp Input}
\label{figure2}
\end{figure}

Step 2: Neutrosophic Inference

The core of INLS is the neutrosophic inference, the principle of which has
already been explained above. Suppose the $k$th rule is fired. 
Let $G^k$ be an interval neutrosophic set to represent the result of
the input and antecedent operation for $k$th rule, then
\begin{eqnarray}
\sup T_{G^k}(x) &=& \sup_{x \in X}\min(\sup T_{\tilde{A}_{1}^k}(x_1), \sup T_{A_
{1}^k}(x_1), \ldots, \sup T_{\tilde{A}_{n}^k}(x_n), \sup T_{A_{n}^k}(x_n)), \\
\inf T_{G^k}(x) &=& \sup_{x \in X}\min(\inf T_{\tilde{A}_{1}^k}(x_1), \inf T_{A_
{1}^k}(x_1), \ldots, \inf T_{\tilde{A}_{n}^k}(x_n), \inf T_{A_{n}^k}(x_n)), \\
\sup I_{G^k}(x) &=& \sup_{x \in X}\max(\sup I_{\tilde{A}_{1}^k}(x_1), \sup I_{A_
{1}^k}(x_1), \ldots, \sup I_{\tilde{A}_{n}^k}(x_n), \sup I_{A_{n}^k}(x_n)), \\
\inf I_{G^k}(x) &=& \sup_{x \in X}\max(\inf I_{\tilde{A}_{1}^k}(x_1), \inf I_{A_
{1}^k}(x_1), \ldots, \inf I_{\tilde{A}_{n}^k}(x_n), \inf I_{A_{n}^k}(x_n)), \\
\sup F_{G^k}(x) &=& \inf_{x \in X}\max(\sup F_{\tilde{A}_{1}^k}(x_1), \sup F_{A_
{1}^k}(x_1), \ldots, \sup F_{\tilde{A}_{n}^k}(x_n), \sup F_{A_{n}^k}(x_n)), \\
\inf F_{G^k}(x) &=& \inf_{x \in X}\max(\inf F_{\tilde{A}_{1}^k}(x_1), \inf F_{A_
{1}^k}(x_1), \ldots, \inf F_{\tilde{A}_{n}^k}(x_n), \inf F_{A_{n}^k}(x_n)),
\end{eqnarray}
where $x_i \in X_i$.

Here we 
restate the result of neutrosophic inference: 

\begin{eqnarray}
\sup T_{\tilde{B}^k}(y) &=& 
\min(\sup T_{G_{1}^k}(x), \sup T_{B^k}(y)), \\
\inf T_{\tilde{B}^k}(y) &=& 
\min(\inf T_{G^k}(x), \inf T_{B^k}(y)), \\
\sup I_{\tilde{B}^k}(y) &=& 
\max(\sup I_{G^k}(x), \sup I_{B^k}(y)), \\
\inf I_{\tilde{B}^k}(y) &=& 
\max(\inf I_{G^k}(x), \inf I_{B^k}(y)), \\
\sup F_{\tilde{B}^k}(y) &=& 
\max(\sup F_{G^k}(x), \sup F_{B^k}(y)), \\
\inf F_{\tilde{B}^k}(y) &=& 
\max(\inf F_{G^k}(x), \inf F_{B^k}(y)),
\end{eqnarray}
where $x \in X, y \in Y$.

Suppose that $N$ rules in the neutrosophic rule base are fired, where $N \leq M$, then,
the output interval neutrosophic set $\tilde{B}$ is:
\begin{eqnarray}
\sup T_{\tilde{B}}(y) &=& \max_{k=1}^{N} \sup T_{\tilde{B}^k}(y), \\
\inf T_{\tilde{B}}(y) &=& \max_{k=1}^{N} \inf T_{\tilde{B}^k}(y), \\
\sup I_{\tilde{B}}(y) &=& \min_{k=1}^{N} \sup I_{\tilde{B}^k}(y), \\
\inf I_{\tilde{B}}(y) &=& \min_{k=1}^{N} \inf I_{\tilde{B}^k}(y), \\
\sup F_{\tilde{B}}(y) &=& \min_{k=1}^{N} \sup T_{\tilde{B}^k}(y), \\
\inf T_{\tilde{B}}(y) &=& \min_{k=1}^{N} \inf T_{\tilde{B}^k}(y),
\end{eqnarray}
where $y \in Y$.

Step 3: Neutrosophic type reduction

After neutrosophic inference, we will get an interval neutrosophic set $\tilde{B}$ with $T_{\tilde{B}}(y), I_{\tilde{B}}(y), F_{\tilde{B}}(y) \subseteq [0, 1]$. Then, we do the neutrosophic type reduction to transform each interval 
into one number. There are many ways to do it, here, we give one method:
\begin{eqnarray}
T_{\tilde{B}}^{'}(y) &=& (\inf T_{\tilde{B}}(y) + \sup T_{\tilde{B}}(y)) / 2, \\
I_{\tilde{B}}^{'}(y) &=& (\inf I_{\tilde{B}}(y) + \sup I_{\tilde{B}}(y)) / 2, \\
F_{\tilde{B}}^{'}(y) &=& (\inf F_{\tilde{B}}(y) + \sup F_{\tilde{B}}(y)) / 2,
\end{eqnarray} 
where $y \in Y$.

So, after neutrosophic type reduction, we will get an ordinary neutrosophic set
(a type-1 neutrosophic set)
$\tilde{B}$. Then we need to do the deneutrosophication to get a crisp output.

Step 4: Deneutrosophication

The purpose of 
deneutrosophication is to convert an ordinary neutrosophic set (a type-1 neutrosophic set) obtained by
neutrosophic type reduction to a single real number which represents the
real output. 
Similar to defuzzification~\cite{KY95}, there are 
many deneutrosophication methods according to different applications. Here
we give one method. The deneutrosophication process consists of two steps. 

Step 4.1: \emph{Synthesization}: It is the process to transform an ordinary neutrosophic set (a type-1 neutrosophic set) $\tilde{B}$ into a fuzzy set $\bar{B}$.
It can be expressed using the following function: 

\begin{equation}
f(T_{\tilde{B}}^{'}(y), I_{\tilde{B}}^{'}(y), F_{\tilde{B}}^{'}(y)) : [0,1] \times [0,1] \times [0,1] \rightarrow [0,1]
\end{equation}

Here we give one definition of $f$:

\begin{equation}
T_{\bar{B}}(y) = a*T_{\tilde{B}}^{'}(y) + b*(1-F_{\tilde{B}}^{'}(y)) + c*I_{\tilde{B}}^{'}(y)/2+d*(1-I_{\tilde{B}}^{'}(y)/2),
\end{equation}
where $0 \leq a,b,c,d \leq 1, a+b+c+d = 1$.

The purpose of synthesization is to calculate the overall truth degree according to three components: truth-membership function, indeterminacy-membership function and falsity-membership function. The component--truth-membership function
gives the direct information about the truth-degree, so we use it directly in
the formula; The component--falsity-membership function gives the indirect
information about the truth-degree, so we use $(1 - F)$ in the formula. To
understand the meaning of indeterminacy-membership function $I$, 
we give an example: a statement is ``The quality of service is good", now firstly a person has to select a decision among $\{T, I, F\}$, 
secondly he or she has to answer the degree of the decision in $[0, 1]$. 
If he or she chooses $I = 1$, it means $100\%$ ``not sure" about the statement, 
i.e., $50\%$ true and $50\%$ false for the statement ($100\%$ balanced), 
in this 
sense, $I = 1$ contains the potential truth value $0.5$. If he or she chooses
$I = 0$, it means $100\%$ ``sure" about the statement, i.e., either $100\%$
true or $100\%$ false for the statement ($0\%$ balanced), in this sense, 
$I = 0$
is related to two extreme cases, but we do not know which one is in his or 
her mind.
So we have to consider both at the same time: $I = 0$ contains the potential
truth value that is either $0$ or $1$. If $I$ decreases from $1$ to $0$, then 
the potential truth value changes from one value $0.5$ to two different possible values gradually to the final possible ones $0$ and 
$1$ (i.e., from $100\%$ balanced to $0\%$ balanced), 
since he or she does not choose either $T$ or $F$ but $I$, we do not know 
his or her
final truth value. Therefore, the formula has  to consider two potential 
truth 
values implicitly represented by $I$ with different weights ($c$ and $d$) 
because of lack of 
his or her final decision information after he or she has chosen $I$.
Generally, $a > b > c, d$; $c$ and $d$ could be decided subjectively or 
objectively as long as enough information is available.
The parameters $a,b,c$ and $d$ can be tuned using learning algorithms such as
neural networks and genetic algorithms in the 
development of 
application to
improve the performance of the INLS.

Step 4.2: \emph{Calculation of a typical neutrosophic value}: 
Here we introduce one method of calculation of center of area. The method is sometimes called 
the 
\emph{center of gravity method} or \emph{centroid method}, the 
deneutrosophicated value, 
$dn(T_{\bar{B}}(y))$ is 
calculated by the formula
\begin{equation}
  dn(T_{\bar{B}}(y)) = \frac{\int_{\alpha}^{\beta} T_{\bar{B}}(y)y dy}{\int_{\alpha}^{\beta} T_{\bar{B}} dy}. 
\end{equation} 

END.

\section{Conclusions}
\label{conclusion}
In this chapter, we give the formal definitions of interval neutrosophic logic
which are extension of many other classical logics such as fuzzy
logic, intuitionistic fuzzy logic and paraconsistent logics, etc. 
Interval neutrosophic logic include
interval neutrosophic propositional logic and first order interval 
neutrosophic predicate logic. We call them classical (standard) neutrosophic 
logic.
In the future, we also will discuss and explore the non-classical (non-standard) neutrosophic logic such as modal interval neutrosophic logic, temporal interval neutrosophic logic, etc. Interval neutrosophic logic can not only handle
imprecise, fuzzy and incomplete propositions but also inconsistent propositions without the danger of trivialization. The chapter also give one application based
on the semantic notion of interval neutrosophic logic -- the Interval Neutrosophic Logic Systems (INLS) which
is the generalization of classical FLS and interval valued fuzzy FLS. 
Interval neutrosophic logic will have a lot of potential applications in 
computational Web intelligence~\cite{ZKL04}. For example, current fuzzy Web 
intelligence techniques can be improved by using more reliable interval neutrosophic logic methods because $T, I$ and $F$ are all used in decision making. In large, such robust interval neutrosophic logic methods can also be used in other applications such as medical informatics, bioinformatics and human-oriented decision-making under uncertainty. In fact, interval neutrosophic sets and interval neutrosophic
logic could be applied in the fields that fuzzy sets and fuzz logic are suitable for, also the fields that paraconsistent logics are suitable for.

%% file: chaptr3.tex
\chapter{Neutrosophic Relational Data Model} 
\label{NRDM} 

{\small
In this chapter, we present a generalization of the relational data
model based on interval neutrosophic sets.
Our data model is capable of manipulating incomplete as well as
inconsistent information. 
Fuzzy relation or intuitionistic fuzzy relation
can only handle incomplete information.
Associated with each relation are two membership functions one is 
called truth-membership function $T$ which
keeps track of the extent to which we believe the tuple is in the
relation, another is called falsity-membership function which keeps 
track of the extent to which we believe that it is not in the relation.
A neutrosophic relation is inconsistent if there
exists one tuple $a$ such that $T(a) + F(a) > 1$.
In order to handle inconsistent
situation, we propose an operator called ``split'' to transform inconsistent
neutrosophic relations into pseudo-consistent
neutrosophic relations and do the set-theoretic and
relation-theoretic operations on them and finally use another operator called
``combine'' to transform the result back to neutrosophic
relation.
For this model, we define algebraic operators that are generalisations
of the usual operators such as interesection, union, selection, join on fuzzy
relations.
Our data model can underlie any database and knowledge-base management 
system that deals
with incomplete and inconsistent information.
}

\section{Introduction} \label{Intro}

Relational data model was proposed by Ted Codd's pioneering paper \cite{cdd70}.
Since then, relational database systems have been extensively studied and 
a lot of commercial relational database systems are currently 
available \cite{en2000,sks96}. This data model usually takes care of only
well-defined and unambiguous data. However, imperfect information is 
ubiquitous
-- almost all the information that we have about the real world is not certain,
complete and precise~\cite{prs96}. Imperfect information can be classified as:
incompleteness, imprecision, uncertainty, inconsistency. Incompleteness arises
from the absence of a value, imprecision from the existence of a value which
cannot be measured with suitable precision, uncertainty from the fact that
a person has given a subjective opinion about the truth of a fact which he/she  
does not know for certain, and inconsistency from the fact that there are
two or more conflicting values for a variable.

In order to represent and manipulate various forms of incomplete information in
relational databases, several extensions of the classical relational model have
been proposed \cite{bsk83,bd84,cdd79,lip79,lps:datinc,mai83}. In some of these extensions, a variety of "null values" 
have been introduced to model unknown or not-applicable data values. Attempts 
have also been made to generalize operators of relational algebra to manipulate
such extended data models \cite{bsk83,cdd79,mai83,ls90,ls91}. The fuzzy set theory and fuzzy logic 
proposed by Zadeh \cite{zdh65} provide a requisite mathematical framework for 
dealing with incomplete and imprecise information. 
 Later on, the concept of
interval-valued fuzzy sets was proposed to capture the fuzziness of grade
of membership itself~\cite{TUR86}. In 1986, Atanassov introduced the
intuitionistic fuzzy set~\cite{ATA86} which is a generalization of fuzzy
set and provably equivalent to interval-valued fuzzy set. The intuitionistic
fuzzy sets consider both truth-membership $T$ and falsity-membership $F$ with
$T(a), F(a) \in [0, 1]$ and $T(a) + F(a) \leq 1$. Because of the restriction,
the fuzzy set,
interval-valued fuzzy set and
intuitionistic fuzzy set cannot handle inconsistent information.
Some authors \cite{nr84,bld83,bp82,ck78,kz86,prd84,rm88} have
studied relational databases in the light of fuzzy set theory with an objective
to accommodate a wider range of real-world requirements and to provide closer
man-machine interactions. Probability, possibility and Dempster-Shafer theory 
have been proposed to deal with uncertainty. Possibility theory ~\cite{zdh78}
is built upon the idea of a fuzzy restriction. That means a variable could only
take its value from some fuzzy set of values and any value within that set is 
a possible value for the variable. Because values have different
degrees of membership in the set, they are possible to different degrees.
Prade and Testemale ~\cite{prt84} initially suggested using 
possibility theory to deal with incomplete and uncertain information in 
database. Their work is extended in ~\cite{prt87} to cover multivalued
attributes. Wong~\cite{wng82} proposes a method that quantifies the uncertainty in a database using probabilities. His method maybe is the simplest one which attached a probability to every member of a relation, and to use these values
to provide the probability that a particular value is the correct answer to
a particular query. Carvallo and Pittarelli ~\cite{cvp87}
also use probability theory to model uncertainty in relational databases systems. Their method augmented projection an join operations with probability measures.  

However, unlike incomplete, imprecise and uncertain information, 
inconsistent information has not
enjoyed enough research attention. In fact, inconsistent information exists
in a lot of applications. For example, in data warehousing application, 
inconsistency will appear when trying to integrate the data from many different
sources. Another example is that in the expert system, there exist facts which
are inconsistent with each other. 
Generally, two basic approaches have been
followed in solving the inconsistency problem in knowledge bases: belief revision and paraconsistent logic. The goal of the first approach is to make an 
inconsistent theory consistent, either by revising it or by representing it
by a consistent semantics. On the other hand, the paraconsistent approach
allows reasoning in the presence of inconsistency, and contradictory information can be derived or introduced without trivialization \cite{mcm2002}. 
Bagai and Sunderraman \cite{bgs95a, sb95} proposed a paraconsistent
relational data model to deal with incomplete and inconsistent information.
The data model has been applied to compute the well-founded and fitting model of logic programming \cite{bs96a, bs96b}.
This data model is based on paraconsistent logics which were studied in detail
by de Costa \cite{cst77} and Belnap \cite{bln77}.

\begin{figure}[htbp]
\begin{center}
\includegraphics{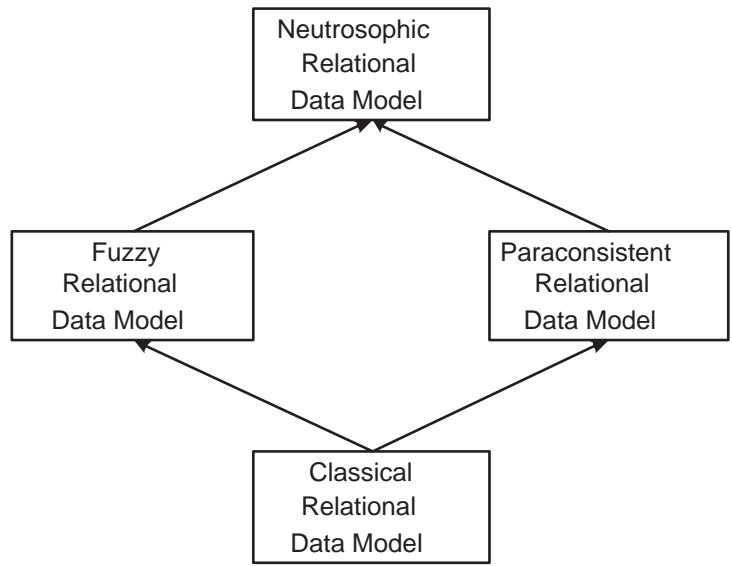}
\end{center}
\caption{Relationship Among FRDM, PRDM, NRDM and RDM}
\label{chapter3:fig1}
\end{figure}

In this chapter, we present a new relational data model -- neutrosophic relational data
model (NRDM).
Our model is based on the neutrosophic set theory 
which is an extension of intuitionistic fuzzy set theory\cite{gb93} and 
is capable of 
manipulating incomplete as well as inconsistent information. We use 
both truth-membership function grade $\alpha$ and falsity-membership 
function grade $\beta$ 
to denote the
status of a tuple of a certain relation with $\alpha, \beta \in [0,1]$
and $\alpha + \beta \leq 2$.  
NRDM is the 
generalization of fuzzy relational data model (FRDM). 
That is, when $\alpha + \beta = 1$, neutrosophic 
relation is the ordinary fuzzy relation. This model is distinct with 
paraconsistent relational
data model (PRDM), in fact it can be easily shown that PRDM is
a special case of PIFRDM. That is when $\alpha, \beta = 0 \mbox{ or } 1$, 
neutrosophic relation is just paraconsistent relation. 
We can use Figure ~\ref{chapter3:fig1} to express
the relationship among FRDM, PRDM and PIFRDM. 

We introduce neutrosophic relations, which are the 
fundamental mathematical structures
underlying our model. These structures are strictly more general than 
classical
fuzzy relations and intuitionistic fuzzy relations 
(interval-valued fuzzy relations), in that for any fuzzy relation or
intuitionistic fuzzy relation (interval-valued fuzzy relation) there 
is a neutrosophic relation with
the same information content, but not {\em vice versa}. The claim is also true 
for the relationship between neutrosophic relations and paraconsistent relations. We define algebraic
operators over neutrosophic relations that 
extend the standard operators such as
selection, join, union over fuzzy relations.

There are many potential applications of our new data model. 
Here are some examples:
\begin{description}
\item[(a)] Web mining. Essentially the data and documents on the Web are 
heterogeneous, inconsistency is unavoidable. Using the presentation and
reasoning method of our data model, it is easier to capture imperfect 
information on the Web which will provide more potentially value-added 
information. 
\item[(b)] Bioinformatics. There is a proliferation of data sources. Each
research group and each new experimental technique seems to generate yet
another source of valuable data. But these data can be incomplete and 
imprecise and even inconsistent. We could not simply throw away one data 
in favor of
other data. So how to represent and extract useful information
from these data will be a challenge problem. 
\item[(c)] Decision Support System. In decision support system, we need to
combine the database with the knowledge base. There will be a lot of 
uncertain
and inconsistent information, so we need an efficient data model to 
capture
these information and reasoning with these information.
\end{description}

The chapter is organized as follow. Section \ref{Fuzzy_Relation} 
of the chapter deals with some of the
basic definitions and concepts of fuzzy relations and operations. 
Section \ref{Vague_Relation}  
introduces neutrosophic relations and two notions of 
generalising the fuzzy relational
operators such as union, join, projection for these relations. 
Section \ref{Generalized} 
presents some actual generalised algebraic operators for neutrosophic relations.
These operators can be used for sepcifying queries for database systems 
built
on such relations. 
Section \ref{Application}  gives an illustrative application of these 
operators. 
Finally, Section \ref{Conclusion} contains some concluding remarks and 
directions
for future work. 

\section{Fuzzy Relations and Operations} \label{Fuzzy_Relation}

In this section, we present the essential concepts
of a fuzzy relational database. Fuzzy relations associate 
a value between 0 and 1 with every tuple representing the
degree of membership of the tuple in the relation.
We also present several useful query operators on
fuzzy relations.

Let a {\em relation scheme} (or just {\em scheme})
$\Sigma$ be a finite set of {\em attribute names},
where for any attribute name
$A \in \Sigma$, $\mbox{\em dom}(A)$ is a non-empty
{\em domain} of values for $A$.
A {\em tuple} on $\Sigma$ is any map $t:\Sigma \rightarrow \cup_{A
\in \Sigma}
\mbox{\em dom}(A)$, such that $t(A) \in \mbox{\em dom}(A)$, for each
$A \in \Sigma$.
Let $\tau(\Sigma)$ denote the set of all tuples on $\Sigma$.

\begin{definition}
A {\em fuzzy relation} on scheme $\Sigma$ is any 
map $R: \tau(\Sigma) \rightarrow [0,1]$.
We let ${\cal F}(\Sigma)$ be the set of all fuzzy relations on
$\Sigma$.
$\Box$
\end{definition}                                                                

If $\Sigma$ and $\Delta$ are relation schemes such that $\Delta  
\subseteq \Sigma$, then for any tuple $t \in \tau(\Delta)$, 
we let $t^\Sigma$ denote the set $\{t' \in \tau(\Sigma) ~|~ t'(A) = 
t(A)\mbox{, for all  $A \in \Delta$}\}$ of all extensions of $t$.
We extend this notion for any $T \subseteq \tau(\Delta)$ by defining
$T^\Sigma = \cup_{t \in T} ~ t^\Sigma$.

\subsection{Set-theoretic operations on Fuzzy relations}

\begin{definition}
{\bf Union:} Let $R$ and $S$ be fuzzy relations on scheme $\Sigma$.
Then, $R \cup S$ is a fuzzy relation on scheme $\Sigma$ given by
\[ 
 (R \cup S)(t) =  \max \{ R(t), S(t) \}, \mbox{for any } t \in \tau(\Sigma).\Box
\]

\end{definition}

\begin{definition}
{\bf Complement:} Let $R$ be a fuzzy relation on scheme $\Sigma$.
Then, $-R$ is a fuzzy relation on scheme $\Sigma$ given by
\[
 (-R)(t) = 1 - R(t), \mbox{for any } t \in \tau(\Sigma).\Box
\]
\end{definition}

\begin{definition}
{\bf Intersection:} Let $R$ and $S$ be fuzzy relations on scheme $\Sigma$.
Then, $R \cap S$ is a fuzzy relation on scheme $\Sigma$ given by
\[
 (R \cap S)(t) = \min\{ R(t), S(t) \}, \mbox{for any }t \in \tau(\Sigma).\Box
\]
\end{definition}

\begin{definition}
{\bf Difference:} Let $R$ and $S$ be fuzzy relations on scheme $\Sigma$.
Then, $R - S$ is a fuzzy relation on scheme $\Sigma$ given by
\[
 (R - S)(t) = \min\{ R(t), 1 - S(t) \}, \mbox{for any } t \in \tau(\Sigma).\Box
\]
\end{definition}

\subsection{Relation-theoretic operations on Fuzzy relations}

\begin{definition}
Let $R$ and $S$ be fuzzy relations on schemes $\Sigma$ and  
$\Delta$, respectively.
Then, the {\em natural join} (or just {\em join}) of $R$ and $S$,
denoted $R \Join S$, is a fuzzy relation on scheme  
$\Sigma \cup \Delta$, given by
\[
 (R \Join S)(t) = \min \{ R(\pi_{\Sigma}(t)), S(\pi_{\Delta}(t)) \}, 
\mbox{for any }t \in \tau(\Sigma \cup \Delta).\Box
\]
\end{definition}

\begin{definition}
Let $R$ be a fuzzy relation on scheme $\Sigma$ and let 
$\Delta \subseteq \Sigma$. Then, the projection of $R$ onto
$\Delta$, denoted by $\Pi_{\Delta}(R)$ is a fuzzy relation
on scheme $\Delta$ given by
\[
(\Pi_{\Delta}(R))(t) = \max \{ R(u) | u \in t^{\Sigma} \},
\mbox{for any }t \in \tau(\Delta).\Box
\]
\end{definition}

\begin{definition}
Let $R$ be a fuzzy relation on scheme $\Sigma$,
and let $F$ be any logic formula involving attribute names in  
$\Sigma$,
constant symbols
(denoting values in the attribute domains), equality symbol $=$,
negation symbol $\neg$, and connectives $\vee$ and $\wedge$.
Then, the {\em selection} of $R$ by $F$,
denoted $\dot{\sigma}_{F}(R)$, is a fuzzy relation on scheme  
$\Sigma$,
given by
\[
(\dot{\sigma}_{F}(R))(t) = \left \{ \begin{array}{ll}
R(t) & \mbox{if }t \in \sigma_{F}(\tau(\Sigma)) \\
0 & \mbox{Otherwise}
\end{array} \right. 
\]
where $\sigma_F$ is the usual selection of tuples satisfying $F$.
$\Box$
\end{definition}

\section{Neutrosophic Relations} 
\label{Vague_Relation}

In this section, we generalize
fuzzy relations in such a manner that we are now able to assign
a measure of belief and a measure of doubt to each tuple. 
We shall refer to these
generalized fuzzy relations as 
{\em neutrosophic relations}. 
So, a tuple in a neutrosophic relation is 
assigned a measure 
$\langle \alpha, \beta \rangle$, $0 \leq \alpha, \beta \leq 1$.
$\alpha$ will be referred to as the {\em belief} factor and $\beta$ will
be referred to as the {\em doubt} factor.
The interpretation of this measure is that we believe with 
confidence $\alpha$ and doubt with confidence $\beta$ that the tuple is in 
the relation. The belief and doubt confidence factors for a tuple
need not add to exactly 1. This allows for incompleteness and inconsistency
to be represented. If the belief and doubt factors add up to less than 1, 
we have incomplete information regarding the tuple's status in the relation
and if the belief and doubt factors add up to more than 1, we have 
inconsistent
information regarding the tuple's status in the relation.

In contrast to fuzzy relations where the grade of membership of
a tuple is fixed, neutrosophic relations bound the grade of membership
of a tuple to a subinterval $[\alpha, 1- \beta]$ for the case 
$\alpha + \beta \leq 1$.

The operators on fuzzy relations can also be generalised for  
neutrosophic relations.
However, any such generalization of operators should maintain the belief  
system
intuition behind neutrosophic relations.

This section also develops two different notions of operator
generalisations.

We now formalize the notion of a neutrosophic relation.

Recall that $\tau(\Sigma)$ denotes the set of all tuples on any scheme 
$\Sigma$.

\begin{definition}
A {\em \bf neutrosophic relation} $R$ on scheme $\Sigma$ 
is any subset of 
\[
  \tau(\Sigma) \times [0,1] \times [0,1].
\]
For any $t \in \tau{(\Sigma)}$, we shall denote an element of R as 
$\langle t,R(t)^{+},R(t)^{-} \rangle$, where $R(t)^{+}$ is the belief 
factor assigned to
$t$ by $R$ and $R(t)^{-}$ is the doubt factor assigned to $t$ by $R$. 
Let ${\cal V}(\Sigma)$ be the set of all neutrosophic relations  
on $\Sigma$. 
\end{definition}

\begin{definition}
A neutrosophic relation $R$ on scheme
$\Sigma$ is {\em \bf consistent} if $R(t)^{+} + R(t)^{-} \leq 1$,
for all $t \in \tau(\Sigma)$.
Let ${\cal C}(\Sigma)$ be the set of all consistent neutrosophic relations
on $\Sigma$.
$R$ is said to be {\em \bf complete} if $R(t)^{+} + R(t)^{-} \geq 1$,
for all $t \in \tau(\Sigma)$.  
If $R$ is both consistent and complete,
i.e. $R(t)^{+} + R(t)^{-} = 1$,
for all $t \in \tau(\Sigma)$,
then it is a {\em \bf total} neutrosophic relation,
and let ${\cal T}(\Sigma)$ be the set of all total neutrosophic 
relations on $\Sigma$. 
\end{definition}

\begin{definition}
$R$ is said to be {\em \bf pseudo-consistent} if \\ 
$\max \{b_i | (\exists t \in \tau(\Sigma))(\exists d_i)(\langle t,b_i,d_i \rangle \in R) \} + \max \{d_i | (\exists t \in \tau(\Sigma))(\exists b_i)(\langle t,b_i,d_i \rangle \in R) \} > 1, \mbox{ where for these }$ \\ $\langle t,b_i,d_i \rangle, b_i+d_i=1$.
Let ${\cal P}(\Sigma)$ be the set of all pseudo-consistent neutrosophic
relations on $\Sigma$.  
\end{definition}

\begin{example}
Neutrosophic relation $R = \{\langle a,0.3,0.7 \rangle,$
\\ $\langle a,0.4,0.6 \rangle,\langle b,0.2,0.5 \rangle,\langle c,0.4,0.3 \rangle\}$ 
is
pseudo-consistent. Because for $t=a,$ $\max \{0.3,0.4\}+\max \{0.7,0.6\}=1.1 > 1$.
\end{example}

It should be observed that total neutrosophic relations
are essentially fuzzy relations where the uncertainty in the
grade of membership is eliminated.
We make this relationship explicit by defining a one-one  
correspondence
$\lambda_\Sigma:{\cal T}(\Sigma) \rightarrow {\cal F}(\Sigma)$, given  
by
$\lambda_\Sigma(R)(t) = R(t)^{+}$, for all $t \in \tau(\Sigma)$.
This correspondence is used frequently in the following discussion.

\subsection*{Operator Generalisations}

It is easily seen that neutrosophic relations are a 
generalization  of
fuzzy relations, in that for each fuzzy relation there is a
neutrosophic relation with the same information content, but not  
{\em vice versa}.
It is thus natural to think of generalising the operations on  
fuzzy relations such as union, join, projection etc. to neutrosophic relations.
However, any such generalization should be intuitive with respect to
the belief system model of neutrosophic relations.
We now construct a framework for operators on both kinds of relations
and introduce two different notions of the generalization  
relationship among their operators.

An $n$-ary {\em operator on fuzzy relations with signature
$\langle \Sigma_1,\ldots,\Sigma_{n+1} \rangle$} is a function
$\Theta:{\cal F}(\Sigma_1) \times \cdots \times {\cal F}(\Sigma_n)
\rightarrow {\cal F}(\Sigma_{n+1})$, where
$\Sigma_1,\ldots,\Sigma_{n+1}$ are any schemes.
Similarly, an $n$-ary {\em operator on neutrosophic relations with  
signature
$\langle \Sigma_1,\ldots,\Sigma_{n+1} \rangle$} is a function
$\Psi:{\cal V}(\Sigma_1) \times \cdots \times {\cal V}(\Sigma_n)
\rightarrow {\cal V}(\Sigma_{n+1})$.

\begin{definition}
An operator $\Psi$ on neutrosophic relations
with signature $\langle \Sigma_1,\ldots,\Sigma_{n+1} \rangle$
is {\em totality preserving} if for any
total neutrosophic relations $R_1,\ldots,R_n$ on 
schemes
$\Sigma_1,\ldots,\Sigma_n$, respectively, $\Psi(R_1,\ldots,R_n)$
is also total.
$\Box$
\end{definition}

\begin{definition}
A totality preserving operator
$\Psi$ on neutrosophic relations
with signature 
\begin{center}
$\langle \Sigma_1,\ldots,\Sigma_{n+1} \rangle$
\end{center}
is a {\em weak generalization} of an
operator
$\Theta$ on fuzzy relations with the same signature, if for
any total neutrosophic relations $R_1,\ldots,R_n$ on 
schemes
$\Sigma_1,\ldots,\Sigma_n$, respectively, we have
\[
 \lambda_{\Sigma_{n+1}}(\Psi(R_1,\ldots,R_n)) =
\Theta(\lambda_{\Sigma_1}(R_1),\ldots,\lambda_{\Sigma_n}(R_n)). 
\Box
\]
\end{definition}
The above definition essentially requires $\Psi$ to coincide with
$\Theta$ on total neutrosophic relations 
(which are in one-one
correspondence with the fuzzy relations).
In general, there may be many operators on neutrosophic relations
that are weak generalisations
of a given operator $\Theta$ on fuzzy relations.
The behavior of the weak generalisations of $\Theta$ on even just the
consistent neutrosophic relations may in general vary.
We require a stronger notion of operator generalization under which,  
at
least when restricted to consistent 
intuitionistic fuzzy 
relations, the behavior of all the generalised operators is the
same.
Before we can develop such a notion, we need that of `representations' of  
a neutrosophic relation.

We associate with a consistent neutrosophic relation $R$ the set of
all (fuzzy relations corresponding to) total neutrosophic relations
obtainable from $R$ by filling in the gaps between the belief and
doubt factors for each tuple.
Let the map $\mbox{\bf reps}_\Sigma:{\cal C}(\Sigma) \rightarrow  
2^{{\cal F}
(\Sigma)}$ be given by
\[
\mbox{\bf reps}_\Sigma(R) = \{Q \in {\cal F}(\Sigma) ~|~ 
\bigwedge_{t_{i} \in \tau(\Sigma)}  
(R(t_{i})^{+} \leq Q(t_{i}) \leq 1 - R(t_{i})^{-} ) \}.
\]
The set $\mbox{\bf reps}_\Sigma(R)$ contains all fuzzy relations  
that are `completions' of the consistent neutrosophic relation $R$.
Observe that $\mbox{\bf reps}_\Sigma$ is defined only for consistent  
neutrosophic relations and produces sets of 
fuzzy relations. Then we have following 
observation.

\begin{proposition} \label{prop1}
 For any consistent neutrosophic relation $R$ on 
 scheme $\Sigma$, $\mbox{\bf reps}_\Sigma(R)$ is the singleton 
 $\{\lambda_\Sigma(R)\} \mbox{ iff }R \mbox{ is total}.\Box$
\end{proposition}
\begin{proof}
It is clear from the definition of consistent and total neutrosophic 
relations and from the definition of {\bf reps}
operation.
\end{proof}

We now need to extend operators on fuzzy relations to sets of
fuzzy relations.
For any operator
$\Theta:{\cal F}(\Sigma_1) \times \cdots \times {\cal F}(\Sigma_n)
\rightarrow {\cal F}(\Sigma_{n+1})$
on fuzzy relations, we let ${\cal S}(\Theta):
2^{{\cal F}(\Sigma_1)} \times \cdots \times 2^{{\cal F}(\Sigma_n)}
\rightarrow 2^{{\cal F}(\Sigma_{n+1})}$ be a map on
sets of fuzzy relations defined as follows.
For any sets $M_1,\ldots,M_n$ of fuzzy relations on schemes
$\Sigma_1,\ldots,\Sigma_n$, respectively,
\[
 {\cal S}(\Theta)(M_1,\ldots,M_n) = \{\Theta(R_1,\ldots,R_n) ~|~ R_i  
\in M_i,
\mbox{ for all } i, 1\leq i\leq n\}.
\]
In other words, ${\cal S}(\Theta)(M_1,\ldots,M_n)$ is the set of
$\Theta$-images of all tuples in the cartesian product
$M_1 \times \cdots \times M_n$.
We are now ready to lead up to a stronger notion of operator
generalization.

\begin{definition}
An operator $\Psi$ on neutrosophic relations
with signature $\langle \Sigma_1,\ldots,\Sigma_{n+1} \rangle$
is {\em consistency preserving} if for any
consistent neutrosophic relations $R_1,\ldots,R_n$ on schemes
$\Sigma_1,\ldots,\Sigma_n$, respectively, $\Psi(R_1,\ldots,R_n)$
is also consistent.
$\Box$
\end{definition}

\begin{definition}
A consistency preserving operator
$\Psi$ on neutrosophic relations
with signature $\langle \Sigma_1,\ldots,\Sigma_{n+1} \rangle$
is a {\em strong generalization} of an operator
$\Theta$ on fuzzy relations with the same signature, if
for any consistent neutrosophic relations $R_1,\ldots,R_n$ on schemes
$\Sigma_1,\ldots,\Sigma_n$, respectively,
we have
\[
 \mbox{\bf reps}_{\Sigma_{n+1}}(\Psi(R_1,\ldots,R_n)) =
{\cal S}(\Theta)(\mbox{\bf reps}_{\Sigma_1}(R_1),\ldots,
\mbox{\bf reps}_{\Sigma_n}(R_n)). \Box
\]
\end{definition}

Given an operator $\Theta$ on fuzzy relations, the behavior of a
weak generalization of $\Theta$ is `controlled' only over the total
neutrosophic relations.
On the other hand, the behavior of a strong generalization is
`controlled' over all consistent neutrosophic relations.
This itself suggests that strong generalization is a stronger notion
than weak generalization.
The following proposition makes this precise.

\begin{proposition}
If $\Psi$ is a strong generalization of $\Theta$,
then $\Psi$ is also a weak generalization of $\Theta$.$\Box$
\end{proposition}
\begin{proof}
Let $\langle \Sigma_1, \ldots, \Sigma_{n+1} \rangle$ be the signature
of $\Psi$ and $\Theta$, and let $R_1, \ldots, R_n$ be any total neutrosophic
relations on schemes $\Sigma_1, \ldots, \Sigma_n$, respectively. Since
all total relations are consistent, and $\Psi$ is a strong generalization
of $\Theta$, we have that
\[
 {\bf reps}_{\Sigma_{n+1}}(\Psi(R_1, \ldots, R_n)) = {\cal S}(\Theta)({\bf reps}_{\Sigma_1}(R_1), \ldots, {\bf reps}_{\Sigma_n}(R_n)),
\]
Proposition~\ref{prop1} gives us that for each $i$, $1 \leq i \leq n$,
${\bf reps}_{\Sigma_i}(R_i)$ is the singleton set $\{\lambda_{\Sigma_i}(R_i)\}$.
Therefore, 
${\cal S}(\Theta)({\bf reps}_{\Sigma_1}(R_i), \ldots, {\bf reps}_{\Sigma_n}(R_n))$ 
is just the singleton set:
\begin{center}
$\{\Theta(\lambda_{\Sigma_1}(R_1), \ldots, \lambda_{\Sigma_n}(R_n))\}$. 
\end{center}
Here, $\Psi(R_1, \ldots, R_n)$ is total, and \\
$\lambda_{\Sigma_{n+1}}(\Psi(R_1, \ldots, R_n)) = \Theta(\lambda_{\Sigma_1}(R_1), \ldots, \lambda_{\Sigma_n}(R_n))$, i.e. $\Psi$ is a weak generalization of $\Theta$.
\end{proof}

Though there may be many strong generalisations of an operator on  
fuzzy
relations, they all behave the same when restricted to consistent
neutrosophic relations.
In the next section, we propose strong generalisations for the usual
operators on fuzzy relations.
The proposed generalised operators on neutrosophic relations  
correspond to the
belief system intuition behind neutrosophic relations.

First we will introduce two special operators on neutrosophic 
relations called split and combine to
transform inconsistent neutrosophic
relations into pseudo-consistent neutrosophic relations and transform pseudo-consistent neutrosophic relations into inconsistent neutrosophic relations.

\begin{definition}[Split]
Let $R$ be a neutrosophic
relation on scheme $\Sigma$.
Then, \\
$\triangle(R) = \{\langle t,b,d \rangle | \langle t,b,d \rangle \in R \mbox{ and } b+d \leq 1 \} \cup \{\langle t,b',d' \rangle | \langle t,b,d \rangle \in R \mbox{ and } b+d > 1 \mbox{ and } b'=b \mbox{ and } d'=1-b \} \cup \{\langle t,b',d' \rangle | \langle t,b,d \rangle \in R \mbox{ and } b+d > 1 \mbox{ and } b'=1-d \mbox { and } d'=d \}$. 
\end{definition}

It is obvious that $\triangle(R)$ is
pseudo-consistent if $R$ is inconsistent.

\begin{definition}[Combine]
Let $R$ be a neutrosophic relation 
on scheme $\Sigma$.
Then, \\
$\nabla(R) = \{\langle t,b',d' \rangle | (\exists b)(\exists d)((\langle t,b',d \rangle \in R \mbox{ and } (\forall b_i,d_i)(\langle t,b_i,d_i \rangle \rightarrow b' \geq b_i) \mbox{ and }$ \\ $\langle t,b,d' \rangle \in R \mbox{ and } (\forall b_i)(\forall d_i)(\langle t,b_i,d_i \rangle \rightarrow d' \geq d_i)) \}$. 
\end{definition}

It is obvious that $\nabla(R)$ is inconsistent if $R$ is pseudo-consistent.

Note that strong generalization defined above only holds for consistent or
pseudo-consistent neutrosophic relations. For any
arbitrary paraconsisent intuitionistic fuzzy relations, we should first
use split operation to transform them into non inconsistent neutrosophic 
relations and apply the set-theoretic and 
relation-theoretic
operations on them and finally use combine operation to transform the result
into arbitrary neutrosophic relation. For the simplification
of notation, the following generalized algebra is defined under such assumption.

\section{Generalized Algebra on Neutrosophic Relations} 
\label{Generalized}

In this section, we present one strong generalization each for the  
fuzzy relation operators such as union, join, projection.
To reflect generalization, a hat is placed over
a fuzzy relation operator to
obtain the corresponding neutrosophic relation operator.
For example, $\bowtie$ denotes the natural join among fuzzy  
relations, and
${\bowtie}$ denotes natural join on neutrosophic relations.
These generalized operators maintain the belief system
intuition behind neutrosophic relations.

\subsection*{Set-Theoretic Operators}

We first generalize the two fundamental set-theoretic operators,
union and complement.

\begin{definition}
Let $R$ and $S$ be neutrosophic relations on scheme $\Sigma$.
Then,
\begin{description}
\item[(a)] the {\em union} of $R$ and $S$, denoted $R~\widehat{\cup}~S$,  
is a neutrosophic relation on scheme $\Sigma$, given
by
\[
(R ~\widehat{\cup}~ S)(t) = \langle \max\{R(t)^{+},S(t)^{+}\}, \min\{R(t)^{-},S(t)^{-}\} \rangle,
\mbox{ for any } t \in \tau(\Sigma);
\]
\item[(b)] the {\em complement} of $R$, denoted $\widehat{-}~R$, is a
neutrosophic relation on scheme $\Sigma$, given
by
\[
(\widehat{-}~R)(t) = \langle R(t)^{-},R(t)^{+} \rangle, \mbox{ for any } t \in \tau(\Sigma).
~
\]
\hfill{\space}  $\Box$
\end{description}
\end{definition}
An intuitive appreciation of the union operator can be obtained as follows:
Given a tuple $t$, since we believed that it is present in the relation $R$
with confidence $R(t)^{+}$ and that it is present in the relation $S$
with confidence $S(t)^{+}$, we can now believe that the tuple $t$ is
present in the ``either-$R$-or-$S$'' relation with confidence which is equal to
the larger of $R(t)^{+}$ and $S(t)^{+}$. Using the same logic, we can now
believe in the absence of the tuple $t$ from the ``either-$R$-or-$S$'' relation
with confidence which is equal to the smaller (because $t$ must
be absent from both $R$ and $S$ for it to be absent from the union) of
$R(t)^{-}$ and $S(t)^{-}$. The definition of {\em complement} and
of all the other operators on neutrosophic relations defined
later can (and should) be understood in the same way.

\begin{proposition}
The operators $\widehat{\cup}$ and unary $\widehat{-}$ on neutrosophic  
relations are
strong generalisations of the operators $\cup$ and unary $-$ on
fuzzy relations.
\end{proposition}
\begin{proof}
Let $R$ and $S$ be consistent neutrosophic relations on scheme $\Sigma$. Then
${\bf reps}_{\Sigma}(R~\widehat{\cup}~S)$ is the set
\[
 \{Q ~|~ \bigwedge_{t_i \in \tau(\Sigma)}(\max\{R(t_i)^+,~S(t_i)^+\} \leq Q(t_i) \leq 1 - \min\{R(t_i)^-,~S(t_i)^-\})\} 
\]
This set is the same as the set 
\[
 \{r~\cup~s ~|~ \bigwedge_{t_i \in \tau(\Sigma)}(R(t_i)^+ \leq r(t_i) \leq 1-R(t_i)^-), \bigwedge_{t_i \in \tau(\Sigma)}(S(t_i)^+ \leq s(t_i) \leq 1-S(t_i)^-)\}
\]
which is $S(\cup)({\bf reps}_{\Sigma}(R),~{\bf reps}_{\Sigma}(S))$. 
Such a result for unary $\widehat{-}$ can also be shown similarly.
\end{proof}

For sake of completeness, we define the following two related  
set-theoretic operators:

\begin{definition}
Let $R$ and $S$ be neutrosophic relations on scheme $\Sigma$.
Then,
\begin{description}
\item[(a)] the {\em intersection} of $R$ and $S$, denoted  
$R~\widehat{\cap}~S$,
is a neutrosophic relation on scheme $\Sigma$, given
by
\[
(R ~\widehat{\cap}~ S)(t) = \langle \min\{R(t)^{+},S(t)^{+}\}, \max\{R(t)^{-},S(t)^{-}\} \rangle,
\mbox{ for any } t \in \tau(\Sigma);
\]
\item[(b)] the {\em difference} of $R$ and $S$, denoted  
$R~\widehat{-}~S$, is
a neutrosophic relation on scheme $\Sigma$, given
by
\[
(R~\widehat{-}~S)(t) = \langle \min\{R(t)^{+},S(t)^{-}\}, \max\{R(t)^{-},S(t)^{+}\} \rangle,
\mbox{ for any } t \in \tau(\Sigma);
\]
\hfill{\space}  $\Box$
\end{description}
\end{definition}

The following proposition relates the intersection and difference
operators in terms of the more fundamental set-theoretic operators
union and complement.

\begin{proposition}
For any neutrosophic relations $R$ and $S$ on the same scheme, 
we have
\begin{eqnarray*}
R~\widehat{\cap}~S &=& \widehat{-}(\widehat{-}R~\widehat{\cup}~\widehat{-}S), \mbox{  and  
}\\
R~\widehat{-}~S &=& \widehat{-}(\widehat{-}R~\widehat{\cup}~S).
\end{eqnarray*}
\end{proposition}
\begin{proof}
\begin{eqnarray*}
\mbox{By definition, } \widehat{-}R(t) &=& \langle R(t)^{-}, R(t)^{+} \rangle \\
                     \widehat{-}S(t) &=& \langle S(t)^{-}, S(t)^{+} \rangle \\
\mbox{ and }       (\widehat{-}R~\widehat{\cup}~\widehat{-}S)(t) &=& \langle \max(R(t)^{-}, S(t)^{-}),~\min(R(t)^{+}, S(t)^{+}) \rangle \\
\mbox{ so, }    (\widehat{-}(\widehat{-}R~\widehat{\cup}~\widehat{-}S))(t) &=& \langle \min(R(t)^{+}, S(t)^{+}), \max(R(t)^{-}, S(t)^{-}) \rangle \\
                    &=& R~\widehat{\cap}~S(t).
\end{eqnarray*}
The second part of the result can be shown similarly.
\end{proof}

\subsection*{Relation-Theoretic Operators}

We now define some relation-theoretic algebraic
operators on neutrosophic relations.

\begin{definition}
Let $R$ and $S$ be neutrosophic relations on 
schemes $\Sigma$ and  $\Delta$, respectively.
Then, the {\em natural join} (further for short called {\em join}) 
of $R$ and $S$,
denoted $R~\widehat{\bowtie}~S$, is a neutrosophic relation on scheme  $\Sigma \cup \Delta$,
given by
\[
(R~\widehat{\bowtie}~S)(t) = 
\langle \min \{R(\pi_{\Sigma}(t))^{+}, S(\pi_{\Delta}(t))^{+} \}, 
 \max \{R(\pi_{\Sigma}(t))^{-}, S(\pi_{\Delta}(t))^{-} \} \rangle,
\]
where $\pi$ is the usual projection of a tuple.
\hfill{\space} $\Box$
\end{definition}
It is instructive to observe that, similar to the intersection operator,
the minimum of the belief factors and the maximum of the doubt factors 
are used in the definition of the join operation. 

\begin{proposition} \label{strjoin}
$\widehat{\bowtie}$ is a strong generalization of $\bowtie$.
\end{proposition}
\begin{proof}
Let $R$ and $S$ be consistent neutrosophic relations on schemes $\Sigma$ and $\Delta$,
respectively. Then \\ ${\bf reps}_{\Sigma~\cup~\Delta}(R~\widehat{\bowtie}~S)$ is 
the set
$\{Q \in {\cal F}(\Sigma~\cup~\Delta)~|~\bigwedge_{t_i \in \tau(\Sigma~\cup~\Delta)}(\min \{R_{\pi_\Sigma}(t_i)^+, S_{\pi_\Delta}(t_i)^+ \} \leq Q(t_i) \leq$ \\
$1-\max \{R_{\pi_\Sigma}(t_i)^-, S_{\pi_\Delta}(t_i)^- \}) \}$
and $S(\bowtie)({\bf reps}_\Sigma(R), {\bf reps}_\Delta(S)) = \{ r \bowtie s ~|~r \in {\bf reps}_\Sigma(R), s \in {\bf reps}_\Delta(S) \}$

Let $Q \in {\bf reps}_{\Sigma \cup \Delta}(R~\widehat{\bowtie}~S)$. Then
$\pi_\Sigma(Q)\in {\bf reps}_\Sigma(R)$, where $\pi_\Sigma$ is the usual
projection over $\Sigma$ of fuzzy relations. 
Similarly, $\pi_\Delta(Q) \in {\bf reps}_\Delta(S)$.  
Therefore, $Q \in S(\bowtie)({\bf reps}_\Sigma(R), {\bf reps}_\Delta(S))$.

Let $Q \in S(\bowtie)({\bf reps}_\Sigma(R), {\bf reps}_\Delta(S))$.
Then $Q(t_i) \geq \min \{R_{\pi_\Sigma}(t_i)^+, S_{\pi_\Delta}(t_i)^+ \}$
and $Q(t_i) \leq \min \{1-R{\pi_\Sigma}(t_i)^-, 1-S_{\pi_\Delta}(t_i)^- \} = 1-\max \{R_{\pi_\Sigma}(t_i)^-, S_{\pi_\Delta}(t_i)^- \}$, for any $t_i \in \tau(\Sigma \cup \Delta)$, because $R$ and $S$ are consistent. \\
Therefore, $Q \in {\bf reps}_{\Sigma \cup \Delta}(R~\widehat{\bowtie}~S)$.
\end{proof}

We now present the projection operator.

\begin{definition}
Let $R$ be a neutrosophic relation on scheme $\Sigma$, and
$\Delta \subseteq \Sigma$.
Then, the {\em projection} of $R$ onto $\Delta$,
denoted $\widehat{\pi}_\Delta(R)$, is a neutrosophic relation on  
scheme $\Delta$, given by
\[
(\widehat{\pi}_\Delta(R))(t) = 
\langle \max\{R(u)^{+} | u \in t^{\Sigma} \},
 \min\{R(u)^{-} | u \in t^{\Sigma} \}  \rangle.
\]
\hfill{\space} $\Box$
\end{definition}
The belief factor of a tuple in the projection is the maximum
of the belief factors of all of the tuple's extensions onto the
scheme of the input neutrosophic relation.
Moreover, the doubt factor of a tuple in the projection is the minimum
of the doubt factors of all of the tuple's extensions onto the
scheme of the input neutrosophic relation.

We present the selection operator next.

\begin{definition}
Let $R$ be a neutrosophic relation on scheme $\Sigma$,
and let $F$ be any logic formula involving attribute names in  
$\Sigma$, constant symbols
(denoting values in the attribute domains), equality symbol $=$,
negation symbol $\neg$, and connectives $\vee$ and $\wedge$.
Then, the {\em selection} of $R$ by $F$,
denoted $\widehat{\sigma}_F(R)$, is a neutrosophic relation on
scheme  $\Sigma$, given by

\vspace{.15in}

\begin{tabular}{l}
$(\widehat{\sigma}_F(R))(t) = \langle \alpha,\beta \rangle$, where \\[.15in]
$\alpha = \left\{ \begin{array}{ll}
                    R(t)^{+} & \mbox{ if $t \in \sigma_{F}(\tau(\Sigma))$} \\
                    0        & \mbox{otherwise}
                   \end{array}
           \right. $  \hspace{0.25in} and \hspace{0.25in}
$\beta  = \left\{ \begin{array}{ll}
                    R(t)^{-} & \mbox{ if $t \in \sigma_{F}(\tau(\Sigma))$} \\
                    1        & \mbox{otherwise}
                   \end{array}
           \right. $
\end{tabular} \\[.15in] 
where $\sigma_F$ is the usual selection of tuples satisfying $F$ from
ordinary relations.
\hfill{\space}$\Box$
\end{definition}
If a tuple satisfies the selection criterion, it's belief and doubt factors
are the same in the selection as in the input neutrosophic
relation. In the case where the tuple does not satisfy the selection 
criterion, its belief factor is set to 0 and the doubt factor is set
to 1 in the selection.

\begin{proposition}
The operators $\widehat{\pi}$ and $\widehat{\sigma}$ are strong  
generalisations of
$\pi$ and $\sigma$, respectively.
\end{proposition}
\begin{proof}
Similar to that of Proposition~\ref{strjoin}.
\end{proof}

\begin{example}
Relation schemes are sets of attribute names, but in this example we treat 
them as ordered sequences of attribute names (which can be obtained through
permutation of attribute names), so tuples can be viewed as the usual lists 
of values. Let $\{a, b, c\}$ be a common domain for all attribute names, and 
let $R$ and $S$ be the following
neutrosophic relations on schemes $\langle X,Y \rangle$ and $\langle Y,Z \rangle$
respectively.
\begin{center}
\begin{tabular}{|c|c|} \hline
$t$        & $R(t)$ \\ \hline
$(a,a)$    & $\langle 0,1 \rangle$ \\ 
$(a,b)$    & $\langle 0,1 \rangle$ \\
$(a,c)$    & $\langle 0,1 \rangle$ \\
$(b,b)$    & $\langle 1,0 \rangle$ \\
$(b,c)$    & $\langle 1,0 \rangle$ \\ 
$(c,b)$    & $\langle 1,1 \rangle$ \\ \hline 
\end{tabular}
\hspace{0.5in}
\begin{tabular}{|c|c|} \hline
$t$        & $S(t)$ \\ \hline
$(a,c)$    & $\langle 1,0 \rangle$ \\
$(b,a)$    & $\langle 1,1 \rangle$ \\
$(c,b)$    & $\langle 0,1 \rangle$ \\ \hline
\end{tabular}
\end{center}
For other tuples which are not in the neutrosophic relations $R(t)$ and $S(t)$, their 
$\langle \alpha,\beta \rangle = \langle 0,0 \rangle$ which means no any 
information available. Because $R$ and $S$ are inconsistent, we first
use split operation to transform them into pseudo-consistent and apply the
relation-theoretic operations on them and transform the result back to
arbitrary neutrosophic set using combine operation.
Then, $T_{1} =\nabla(\triangle(R)~\widehat{\bowtie}~\triangle(S))$ is a 
neutrosophic 
relation on scheme $\langle X,Y,Z \rangle$ and 
$T_{2} = \nabla(\widehat{\pi}_{\langle X,Z \rangle}(\triangle(T_{1})))$  and 
$T_{3} = \widehat{\sigma}_{X \neg = Z}(T_{2})$ 
are neutrosophic relations on scheme 
$\langle X,Z \rangle$. $T_{1}$, $T_{2}$ and $T_{3}$ are 
shown below:
\begin{center}
\begin{tabular}{|c|c|} \hline
$t$       & $T_{1}(t)$ \\ \hline
$(a,a,a)$ & $\langle 0,1 \rangle$    \\
$(a,a,b)$ & $\langle 0,1 \rangle$    \\
$(a,a,c)$ & $\langle 0,1 \rangle$    \\
$(a,b,a)$ & $\langle 0,1 \rangle$    \\
$(a,b,b)$ & $\langle 0,1 \rangle$    \\
$(a,b,c)$ & $\langle 0,1 \rangle$    \\
$(a,c,a)$ & $\langle 0,1 \rangle$    \\
$(a,c,b)$ & $\langle 0,1 \rangle$    \\
$(a,c,c)$ & $\langle 0,1 \rangle$    \\ 
$(b,b,a)$ & $\langle 1,1 \rangle$    \\
$(b,c,b)$ & $\langle 0,1 \rangle$    \\
$(c,b,a)$ & $\langle 1,1 \rangle$    \\
$(c,b,b)$ & $\langle 0,1 \rangle$    \\
$(c,b,c)$ & $\langle 0,1 \rangle$    \\
$(c,c,b)$ & $\langle 0,1 \rangle$    \\ \hline
\end{tabular}
\hspace{.4in}
\begin{tabular}{|c|c|} \hline
$t$       &   $T_2(t)$ \\ \hline
$(a,a)$   &   $\langle 0,1 \rangle$ \\
$(a,b)$   &   $\langle 0,1 \rangle$ \\ 
$(a,c)$   &   $\langle 0,1 \rangle$ \\
$(b,a)$   &   $\langle 1,0 \rangle$ \\ 
$(c,a)$   &   $\langle 1,0 \rangle$ \\ \hline
\end{tabular}
\hspace{0.4in}
\begin{tabular}{|c|c|} \hline
$t$       &   $T_3(t)$ \\ \hline
$(a,a)$   &   $\langle 0,1 \rangle$ \\
$(a,b)$   &   $\langle 0,1 \rangle$ \\
$(a,c)$   &   $\langle 0,1 \rangle$ \\
$(b,a)$   &   $\langle 1,0 \rangle$ \\
$(b,b)$   &   $\langle 0,1 \rangle$ \\
$(c,a)$   &   $\langle 1,0 \rangle$ \\
$(c,c)$   &   $\langle 0,1 \rangle$ \\ \hline
\end{tabular}
\end{center}
\hfill{\space} $\Box$
\end{example}

\section{An Application} \label{Application}

Consider the target recognition example
presented in \cite{sbr94}. Here,
an autonomous vehicle needs to identify
objects in a hostile environment such as a military battlefield.
The autonomous vehicle is equipped with a number of sensors which are used
to collect data, such as speed and size of
the objects (tanks) in the battlefield. 
Associated with each sensor, we have a set of rules that describe the type
of the object based on the properties detected by the sensor.

Let us assume that the autonomous vehicle is equipped with three sensors
resulting in data collected about radar readings, 
of the tanks, their gun characteristics and their speeds.
What follows is a set of rules that associate the type of object with
various observations.

\vspace*{.15in}

\noindent
{\bf Radar Readings:} 
\begin{itemize}
\item Reading $r_1$ indicates that the object is a T-72 tank
with belief factor 0.80 and doubt factor 0.15.
\item Reading $r_2$ indicates that the object is a T-60 tank
with belief factor 0.70 and doubt factor 0.20.
\item Reading $r_3$ indicates that the object is not a T-72 tank
with belief factor 0.95 and doubt factor 0.05.
\item Reading $r_4$ indicates that the object is a T-80 tank
with belief factor 0.85 and doubt factor 0.10.
\end{itemize}

\noindent
{\bf Gun Characteristics:}
\begin{itemize}
\item Characteristic $c_1$ indicates that the object is a T-60 tank
with belief factor 0.80 and doubt factor 0.20.
\item Characteristic $c_2$ indicates that the object is not
a T-80 tank with belief factor 0.90 and doubt factor 0.05.
\item Characteristic $c_3$ indicates that the object is a T-72 tank
with belief factor 0.85 and doubt factor 0.10.
\end{itemize}

\noindent
{\bf Speed Characteristics:}
\begin{itemize}
\item Low speed indicates that the object is a T-60 tank
with belief factor 0.80 and doubt factor 0.15.
\item High speed indicates that the object is not a T-72 tank
with belief factor 0.85 and doubt factor 0.15.
\item High speed indicates that the object is not a T-80 tank
with belief factor 0.95 and doubt factor 0.05.
\item Medium speed indicates that the object is not a T-80 tank
with belief factor 0.80 and doubt factor 0.10.
\end{itemize}

\noindent
These rules can be captured in the following three neutrosophic relations:

\begin{center}
\begin{tabular}{|c|c||c|}
\multicolumn{3}{c}{Radar Rules} \\ \hline
Reading   & Object   & Confidence Factors \\ \hline \hline
$r_1$  & T-72 & $\langle 0.80,0.15 \rangle$ \\ \hline
$r_2$  & T-60 & $\langle 0.70,0.20 \rangle$ \\ \hline
$r_3$  & T-72 & $\langle 0.05,0.95 \rangle$ \\ \hline
$r_4$  & T-80 & $\langle 0.85,0.10 \rangle$ \\ \hline
\end{tabular} \\[0.25in]
\begin{tabular}{|c|c||c|}
\multicolumn{3}{c}{Gun Rules} \\ \hline
Reading   & Object   & Confidence Factors \\ \hline \hline
$c_1$  & T-60 & $\langle 0.80,0.20 \rangle$ \\ \hline
$c_2$  & T-80 & $\langle 0.05,0.90 \rangle$ \\ \hline
$c_3$  & T-72 & $\langle 0.85,0.10 \rangle$ \\ \hline
\end{tabular} \\[0.25in]
\begin{tabular}{|c|c||c|}
\multicolumn{3}{c}{Speed Rules} \\ \hline
Reading   & Object   & Confidence Factors \\ \hline \hline
low     & T-60 & $\langle 0.80,0.15 \rangle$ \\ \hline
high    & T-72 & $\langle 0.15,0.85 \rangle$ \\ \hline
high    & T-80 & $\langle 0.05,0.95 \rangle$ \\ \hline
medium  & T-80 & $\langle 0.10,0.80 \rangle$ \\ \hline
\end{tabular}
\end{center}

The autonomous vehicle uses the sensors to make observations
about the different objects and then uses the rules
to determine the type of each object in the battlefield.
It is quite possible that two different sensors may identify the same
object as of different types, thereby introducing inconsistencies.

Let us now consider three objects $o_1$, $o_2$ and $o_3$ which need to be identified by the autonomous vehicle. Let us assume the following
observations made by the three sensors about the three objects.
Once again, we assume certainty factors (maybe derived from the accuracy of the sensors) are associated with each observation.

\begin{center}
\begin{tabular}{|c|c||c|} 
\multicolumn{3}{c}{Radar Data} \\ \hline
Object-id & Reading & Confidence Factors \\ \hline \hline
$o_1$ & $r_3$ & $\langle 1.00,0.00 \rangle$ \\ \hline
$o_2$ & $r_1$ & $\langle 1.00,0.00 \rangle$ \\ \hline
$o_3$ & $r_4$ & $\langle 1.00,0.00 \rangle$ \\ \hline
\end{tabular} \\[0.25in]
\begin{tabular}{|c|c||c|} 
\multicolumn{3}{c}{Gun Data} \\ \hline
Object-id & Reading & Confidence Factors \\ \hline \hline
$o_1$ & $c_3$ & $\langle 0.80,0.10 \rangle$ \\ \hline
$o_2$ & $c_1$ & $\langle 0.90,0.10 \rangle$ \\ \hline
$o_3$ & $c_2$ & $\langle 0.90,0.10 \rangle$ \\ \hline
\end{tabular} \\[.25in]
\begin{tabular}{|c|c||c|} 
\multicolumn{3}{c}{Speed Data} \\ \hline
Object-id & Reading & Confidence Factors \\ \hline \hline
$o_1$ & high   & $\langle 0.90,0.10 \rangle$ \\ \hline
$o_2$ & low    & $\langle 0.95,0.05 \rangle$ \\ \hline
$o_3$ & medium & $\langle 0.80,0.20 \rangle$ \\ \hline
\end{tabular}
\end{center}

\noindent
Given these observations and the rules, we can use the following algebraic
expression to identify the three objects:

\begin{center}
\begin{tabular}{l}
$\widehat{\pi}_{\mbox{Object-id,Object}}(\mbox{Radar Data} ~\widehat{\bowtie}~ \mbox{Radar Rules}) ~\widehat{\cap}~$ \\
$\widehat{\pi}_{\mbox{Object-id,Object}}(\mbox{Gun Data}  ~\widehat{\bowtie}~
\mbox{Gun Rules}) ~\widehat{\cap}~$ \\
$\widehat{\pi}_{\mbox{Object-id,Object}}(\mbox{Speed Data} ~\widehat{\bowtie}~
\mbox{Speed Rules})$
\end{tabular}
\end{center}

\noindent
The intuition behind the intersection is that we would like to 
capture the common (intersecting) information among the three
sensor data.
Evaluating this expression, we get the following neutrosophic relation:

\begin{center}
\begin{tabular}{|c|c||c|} \hline
Object-id & Object & Confidence Factors \\ \hline \hline
$o_1$ & T-72 & $\langle 0.05,0.0 \rangle$ \\ \hline
$o_2$ & T-80 & $\langle 0.0,0.05 \rangle$ \\ \hline
$o_3$ & T-80 & $\langle 0.05,0.0 \rangle$ \\ \hline
\end{tabular}
\end{center}

\noindent
It is clear from the result that by the given information,
we could not infer any useful information that is we could
not decide the status of objects $o_1, o_2$ and $o_3$.

\section{An Infinite-Valued Tuple Relational Calculus}
\label{IVTRC}

As an example, suppose in the e-shopping environment, there are two items 
 $I_1$ and $I_2$, which are evaluated by customers for some
categories of quality $q_1, q_2$ and $q_3$. Let the evaluation results be 
captured by the 
following neutrosophic relation EVAL on scheme $\{I,Q\}$: 

\begin{table}
\caption{EVAL}
\begin{center}
\begin{tabular}{|c|c|c|}
\hline
$I_1$ & $q_1$ & $\langle 0.9,0.2 \rangle$ \\
\hline
$I_1$ & $q_2$ & $\langle 1.0,0.0 \rangle$ \\
\hline
$I_1$ & $q_3$ & $\langle 0.1,0.8 \rangle$ \\
\hline
$I_2$ & $q_1$ & $\langle 1.0,1.0 \rangle$ \\
\hline
$I_2$ & $q_3$ & $\langle 0.8,0.3 \rangle$ \\
\hline 
\end{tabular}
\end{center}
\end{table}

The above neutrosophic relation contains the information that the confidence of 
item $I_1$ was evaluated 
"good" for category $q_1$ is 0.9 and the doubt is 0.2. The confidence of
item $I_1$ was evaluated "good" for category $q_2$ is 1.0 and the doubt is
0.0. The confidence of item $I_1$ was evaluated "poor" for category $q_3$ is
0.8 and the doubt is 0.1. Also, the confidence of item $I_2$ was evaluated
"good" for category $q_1$ is 1.0 and the doubt is 1.0 (similarly, the 
confidence of item $I_2$ was evaluated "poor" for category $q_1$ is 1.0 and 
the doubt is 1.0). The confidence of $I_2$ was evaluated "good" for category 
$q_3$ is 0.8 and the doubt is 0.3. Note that the evaluation results of item 
$I_2$
for category $q_2$ is unknown. The above information contains fuzziness,
incompleteness and inconsistency. Such information may be due to various 
reasons, such as evaluation not conducted, or evaluation results not yet 
available, the
evaluation is not reliable, and different evaluation results for the same 
category producing
different results, etc. 

We define a infinite-valued membership function of a neutrosophic relation, 
which maps tuples to 
the pair of values $\langle \alpha, \beta, \mbox{ with }$ 
$0 \leq \alpha+\beta \leq 2$. We use the symbol {\bf I} to denote the set of
these values, i.e. {\bf I} = $\{\langle \alpha, \beta \rangle \}$. Now, for
a neutrosophic relation $R = \langle t, R(t)^+, R(t)^- \rangle$ on scheme $\Sigma$, 
its
membership function is a 
{\em infinite-valued predicate} $\Phi_{R} : \tau(\Sigma) \rightarrow {\bf I}$, 
given by

\[
\Phi_{R}(t) = \langle R(t)^+, R(t)^- \rangle.
\] 
 
In~\cite{bagai00}, it proposed a 4-valued characteristic function of neutrosophic relation, which maps tuples to one of the following values: $\top$ (for contradiction), {\em t} (for true), {\em f}(for false) and $\bot$ (for unknown).
It can be easily verified that when $R(t)^+ = R(t)^- = 1$, it corresponds to
$\top$; when $R(t)^+ = 1, R(t)^- = 0$, it corresponds to {\em t}; when 
$R(t)^+ = 0, R(t)^- = 1$, it corresponds to {\em f}; and when $R(t)^+ = R(t)^- = 0$, it
corresponds to $\bot$.  

The tuple relational calculus provides a very natural, set-theoretic, 
declarative notation for querying ordinary relational database management
systems. A tuple calculus expression has the form:
\[
\{t \mbox{ of } \Sigma | P(t)\},
\]
where $t$ is a tuple variable, $\Sigma$ a scheme, and $P$ is some 2-valued
predicate on tuples in $\tau(\Sigma)$. The expression denotes the set of all
tuple values $T$ (from $\tau(\Sigma)$) of the variable $t$ for which the 
predicate $P(T)$ is true.

We retain the above simple syntax in the generalised tuple calculus expression
for neutrosophic databases. However, the predicate $P$
is now interpreted as a {\em infinite-valued} predicate on tuples. Moreover,
the entire expression now denotes a neutrosophic relation (of which $P$ is the membership
function).

In this section we define the syntax and semantics of legal infinite-valued
predicate expressions. They are defined in relation to a given set of binary
comparators on domains associated with the attribute names appearing in 
schemes. Most intuitive binary comparators, like $<$ and $\leq$, produce 
2-valued results, but in principle infinite-valued comparators are possible.
The basic building blocks of formulas are {\em atoms}, of which there are 
four kinds:

\begin{enumerate}
\item For any tuple variable $t$ and relation $R$ on the same scheme, 
$t \mbox{ } \tilde{\in} \mbox{ } R$ is an atom. For any tuple value $T$ for the variable 
$t$, the atom
$t \mbox{ } \tilde{\in} \mbox{ } R$ denotes the value $\Phi_R(T)$.

\item For any tuple variable $t_1$ and $t_2$, attribute names $A$ and $B$ in 
the schemes of $t_1$ and $t_2$ respectively, and binary comparator $\Theta$
such that $A$ and $B$ are $\Theta$-comparable, $t_1.A \mbox{ } \Theta \mbox{ } t_2.B$ is an atom. For any tuple values $T_1$ and $T_2$ for the variables $t_1$
and $t_2$ respectively, the atom $t_1.A \mbox{ } \Theta \mbox{ } t_2.B$ denotes
the value $T_1(A) \mbox{ } \Theta \mbox{ } T_2(B)$.

\item For any tuple variable $t$, constant $c$, and attribute names $A$ and
$B$ such that $A$ is in the scheme of $t$, $c \mbox{ } \in \mbox{ } dom(B)$,
and $A$ and $B$ are $\Theta$-comparable, $t.A \mbox{ } \Theta \mbox{ } c$ is
an atom. For any tuple value $T$ for the variable $t$, the atom 
$t.A \mbox{ } \Theta \mbox{ } c$ denotes the value $T(A) \mbox{ } \Theta \mbox{ } c$.

\item For any constant $c$, tuple variable $t$, and attribute names $A$ and
$B$ such that $c \mbox{ } \in \mbox{ } dom(A)$, $B$ is in the scheme of $t$,
and $A$ and $B$ are $\Theta$-comparable, $c \mbox{ } \Theta \mbox{ } t.B$ is
an atom. For any tuple value $T$ for the variable $t$, the atom 
$c \mbox{ } \Theta \mbox{ } t.B$ denotes the value $c \mbox{ } \Theta \mbox{ } T(B)$.
\end{enumerate}

We use infinite-valued connectives $\tilde{\neg}$ (not), $\tilde{\wedge}$ (and),
$\tilde{\vee}$ (or), $\tilde{\exists}$ (there exists) and $\tilde{\forall}$ (for all)
to recursively build {\em formulas} from atoms. Any atom is a formula, where
the formula denotes the same value as the atom.

If $f$ and $g$ are formulas, and $f^+$ is truth-degree of the $f$, $f^-$ is
falsity-degree of $f$, then $\tilde{\neg} \mbox{ } f$, $f \mbox{ } \tilde{\wedge} \mbox{ } g$ and $f \mbox{ } \tilde{\vee} \mbox{ } g$ are also formulas. The
values of such formulas are given as the following: \\

\begin{equation}
\tilde{\neg} \mbox{ } f = \langle f^-, f^+ \rangle 
\end{equation}

\begin{equation}
f \mbox{ } \tilde{\wedge} \mbox{ } g = \langle \min(f^+,g^+), \max(f^-,g^-) \rangle
\end{equation}

\begin{equation}
f \mbox{ } \tilde{\vee} \mbox{ } g = \langle \max(f^+,g^+), \min(f^-,g^-) \rangle
\end{equation}

An intuitive appreciation of the disjunctive connective can be obtained as
follows: Given a tuple $t$, since we believed that it is present in the 
relation $R$ with confidence $R(t)^+$ and that it is present in the relation
$S$ with confidence $S(t)^+$, we can now believe that the tuple $t$ is present
in the ``either-$R$-or-$S$" relation with confidence which is equal to the
larger of $R(t)^+$ and $S(t)^+$. Using the same logic, we can now believe in
the absence of the tuple $t$ from the ``either-$R$-or-$S$" relation with
confidence which is equal to the smaller (because $t$ must be absent from
both $R$ and $S$ for it to be absent from the disjunction) of $R(t)^-$ and
$S(t)^-$. The definition of {\em negation} and {\em conjunction} can be 
understood in the same way.

The duality of $\tilde{\wedge}$ and $\tilde{\vee}$ is evident from the above formulas. It is interesting to note the algebraic laws shown in Table~\ref{chapter3:table2} 
that are exhibited by these connectives.

\begin{table}[htb!]
\caption{Albegraic Properties of Infinite-Valued Propositional Connectives}
\label{chapter3:table2}
\begin{center}
\begin{tabular}{|ll|}
\hline
(commutative laws) & $f \mbox{ } \tilde{\vee} \mbox{ } g = g \mbox{ } \tilde{\vee} \mbox{ } f$ \\
& $f \mbox{ } \tilde{\wedge} \mbox{ } g = g \mbox{ } \tilde{\wedge} \mbox{ } f$ \\
\hline 
(associative laws) & $(f \mbox{ } \tilde{\vee} \mbox{ } g) \mbox{ } \tilde{\vee} \mbox{ } h = f \mbox{ } \tilde{\vee} \mbox{ } (g \mbox{ } \tilde{\vee} \mbox{ } h)$ \\
& $(f \mbox{ } \tilde{\wedge} \mbox{ } g) \mbox{ } \tilde{\wedge} \mbox{ } h = f \mbox{ } \tilde{\wedge} \mbox{ } (g \mbox{ } \tilde{\wedge} \mbox{ } h)$ \\
\hline 
(distributive laws) & $f \mbox{ } \tilde{\vee} \mbox{ } (g \mbox{ } \tilde{\wedge} \mbox{ } h) = (f \mbox{ } \tilde{\vee} \mbox{ } g) \mbox{ } \tilde{\wedge} \mbox{ } (f \mbox{ } \tilde{\vee} \mbox{ } h)$ \\
& $f \mbox{ } \tilde{\wedge} \mbox{ } (g \mbox{ } \tilde{\vee} \mbox{ } h) = (f \mbox{ } \tilde{\wedge} \mbox{ } g) \mbox{ } \tilde{\vee} \mbox{ } (f \mbox{ } \tilde{\wedge} \mbox{ } h)$ \\
\hline
(idempotent laws) & $f \mbox{ } \tilde{\vee} \mbox{ } f = f$ \\
& $f \mbox{ } \tilde{\wedge} \mbox{ } f = f$ \\
\hline
(identity laws) & $f \mbox{ } \tilde{\vee} \mbox{ } {\bf f} = f$ \\
& $f \mbox{ } \tilde{\wedge} \mbox{ } {\bf t} = f$ \\
\hline
(double complementation) & $\tilde{\neg} \mbox{ } (\tilde{\neg} \mbox{ } f) = f$ \\
\hline
(De Morgan laws) & $\tilde{\neg} \mbox{ } (f \mbox{ } \tilde{\vee} \mbox{ } g) = \tilde{\neg} \mbox{ } f \mbox{ } \tilde{\wedge} \mbox{ } \tilde{\neg} \mbox{ } g$ \\
& $\tilde{\neg} \mbox{ } (f \mbox{ } \tilde{\wedge} \mbox{ } g) = \tilde{\neg} \mbox{ } f \mbox{ } \tilde{\vee} \mbox{ } \tilde{\neg} \mbox{ } g$ \\
\hline
\end{tabular}
\end{center}
\end{table}

If $t$ is a tuple variable, $\Sigma$ a scheme, and $P$ an infinite-valued
predicate on tuples in $\tau(\Sigma)$, then $\tilde{\exists} t \mbox{ of } \Sigma | P(t)$ and $\tilde{\forall} t \mbox{ of } \Sigma | P(t)$ are formulas. If $P$ is
the membership function of the neutrosophic relation
$R$, then the values denoted by these formulas are given by \\

\begin{equation}
\tilde{\exists}t \mbox{ of } \Sigma | P(t) = \langle t_{\tilde{\exists}}, f_{\tilde{\exists}} \rangle, 
\end{equation}

$\mbox{where } t_{\tilde{\exists}} = \max\{R(t)^+\}, \mbox{ for all } t \in \tau(\Sigma), f_{\tilde{\exists}} = \min\{R(t)^-\}, \mbox{ for all } t \in \tau(\Sigma)$. 

\begin{equation}
\tilde{\forall}t \mbox{ of } \Sigma | P(t) = \langle t_{\tilde{\forall}}, f_{\tilde{\forall}} \rangle, 
\end{equation}

$\mbox{where } t_{\tilde{\forall}} = \min\{R(t)^+\}, \mbox { for all } t \in \tau(\Sigma), f_{\tilde{\exists}} = \max\{R(t)^+\}, \mbox{ for all } t \in \tau(\Sigma)$. \\

The extended De Morgan laws can be verified to continue to hold for our
generalized infinite-valued semantics for quantifiers, i.e. the following
pairs of formulas are equivalent:
\[
\tilde{\exists}t \mbox{ of } \Sigma | P(t) \equiv \tilde{\neg} \mbox{ } (\tilde{\forall}t \mbox{ of } \Sigma | \tilde{\neg} \mbox{ } P(t))
\]
\[
\tilde{\forall}t \mbox{ of } \Sigma | P(t) \equiv \tilde{\neg} \mbox{ } (\tilde{\exists}t \mbox{ of } \Sigma | \tilde{\neg} \mbox{ } P(t))
\]
\[
\tilde{\exists}t \mbox{ of } \Sigma | (P(t) \mbox{ } \tilde{\wedge} \mbox{ } Q(t)) \equiv \tilde{\neg} \mbox{ }(\tilde{\forall}t \mbox{ of } \Sigma | \tilde{\neg} \mbox{ } P(t) \mbox{ } \tilde{\vee} \mbox{ } \tilde{\neg} \mbox{ } Q(t))
\]
\[
\tilde{\exists}t \mbox{ of } \Sigma | (P(t) \mbox{ } \tilde{\vee} \mbox{ } Q(t)) \equiv \tilde{\neg} \mbox{ }(\tilde{\forall}t \mbox{ of } \Sigma | \tilde{\neg} \mbox{ } P(t) \mbox{ } \tilde{\wedge} \mbox{ } \tilde{\neg} \mbox{ } Q(t))
\]
\[
\tilde{\forall}t \mbox{ of } \Sigma | (P(t) \mbox{ } \tilde{\wedge} \mbox{ } Q(t)) \equiv \tilde{\neg} \mbox{ } (\tilde{\exists}t \mbox{ of } \Sigma | \tilde{\neg} \mbox{ } P(t) \mbox{ } \tilde{\vee} \mbox{ } \tilde{\neg} \mbox{ } Q(t))
\]
\[
\tilde{\forall}t \mbox{ of } \Sigma | (P(t) \mbox{ } \tilde{\vee} \mbox{ } Q(t)) \equiv \tilde{\neg} \mbox{ } (\tilde{\exists}t \mbox{ of } \Sigma | \tilde{\neg} \mbox{ } P(t) \mbox{ } \tilde{\wedge} \mbox{ } \tilde{\neg} \mbox{ } Q(t))
\]

It is worth mentioning that in ordinary 2-valued relational calculus caution
needs to be exercised in mixing negation and quantifiers in a {\em safe} manner
as the resulting expressions have the potential of denoting infinite relations,
even if all components denote finite relations. Fortunately, as neutrosophic 
databases are by nature capable of handling infinite
relations, safety of expressions is not an issue in infinite-valued calculus.

\subsection{An Example}
\label{example}
Let us now consider an example illustrating some infinite-valued computations.
We use the neutrosophic relation EVAL on scheme 
$\{I,Q\}$ of the item-category evaluation as example.

Consider the query: \\

{\em What items showed contradictory evaluation results for some category?} \\

In ordinary relational databases, it is impossible to store contradictory
information, let alone entertaining queries about contradiction.

Let $\Delta = \{I\}$, and $\Sigma = \{I,Q\}$ be schemes. A tuple calculus
expression for this query is: 
\[
\{d \mbox{ of } \Delta | (\tilde{\exists}t \mbox{ of } \Sigma | t.I = d.I \mbox{ } \tilde{\wedge} \mbox{ } t \mbox{ } \tilde{\in} \mbox{ } \mbox{EVAL} \mbox{ } 
\mbox{          }\tilde{\wedge} \mbox{ } \tilde{\neg} \mbox{ } t \mbox{ } \tilde{\in} \mbox{ EVAL})\}
\]
In the ordinary 2-valued logic the above query will produce an empty answer
due to the condition for the tuple $t$ to simultaneously be in EVAL as
well as not be in EVAL. In infinite-valued logic, however, the query
denotes that neutrosophic relation on scheme $\Delta$
whose membership function is denoted by the infinite-valued predicate 
expression

\begin{equation}
\label{eqn1}
\tilde{\exists}t \mbox{ of } \Sigma | t.I = d.I \mbox{ } \tilde{\wedge} \mbox{ } t \mbox{ } \tilde{\in} \mbox{ } \mbox{EVAL} \mbox{ } \\
\tilde{\wedge} \mbox{ } \tilde{\neg} \mbox{ } t \mbox{ } \tilde{\in} \mbox{ EVAL}
\end{equation}

That function can be computed by determining the value of the above expression
for all possible values of its free variable $d$, namely $I_1$ and $I_2$.

For the value $d = I_1$, the expression (~\ref{eqn1}) can be seen to reduce
to the value $\langle 0.2,0.8 \rangle$. For the value $d = I_2$, the expression
 (~\ref{eqn1}) can be seen to reduce to the value $\langle 1.0,0.8 \rangle$.
The result is the neutrosophic relation:

\begin{table}[htb!]
\begin{center}
\begin{tabular}{|c|c|}
\hline
$I_1$ & $\langle 0.2, 0.8 \rangle$ \\
\hline
$I_2$ & $\langle 1.0, 0.8 \rangle$ \\
\hline
\end{tabular}
\end{center}
\end{table}

The result states that $I_1$ showed contradictory evaluation result
for some category with confidence is 0.2 and doubt is 0.8, so it is safe to
conclude that $I_1$ did not show contradictory evaluation resulti, but
$I_2$ showed contradictory evaluation result for some category with confidence
1.0 and doubt is 0.8, the explanation is that $I_2$ did show contradictory
result for some category and did not show contradictory for other category at
the same times.

\section{A Generalized SQL Query Construct for Neutrosophic Relations}
The most popular construct for information retrieval from most commercial systems is the SQL's SELECT statement. While the statement has many options and extensions to its basic form, here we just present an infinite-valued generalization to the basic form, as generalizing the options then just becomes a trivial matter of detail. The basic form of the statement contains three clauses select, from and where, and has the following format:
\begin{center}
                                select $A_1,A_2, \ldots A_m$
                                from $R_1,R_2, \ldots R_n$
                                where $C$
\end{center}
where
\begin{enumerate}
\item $A_1,A_2, \ldots A_m$ is a list of attribute names whose values are to be retrieved by the query,
\item  $R_1,R_2, \ldots R_n$ is a list of relation names required to process the query, and
\item $C$ is a boolean expression that identifies the tuples to be retrieved by the query.
\end{enumerate}

Without loss of generality, we assume that each attribute name occurs in exactly one relation, because if some attribute $A_i$  occurs in more than one relation, we require, instead of simply the attribute Ai, a pair of the form $R_j.A_i$ qualifying that attribute.
The result of the SELECT statement is a relation with attributes $A_1,A_2, \ldots A_m$ chosen from the attributes of $R_1 \times R_2 \times \cdots \times R_n$ for tuples that satisfy the boolean condition $C$, i.e.

\[
\pi_{A_1,A_2, \ldots A_m}(\sigma_C(R_1 \times R_2 \times \cdots \times R_n)),
\]

where $\pi, \sigma,$ and $\times$ are the projection, selection and product operations, respectively, on ordinary relations.
We retain the above syntax in the generalized SELECT statement for the neutrosophic relations. However, the relation names $R_1,R_2, \ldots R_n$
now represent some neutrosophic relations and C is some infinite-valued condition. The result of the generalized SELECT statement is then the value of the algebraic expression:

\[
{\pi}_{A_1,A_2,\ldots,A_m}(\widehat{\sigma}_C(R_1 \widehat{\times} R_2 \widehat{\times} \cdots \widehat{\times} R_n)),
\]

where $\widehat{\pi}, \widehat{\sigma},$ and $\widehat{\times}$ are, respectively, the projection, selection and product operations on neutrosophic relations constructed in the next section. Furthermore, the result of the generalized SELECT statement is also a 
neutrosophic relation.

\subsection{Infinite-Valued Conditions}
In the generalized SELECT statement, we let the condition occurring in the where clause be infinite-valued. The infinite values, except $\langle 1, 0 \rangle$ and $\langle 0, 1 \rangle$, arise essentially due to any nested subqueries. For any arithmetic expressions $E_1$ and $E_2$, comparisons such as 
$E_1 \leq E_2$ are simply 2-valued conditions ($\langle 1,0 \rangle$ or $\langle 0,1 \rangle$).
Let $\xi$ be a subquery of the form

\begin{center}
(select $\ldots$ from $\ldots$ where $\ldots$)
\end{center}

occurring in the where clause of a SELECT statement. And let $R$ be the neutrosophic 
relation on scheme $\Sigma$ that the subquery $\xi$ evaluates to. Then, conditions involving the subquery $\xi$ evaluate as follows.

\begin{enumerate}
\item The condition
\begin{center}
exists $\xi$
\end{center}

evaluates to $\langle \alpha, \beta \rangle$,
\[
\alpha = \max\{a\}, a = R(t)^+, \mbox{ for all } t \in \tau(\Sigma),
\]
\[
\beta = \min\{b\}, b = R(t)^-, \mbox{ if } R(t)^+ + R(t)^- \leq 1, b = 1 - R(t)^+, \mbox{ if } R(t)^+ + R(t)^- > 1, \mbox{ for all } t \in \tau(\Sigma).
\]

\item For any tuple $t \in \tau(\Sigma)$, the condition
\begin{center}
$t$ \mbox{ in } $\xi$ 
\end{center}
evaluates to $\phi_R(t)$.

\item If $\Sigma$ contains exactly one attribute, then for any (scalar value) 
$t \in \tau(\Sigma)$, the condition

\[
t >\mbox{any } \xi
\]

evaluates to $\langle \alpha, \beta \rangle$, \\
$
\alpha = \max\{a\}, a = R(k)^+, \mbox{ if } t > k, \mbox{ for some } k \in R,
(\beta = \min\{b\}, b = R(k)^-, \mbox{ if } R(k)^+ + R(k)^- \leq 1,$ \\
$b = 1-R(k)^+,  
\mbox{ if } R(k)^+ + R(k)^- > 1), \mbox{ if } t > k, \mbox{ for some } k \in R;
$ \\
$
\alpha = 0, \beta = 1, \mbox{ otherwise}.
$

An infinite-valued semantics for other operators, such as $\geq$any, =any, can be defined similarly. Note that conditions involving such operators never evaluate to the value 
$\alpha, \beta$, such that $\alpha + \beta > 1$.

\item If $\Sigma$ contains exactly one attribute, then for any (scalar value)
$t \in \tau(\Sigma)$, the condition
\[
t >\mbox{all } \xi
\]
evaluates to $\langle \alpha, \beta \rangle$, \\
$
(\alpha = \min\{a\}, a = R(k)^-, \mbox{ if } R(k)^+ + R(k)^- \leq 1, a = 1-R(k)^+,  
\mbox{ if } R(k)^+ + R(k)^- > 1), \mbox{ if } t \leq k,$ \\
$\mbox{ for some } k \in R,
\beta = \max\{b\}, b = R(k)^+, \mbox{ if } t \leq k, \mbox{ for some } k \in R;
$ \\
$
\alpha = 1, \beta = 0, \mbox{ otherwise}.
$

An infinite-valued semantics for other operators, such as $\geq$all, $=$all, can be defined similarly. Note that conditions involving such operators never evaluate to the value 
$\langle \alpha, \beta \rangle$, such that $\alpha + \beta > 1$.

\end{enumerate}

We complete our infinite-valued semantics for conditions by defining the {\bf not, and} and {\bf or} operators on them. Let $C$ and $D$ be any conditions, and value of $C = \langle t_c, f_c \rangle$ and value of $D = \langle t_d, f_d \rangle$. Then, the value of the condition {\bf not} $C$ is given by
\[
\mbox{{\bf not }} C = \langle f_c, t_c \rangle
\]
while the value of the condition $C$ {\bf and} $D$ is given by
\[
C \mbox{ {\bf and }} D = \langle \min{t_c, t_d}, \max{f_c, f_d} \rangle
\]
and that of the condition $C$ {\bf or} $D$ is given by
\[
C \mbox{ {\bf or }} D = \langle \max{t_c, t_d}, \min{f_c, f_d} \rangle
\]
The duality of {\bf and} and {\bf or} is evident from their formulas. It is interesting to note the following algebraic laws exhibited by the above infinite-valued operators:

\begin{enumerate}
\item Double Complementation Law:
\[
\mbox{ {\bf not }} (\mbox{ {\bf not }} C) = C
\]

\item Identity and Idempotence Laws:
\[
C \mbox{ {\bf and }} \langle 1, 0 \rangle = C \mbox{ {\bf and }} C = C
\]
\[
C \mbox{ {\bf or }} \langle 0, 1 \rangle = C \mbox{ {\bf or }} C = C
\]

\item Commutativity Laws:
\[
C \mbox{ {\bf and }} D = D \mbox{ {\bf and }} C
\]
\[
C \mbox{ {\bf or }} D = D \mbox{ {\bf or }}C
\]

\item Associativity Laws:
\[
C \mbox{ {\bf and }} (D \mbox{ {\bf and }} E) = (C \mbox{ {\bf and }} D) \mbox{ {\bf and }} E
\]
\[
C \mbox{ {\bf or }} (D \mbox{ {\bf or }} E) = (C \mbox{ {\bf or }} D) \mbox{ {\bf or }} E
\]

\item Distributivity Laws:
\[
C \mbox{ {\bf and }} (D \mbox{ {\bf or }} E) = (C \mbox{ {\bf and }} D) \mbox{ {\bf or }} (C \mbox{{\bf and }} E)
\]
\[
C \mbox{ {\bf or }} (D \mbox{ {\bf and }} E) = (C \mbox{ {\bf or }} D) \mbox{ {\bf and }} (C \mbox{{\bf or }} E)
\]

\item De Morgan Laws:
\[
\mbox{ {\bf not }} (C \mbox{ {\bf and }} D) = (\mbox{ {\bf not }} C) \mbox{ {\bf or }}(\mbox{ {\bf not }} D)
\]
\[
\mbox{ {\bf not }} (C \mbox{ {\bf or }} D) = (\mbox{ {\bf not }} C) \mbox{ {\bf and }} (\mbox{ {\bf not }} D)
\]

\end{enumerate}

We are now ready to define the selection operator on neutrosophic relations.

Let $R$ be a neutrosophic relation on scheme $\Sigma$, and $C$ be an infinite-valued condition on tuples of $\Sigma$ denoted $\langle t_C(t), f_C(t) \rangle$. 
Then, the selection of $R$ by $C$, denoted $\widehat{\sigma}_C(R)$, is a neutrosophic
relation on scheme $\Sigma$, given by

\[
(\widehat{\sigma}_C(R))(t) = \langle \min{R(t)^+, t_C(t)}, \max{R(t)^-, f_C(t)} \rangle.
\]

The above definition is similar to that of the {\bf and }operator given earlier.

Since performing a simple union is impossible within a SELECT statement, SQL provides a union operator among subqueries to achieve this. We end this section with an infinite-valued semantics of union.

Let $\xi_1$ and $\xi_2$ be subqueries that evaluate, respectively, to neutrosophic relations 
$R_1$ and $R_2$ on scheme $\Sigma$. Then, the subquery
\[
\xi_1 \mbox{ {\bf union }} \xi_2
\]
evaluates to the neutrosophic relation $R$ on scheme $\Sigma$ given by
\[
R(t) = \langle \max{R1(t)^+, R2(t)^+}, \min{R1(t)^-, R2(t)^-} \rangle
\]
An intuitive appreciation of the union operator can be obtained as follows: Given a tuple $t$, since we believed that it is present in the relation $R_1$ with confidence $R_1(t)^+$ and that it is present in the relation $R_2$ with confidence $R_2(t)^+$, we can now believe that the tuple t is present in the "either-$R_1$-or-$R_2$" relation with confidence which is equal to the larger of $R_1(t)^+$ and $R_2(t)^+$. Using the same logic, we can now believe in the absence of the tuple t from the "either-$R_1$-or-$R_2$" relation with confidence which is equal to the smaller (because $t$ must be absent from both $R_1$ and $R_2$ for it to be absent from the union) of $R_1(t)^-$ and $R_2(t)^-$. 

\subsection{An Example}
Let us now consider an example illustrating some infinite-valued computations. 
We use the neutrosophic relation EVAL on scheme 
$\{I,Q\}$ of the item-category evaluation as example.

Consider the query:
\begin{center}
{\emph What items showed contradictory evaluation of some category of quality?}
\end{center}

A SELECT statement for this query is:
\begin{center}
select $I$ \\
from EVAL
where {\bf not } (($I, Q$) {\bf in } EVAL)
\end{center}

One possible evaluation method for the above query in ordinary 2-valued SQL is to produce the $I$ attribute of those rows of EVAL that satisfy the {\bf where } condition. Since the {\bf where } condition in the above case is exactly that row not be in EVAL, in 2-valued logic the above query will produce an empty answer.

In infinite-valued logic, however, the where condition needs to be evaluated, to one of infinite possible values, for every possible row with attributes $\Sigma = (I, Q)$. The result is then combined with EVAL according to the semantics of  p, on which $\widehat{\sigma}$ is performed to produce the resulting neutrosophic relation.

Therefore, for each of the 6 rows in $\tau(\Sigma)$, we first evaluate the where condition $C$:

\begin{table}
\begin{center}
\begin{tabular}{|c|c|}
\hline
$(I,Q)$ & $C = \mbox{ {\bf not }}((I,Q) \mbox{ {\bf in}} EVAL)$ \\
\hline
$(I_1,q_1)$ & $\langle 0.2,0.9 \rangle$ \\
\hline
$(I_1,q_2)$ & $\langle 0.0,1.0 \rangle$ \\
\hline
$(I_1,q_3)$ & $\langle 0.8,0.1 \rangle$ \\
\hline
$(I_2,q_1)$ & $\langle 1.0,1.0 \rangle$ \\
\hline
$(I_2,q_2)$ & $\langle 0.0,0.0 \rangle$ \\
\hline
$(I_2,q_3)$ & $\langle 0.3,0.8 \rangle$ \\
\hline 
\end{tabular}
\end{center}
\end{table}

Now, $\widehat{\sigma}(EVAL)$ according to the definition of $\widehat{\sigma}$ evaluates to the neutrosophic relation:
\begin{table}
\begin{center}
\begin{tabular}{|c|c|c|}
\hline
$I_1$ & $q_1$ & $\langle 0.2,0.9 \rangle$ \\
\hline
$I_1$ & $q_2$ & $\langle 0.0,0.1 \rangle$ \\
\hline
$I_1$ & $q_3$ & $\langle 0.1,0.8 \rangle$ \\
\hline
$I_2$ & $q_1$ & $\langle 1.0,1.0 \rangle$ \\
\hline
$I_2$ & $q_3$ & $\langle 0.3,0.8 \rangle$ \\
\hline 
\end{tabular}
\end{center}
\end{table}

Finally, $\widehat{\pi}$ of the above is the neutrosophic relation:
\begin{table}
\begin{center}
\begin{tabular}{|c|c|c|}
\hline
$I_1$ & $\langle 0.1,0.8 \rangle$ \\
\hline
$I_2$ & $\langle 1.0,0.0 \rangle$ \\
\hline 
\end{tabular}
\end{center}
\end{table}

\section{Conclusions} \label{Conclusion}

We have presented a generalization of fuzzy relations, 
intuitionistic fuzzy relations (interval-valued fuzzy relations)
and paraconsistent relations,
called neutrosophic relations,  in
which we allow the representation of confidence (belief and doubt)
factors with each tuple. The algebra on fuzzy relations is
appropriately generalized to manipulate neutrosophic
relations. 

Various possibilities exist for further study in this area. 
Recently, there has been some work in extending logic programs
to involve quantitative paraconsistency. Paraconsistent logic programs
were introduced in \cite{blr89} and probabilistic logic programs in 
\cite{ngs92}. Paraconsistent logic programs 
allow negative atoms to appear in the
head of clauses (thereby resulting in the possibility of
dealing with inconsistency), and probabilistic logic programs associate 
confidence measures with literals and with entire clauses. 
The semantics of these
extensions of logic programs have already been presented, but
implementation strategies to answer queries have not been discussed.
We propose to use the model introduced in this chapter 
in computing the
semantics of these extensions of logic programs.
Exploring application areas is another important thrust of
our research.  

We developed two notions of generalising operators on fuzzy  
relations
for neutrosophic relations.
Of these, the stronger notion guarantees that any generalised  
operator is
``well-behaved'' for neutrosophic relation 
operands that contain
consistent information.

For some well-known operators on fuzzy relations, such as union,  
join,
projection, we introduced generalised operators on neutrosophic 
relations.
These generalised operators maintain the belief system intuition  
behind
neutrosophic relations, and are shown to 
be ``well-behaved'' in the  
sense
mentioned above.

Our data model can be used to represent relational information that  
may be incomplete and inconsistent.
As usual, the algebraic operators can be used to construct queries to  
any database systems for retrieving vague information.

%% file: chaptr4.tex
\chapter{Soft Semantic Web Services Agent}
\label{SSWSA}

{\small
Web services technology is critical for the success of business integration and 
other application fields such as bioinformatics. However, there are two 
challenges facing the practicality of Web services: (a) efficient location of the 
Web service registries that contain the requested Web services and (b) efficient
retrieval of the requested services from these registries with high quality of 
service (QoS). The main reason for this problem is that current Web services 
technology is not semantically oriented. Several proposals have been made to 
add semantics to Web services to facilitate discovery and composition of relevant 
Web services. Such proposals are being referred to as Semantic Web services (SWS).
However, most of these proposals do not address the second problem of retrieval of
Web services with high QoS. In this chapter, we propose a framework called Soft Semantic 
Web Services Agent (soft SWS agent) for providing high 
QoS Semantic Web services using soft computing methodology. Since different 
application domains have different requirements for QoS, it is impractical to
use classical mathematical modeling methods to evaluate the QoS of semantic
Web services. We use neutrosophic neural networks with Genetic Algorithms (GA) as our
study case. Simulation results show that the soft SWS agent methodology is extensible and
scalable to handle fuzzy, uncertain and inconsistent QoS metrics effectively. 
}

\section{Introduction}
\label{introduction}
Web services are playing an important role in e-business application 
integration and other application fields such as bioinformatics. So it is
crucial for the success of both service providers as well as service consumers to provide
and invoke the high quality of service (QoS) Web services. Unfortunately,
current Web services technologies such as SOAP (Simple Object Access Protocol) \cite{soap}, 
WSDL (Web Services Description Language) \cite{wsdl}, 
UDDI (Universal Description, Discovery and Integration) \cite{uddi}, 
ebXML (Electronic Business XML Initiative) \cite{ebxml}, XLANG \cite{xlang}, 
WSFL (Web Services Flow Language) \cite{wsfl}, 
BPEL4WS (Business Process Execution Language for Web Services) \cite{bpel4ws}, and 
BSML (Bioinformatic Sequence Markup Language) \cite{bsml} are all syntax-oriented with
little or no semantics associated with them. Computer programs may read and parse them, but 
with little or no semantic information associated with these technologies, the computer
programs can do little to reason and infer knowledge about the Web services.

Current research trend is to add semantics to the Web services 
framework to facilitate the discovery, invocation, composition, and execution
monitoring of Web services. Web services with explicit semantic annotation are
called Semantic Web services (SWS). Several projects are underway to try to
reach such a goal. For example, OWL-S (previously DAML-S \cite{damls} from OWL 
Services Coalition \cite{owls}) uses OWL based ontology for describing Web 
services. 
METEOR-S \cite{svsm} follows the way that relates concepts in WSDL to 
DAML+OIL ontologies in Web services description, and then provides an interface to UDDI
that allows querying based on ontological concepts. The Internet Reasoning
Service (IRS-II) \cite{mdcg} is a Semantic Web services framework, which allows
applications to semantically describe and execute Web services. IRS-II is based
on the UPML framework \cite{ocfb}. The Web Service Modeling Framework (WSMF) \cite{fb}
provides a model for describing the various aspects related to Web services.
Its main goal is to fully enable e-commerce by applying Semantic Web technology
to Web services.

In our vision, with the maturing of semantic Web services technologies, there will
be a proliferation of public and/or private registries for hosting and querying semantic
Web services based on specific ontologies. Currently, there
are many public and private UDDI registries advertising numerous similar Web
services with different QoS. For example, GenBank \cite{genebank}, 
XEMBL \cite{xembl}, and OmniGene \cite{omnigene} all provide similar Web services
with different quality of services. There are two challenges existing for automatic 
discovery and invocation of Web services. One is the efficient location of service registries
advertising requested Web services and the another is the efficient retrieval of
the requested services from these registries with the highest quality of 
service (QoS). The semantic Web services technologies that we mentioned above
can be exploited to solve the first challenge. For the second challenge, we
believe that the QoS of semantic Web services should cover both functional and
non-functional properties. Functional properties include the input, output,
conditional output, pre-condition, access condition, and the effect of service \cite{mn}.
These functional properties can be characterized as the capability of the
service \cite{abh}. Non-functional properties include the availability, 
accessibility, integrity, performance, reliability, regulatory, security,
response time and cost \cite{mn} of the Web service.

Several matchmaking schemes have already been proposed to match the service
requestor's requirements with service provider's advertisement \cite{ctb,skwl,pps}. 
These schemes basically try to solve the capability matching problem. 
Here, we must be aware that on the one hand, the degree of
capability matching and non-functional properties are all fuzzy, and on the 
other hand, different application domains have different requirements on 
non-functional properties. As a consequence, it is not flexible to use classical mathematical
modeling methods to evaluate the QoS of semantic Web services. Although
there are several existing QoS models \cite{csz,fk,gar,ggps,hsuw,rom,sh,scmk,zbs}, 
none of them are suitable for the requirements considered in this chapter.
These QoS models are based on precise QoS metrics and
specific application domains. They cannot handle fuzzy and uncertain QoS 
metrics.

In this chapter, we propose a framework called {\em soft semantic Web services agent} (soft SWS
agent) to provide high QoS semantic Web services based on specific domain ontology
such as gnome. The soft SWS agent could solve the forementioned two challenges
effectively and efficiently. The soft SWS agent itself is implemented as a
semantic Web service and comprises of six components:
(a) Registries Crawler, (b) Repository, (c) Inquiry Server, (d) Publish Server,
(e) Agent Communication Server, and (f) Intelligent Inference Engine. The core of the
soft SWS agent is Intelligent Inference Engine (IIE). It uses soft computing
technologies to evaluate the entire QoS of semantic Web services using both functional and 
non-functional properties. In this chapter, we use semantic Web services for bioinformatics 
as a case study. We employ neutrosophic neural networks with Genetic Algorithms (GA) for the IIE
component of our soft SWS agent. The case study illustrates the flexibility and reliability of 
soft computing methodology for 
handling fuzzy and uncertain linguistic information. For example, capability
of a Web service is fuzzy. It is unreasonable to use crisp values to describe it.
So we can use several linguistic variables such as a "little bit low" and "a little bit high"
to express the capability of services. 

The chapter is organized as follows. In section 2, we present the necessary
background of the QoS model, semantic Web services, and soft computing
methodology. In section 3, we provide the architecture of the
extensible soft SWS agent. In section 4, we present the design of the neutrosophic
neural network
with GA and simulation results. In section 5, we present related work, and
finally, in section 6, we present conclusions and possibilities for future research.

\section{Background}
\label{background}

This section details the background material related to this research. We cover
traditional Web services, semantic Web, semantic Web services, soft computing
methodology, and the QoS model.

\subsection{Traditional Web services}

Web services are modular, self-describing, and self-contained applications that
are accessible over the internet \cite{cnw}. The core components of the Web 
services infrastructure are XML based standards like SOAP, WSDL, and UDDI.
SOAP is the standard messaging protocol for Web services. SOAP messages consist of
three parts: an envelope that defines a framework for describing what is in a
message and how to process it, a set of encoding rules for expressing instances
of application-defined datatypes, and a convention for representing remote
procedure calls and responses. WSDL is an XML format to describe Web services
as collections of communication endpoints that can exchange certain messages.
A complete WSDL service description provides two pieces of information: an
application-level service description (or abstract interface), and the specific
protocol-dependent details that users must follow to access the service at a
specified concrete service endpoint. The UDDI specifications offer users a
unified and systematic way to find service providers through a centralized
registry of services that is roughly equivalent to an 
automated online ``phone directory" of Web services. UDDI provides two basic specifications 
that define a service registry's structure and operation. One is a definition
of the information to provide about each service and how to encode it and the  
other is a publish and query API for the registry that describes how this 
information can be published and accessed.

\subsection{Semantic Web}

The current Web is just a collection of documents which are human readable
but not machine processable. In order to remedy this disadvantage, the 
concept of semantic Web is proposed to add semantics to the Web to facilitate
the information finding, extracting, representing, interpreting and 
maintaining. ``The semantic Web is an extension of the current Web in which
information is given well-defined meaning, better enabling computers and
people to work in cooperation" \cite{lhl01}. The core concept of semantic Web is
ontology. ``Ontology is a set of knowledge terms, including the vocabulary,
the semantic interconnections, and some simple rules of inference and logic
for some particular topic" \cite{hen}. There are many semantic Web technologies
available today, such as RDF \cite{rdf}, RDFS \cite{rdfs}, DAML+OIL \cite{damloil} 
and 
OWL \cite{owl}. The description logics are used as the inference mechanism for
current semantic Web technologies. There are some drawbacks in the description
logics \cite{srt}. It cannot handle fuzziness and uncertainty associated with 
concept membership. The current research trend is to combine soft computing 
with semantic Web \cite{str98,str04,klp,dp}.

\subsection{Semantic Web Services}

The industry is proposing Web services to transform the Web from ``passive
state"--repository of static documents to ``positive state"--repository of
dynamic services. Unfortunately, the current Web services standards are not
semantic-oriented. They are awkward for service discovery, invocation,
composition, and monitoring. So it is natural to combine the semantic Web with 
Web services, the so-called semantic Web services. Several projects have been
initiated to design the framework for semantic Web services such as OWL-S,
IRS-II, WSMF and METEOR-S.

For example, OWL-S 1.0 which is based on OWL is the upper ontology for 
services. It has three subontologies: ServiceProfile, ServiceModel and 
ServiceGrounding. The service profile tells ``what the service does"; this is,
it gives the types of information needed by a service-seeking agent to 
determine whether the service meets its needs. The service model tells
``how the service works"; that is, it describes what happens when the service
is carried out. A service grounding specifies the details of how an agent
can access a service. Typically a grounding will specify a communication
protocol, message formats, and other service-specific details such as port
numbers used in contacting the service. In addition, the grounding must
specify, for each abstract type specified in the ServiceModel, an unambiguous
way of exchanging data elements of that type with the service.

\subsection{Soft Computing Methodology}

``Soft computing differs from conventional (hard) computing in that, unlike
hard computing, it is tolerant of imprecision, uncertainty, partial truth,
and approximations" \cite{zadeh}. The principal constituents of soft computing are
fuzzy logic, neural networks, and generic algorithms. More and more technologies
will join into the soft computing framework in the near future. Fuzzy logic is
primarily concerned with handling imprecision and uncertainty, neural computing
focuses on simulating human being's learning process, and genetic algorithms
simulate the natural selection and evolutionary processes to perform randomized 
global search. Each component of soft computing is complementary to each 
other. Using combinations of several technologies such as fuzzy-neural systems
will generally get better solutions.

\subsection{QoS Model}

Different applications generally have different requirements of QoS dimensions.
Rommel \cite{rom} and Stalk and Hout \cite{sh} investigate the features with 
which successful companies assert 
themselves in the competitive world markets. Their result showed that success
is based on three essential dimensions: time, cost and quality. \cite{gar} 
associates eight dimensions with quality, including performance and 
reliability. Software systems quality of service has been extensively 
studied in \cite{csz,ggps,hsuw,zbs}. For middleware systems, 
Frlund and Koisinen \cite{fk} present a set of practical 
dimensions for distributed object systems reliability and performance, which 
include TTR (time to repair), TTF (time to failure), availability, failure
masking, and server failure. Gardaso, Miller, Sheth and Arnold \cite{gmsa} propose a 
QoS model for workflows and 
Web services processes based on four dimensions: time, cost, reliability and
fidelity. 

In this paper, we construct a QoS model for semantic
Web services. It is composed of the following dimensions: capability, response
time, and trustworthiness.
In order to be more precise, we give our definitions of the three dimensions as
follows:
\begin{enumerate}
\item The {\em capability} of a semantic Web service can be defined as the degree 
to which its functional properties match with the required functional properties of the
semantic Web service requestor; 
\item The {\em response time} of a semantic Web service represents the time that elapses
between service requests arrival and the completion of that service request.
Response time is the sum of waiting time and actual processing time; 
\item The {\em trustworthiness} of a semantic Web services is the extent to which it is
consistent, reliabile, competent, and honest.
\end{enumerate}
 
\section{Architecture of Extensible Soft SWS Agent}

The extensible soft SWS agent can provide high QoS semantic Web services based on specific 
ontology. The extensible SWS agent uses centralized client/server architecture
internally. But itself can also be and should be implemented as a semantic
Web service based on specific service ontology. The extensible soft SWS agent
comprises of six components: (a) Registries Crawler; (b) SWS Repository;
(c) Inquiry Server; (d) Publish Server; (e) Agent Communication Server;
(f) Intelligent Inference Engine. The high level architecture of the extensible
soft SWS agent is shown in Figure 1. Each of the components is described next.

\begin{figure}[htb!]
\begin{center}
\includegraphics{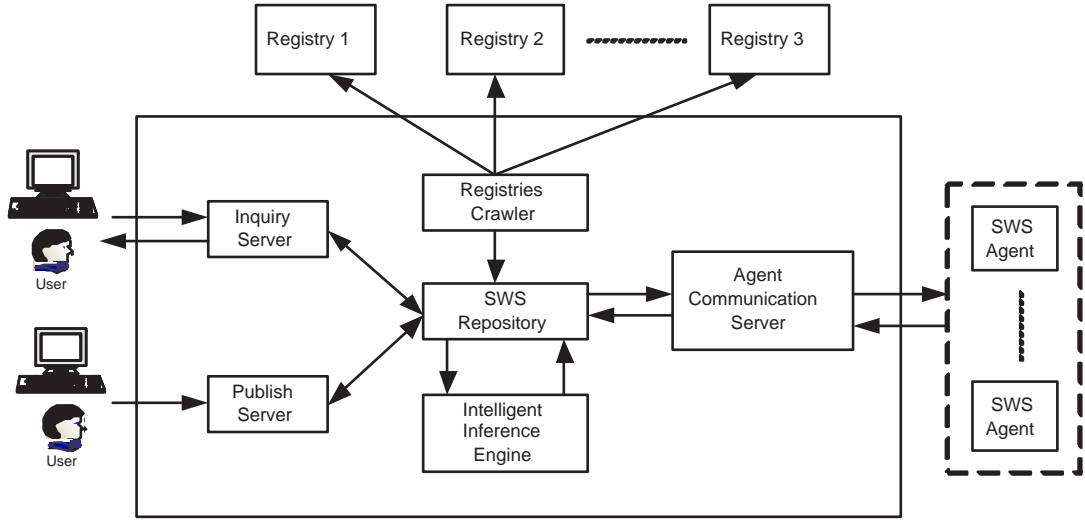}
\caption{Architecture of the Extensible Soft SWS Agent}
\label{figure1}
\end{center}
\end{figure}
 
\subsection{Registries Crawler}

As we pointed out before, the current UDDI registry only supports keyword based
search for the Web services description. Under the Semantic Web environment,
UDDI registry must be extended to be ontology-compatible which supports
semantic matching of semantic Web services' capabilities. One possible way is to
map the OWL-S service profiles into current UDDI registry's data structure.
Semantic Web service providers will publish the service profiles of semantic Web services
in the public or private specific service ontology-oriented UDDI
registries or directly on their semantic Web sites. The specific ontology
based semantic Web services registries crawler has two tasks: 
\begin{enumerate}
\item Accessing these public and private specific service ontology-oriented UDDI registries
using UDDI query API to fetch the service profiles, transforming them into the 
format supported by our repository, and storing them into the repository using the 
publish API of our repository; 
\item Crawling the semantic Web sites hosting the specific
ontology based semantic Web services directly to get the service profiles, 
transforming them into the format supported by the repository, and storing them into
repository using the publish API for the repository. 
\end{enumerate}
The registries crawler should be multithreaded and should be available 24x7. 
The registries crawler must also be provided the information
of highest level specific service ontology before its execution. 

\subsection{SWS Repository}

The specific ontology based semantic Web servcies repository will store service
profiles of semantic Web services. The architecture of repository is shown in Figure 2.

\begin{figure}[htb!] 
\begin{center}
\includegraphics{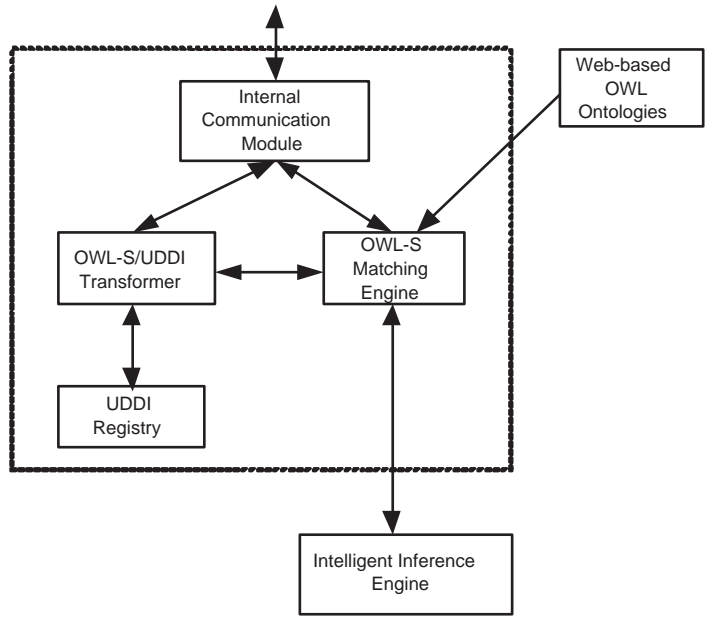}
\caption{Architecture of Repository}
\label{figure2}
\end{center}
\end{figure}

The internal communication module provides the communication interface
between the repository and the registries crawler, inquiry server, publish server, and 
the agent communication server. If a message is an advertisement, the internal
communication module sends it to the OWL-S/UDDI transformer that constructs a
UDDI service description using information about the service provider and the service name. 
The result of publishing with the UDDI is a reference ID of the
service. This ID combined with the capability description and non-functional
properties of the advertisement are sent to the OWL-S matching engine that 
stores the advertisement for capability matching. If a message is a query, the
internal communication module sends the request to the OWL-S matching engine
that performs the capability matching. After calculating the degree of
capability, the OWL-S matching engine will feed the degree of capability and
non-functional properties to the intelligent inference engine to get the
entire Quality of Servie (QoS). The service with highest QoS will be selected.
The result of the selection is the advertisement of the providers selected and
a reference to the UDDI service record. The combination of UDDI records and
advertisements is then sent to the inquiry server. If the required service
does not exist, OWL-S matching engine will transfer the query to the agent
communication server through the internal communication module. The matching
algorithm used by OWL-S matching engine is based on the modified algorithm
described in \cite{pkps}. The modified algorithm considers not only the inputs,
outputs, preconditions and effects, but also service name.
 
\subsection{Inquiry Server}

The specific ontology based semantic Web services inquiry server provides two 
kinds of query interface: a programmatic API to other semantic Web
services or agents and a Web-based interface for the human user. Both
interfaces support keyword oriented query as well as capability oriented 
searches.

For capability oriented query, the inquiry server transforms the service
request profile into the format supported by the repository such as OWL-S
service profile and sends the query message to the internal communication 
module of the repository. The internal communication module sends the service profile to 
the OWL-S matching engine and returns back the requested advertisement to the
inquiry server and then on to the service requestor. The process is shown in Figure \ref{figure5}:

\begin{figure}[htb!]
\begin{center}
\includegraphics{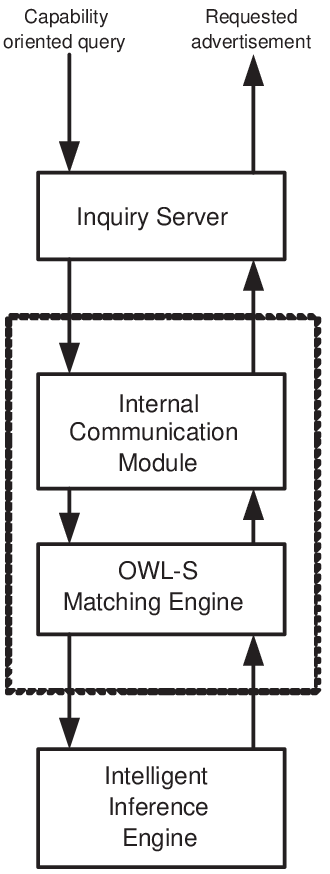}
\caption{Capability oriented query}
\label{figure5}
\end{center}
\end{figure}

For the keyword oriented queries,
the inquiry server will directly send the query string to the internal communication 
module as a query message and the internal communication module sends the query 
string to the UDDI Registry and returns back the requested UDDI records to 
the inquiry server and then on to the service requestor. The
process is shown in Figure \ref{figure6}:  

\begin{figure}[htb!]
\begin{center}
\includegraphics{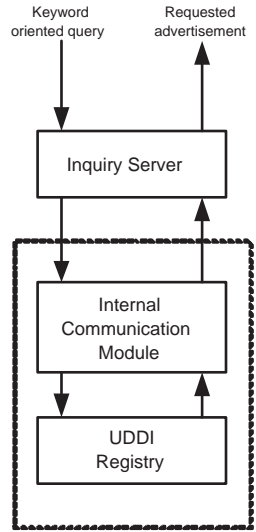}
\caption{Keyword oriented query}
\label{figure6}
\end{center}
\end{figure}

We use SOAP as a communication
protocol between service requestors and the inquiry server.

\subsection{Publish Server}

The specific ontology based semantic Web services publish server provides the 
publishing service for other agents and human users. It has two kinds of interface. 
One is the programmatic API to other semantic Web services or agents and another
is for the human user which is Web-based. The publish server will transform
the service advertisement into the format supported by the repository such as
OWL-S service profile and sends the publish message to the internal communication module. 
The internal communication module sends the transformed OWL-S service
profile to the OWL-S/UDDI transformer. The OWL-S/UDDI transformer will map the
OWL-S service profile into UDDI registries data structure, and store the OWL-S
service profile and reference ID of service into OWL-S matching engine. The process is shown in Figure \ref{figure7}:

\begin{figure}[htb!]
\begin{center}
\includegraphics{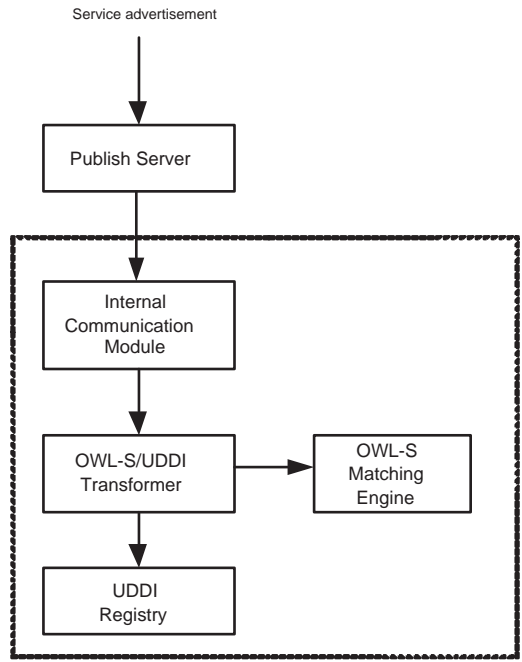}
\caption{Publish service advertisement}
\label{figure7}
\end{center}
\end{figure}

If the
advertised semantic Web services are not in the domain of the soft SWS agent,
the internal communication server will transfer the advertisements to the
agent communication server which will try to publish the advertisements into 
other soft SWS agents. SOAP is used as a communication protocol between service 
publisher and the publisher server.

\subsection{Agent Communication Server}

The soft semantic Web services agent communication server uses a certain
communication protocol such as Knowledge Query and Manipulation Language (KQML)
and Agent Communication Language (ACL) to communicate with other soft SWS agents. 
If the current soft SWS agent cannot fulfill the required services (query and publish), 
the agent communication server is responsible for transfering the requirements to other 
soft SWS agents, getting results back,
and conveying the results back to the service requestors. The current KQML and
ACL should be extended to be ontology-compatible to facilitate the semantic
oriented communication.  

\subsection{Intelligent Inference Engine}

The intelligent inference engine (IIE) is the core of the soft SWS agent. 
The soft SWS agent is extensible because IIE uses soft computing methodology to
calculate the QoS of the semantic Web services with multidimensional QoS metrics. 
IIE gets the degree of capability matching and non-functional properties' 
values from OWL-S matching engine and returns back the whole QoS to OWL-S matching engine. 
In the next section, we show the design of an IIE using neutrosophic logic,
neural networks, and genetic algorithms.
 
\subsection{Design of Intelligent Inference Engine}

This section shows one implementation of IIE based on neutrosophic logic, 
neural
network and genetic algorithm. A schematic diagram of the four-layered 
neutrosophic 
neural network is shown in Figure 3. Nodes in layer one are input nodes
representing input linguistic variables. Nodes in layer two are membership nodes. Membership nodes are truth-membership node, indeterminacy-membership node and
falsity-membership node, which are responsible for mapping an input linguistic 
variable 
into three possibility distributions for that variable. The rule nodes reside in layer three. The last layer contains the output variable nodes \cite{lhl}.

\begin{figure}
\begin{center}
\includegraphics{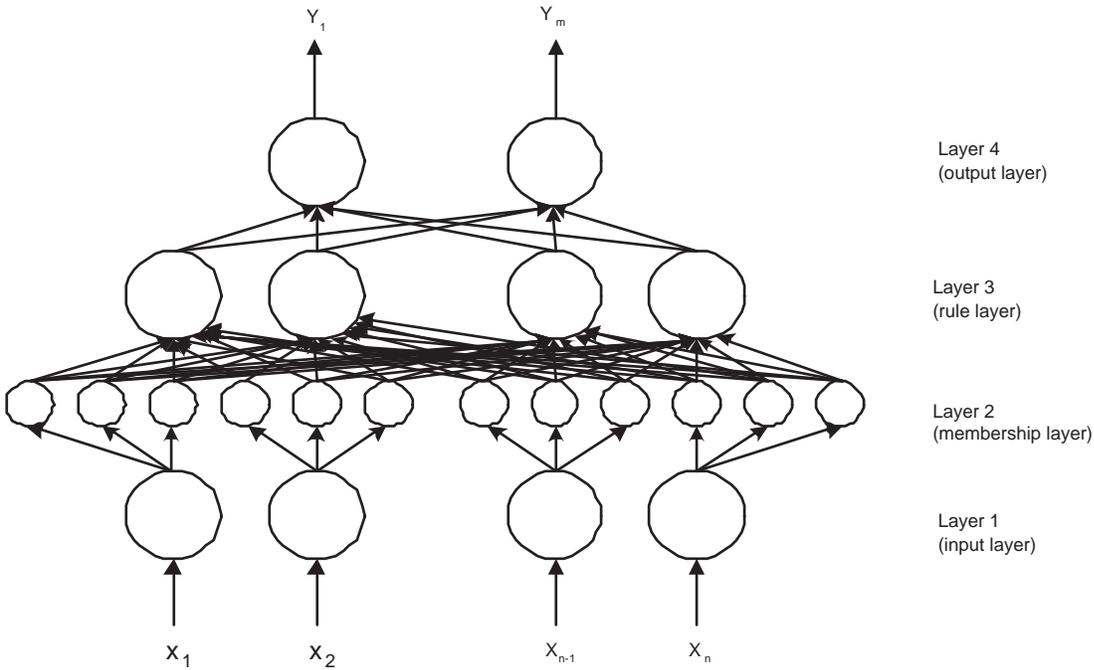}
\caption{Schematic diagram of Neutrosophic Neural Network}
\label{figure3}
\end{center}
\end{figure}

As we mentioned before, the metrics of QoS of Semantic Web services are
multidimensional. For illustration of specific ontology based Semantic Web
services for bioinformatics, we decide to use capability, response time and
trustworthiness as our inputs and whole QoS as output. The neutrosophic logic 
system
is based on TSK model.

\subsection{Input neutrosophic sets}
Let x represent capability, y represent response time and z represent trustworthiness. We scale the capability, response time and trustworthiness to [0,10] respectively. The graphical representation of membership functions of x, y, and z
are shown in Figure 4.

\begin{figure}
\begin{center}
\includegraphics{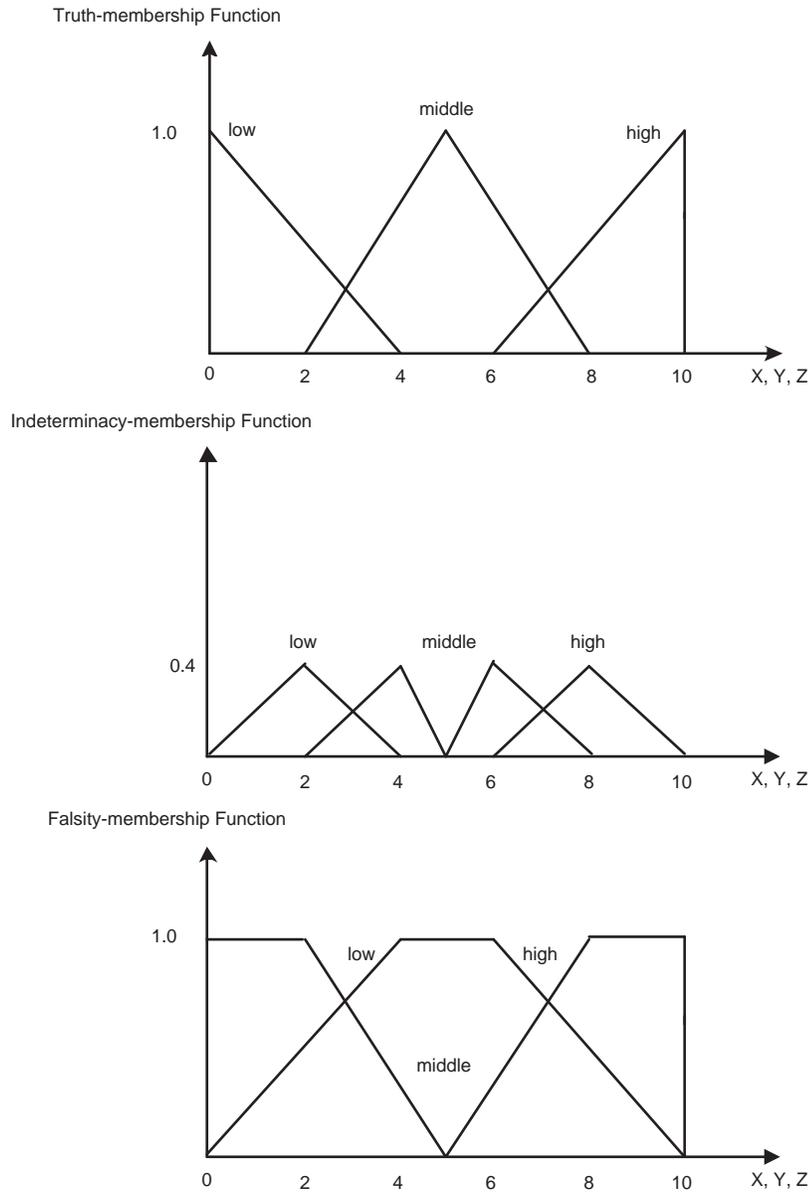}
\caption{Membership functions of inputs}
\label{figure4}
\end{center}
\end{figure}

\subsection{Neutrosophic rule bases}
Here, we design the neutrosophic rule base based on the TSK model. 
A neutrosophic rule is
shown below:

IF $x$ is $I_1$ and $y$ is $I_2$ and $z$ is $I_3$ THEN $O$ is $a_{i,1}*x+a_{i,2}*y+a_{i,3}*Z+a_{i,4}$.

where, $I_1, I_2$ and $I_3$ are in low, middle, and high respectively and $i$
in [1,27]. There are totally 27 neutrosophic rules. The $a_{i,j}$ are 
consequent 
parameters which will be obtained by training phase of neutrosophic neural 
network
using genetic algorithm.
 
\subsection{Design of deneutrosophication}
Suppose, for certain inputs x, y and z, there are m fired neutrosophic rules. To
calculate the firing strength of jth rule, we use the formula:

\begin{equation}
W^j = W_{x}^j*W_{y}^j*W_{z}^j,
\end{equation}
where \\
$W_{x}^j = (0.5*t_x(x)+0.35*(1-f_x(x))+0.025*i_x(x)+0.05)$, \\
$W_{y}^j = (0.5*t_y(y)+0.35*(1-f_y(y))+0.025*i_y(y)+0.05)$, \\
$W_{z}^j = (0.5*t_z(z)+0.35*(1-f_z(z))+0.025*i_z(z)+0.05)$, \\
where $t_{x},f_{x},i_{x}, t_{y},f_{y},i_{y}, t_{z},f_{z},i_{z}$, are the truth-membership, falsity-membership, indeterminacy-membership of 
neutrosophic inputs $x,y,z$, respectively. 

So the crisp output is:

\begin{equation}
O = \sum_{j=1}^mW^j*(a_{j,1}*x+a_{j,2}*y+a_{j,3}*z+a_{j,4})/(\sum_{j=1}^mW^j) 
\end{equation}

\subsection{Genetic algorithms}
GA is a model of machine learning which derives its behavior form a metaphor
of the processes of evolution in nature. This is done by creation within a
machine of a population of individuals represented by chromosomes. Here we
use real-coded scheme. Given the range of parameters (coefficients of linear
equations in TSK model), the system uses the derivate-free random search-GA to
learn to find the near optimal solution by the fitness function through the
training data.

\begin{enumerate}
\item Chromosome: The genes of each chromosome are 108 real numbers (there are
108 parameters in the neutrosophic rule base) which are initially generated randomly
in the given range. So each chromosome is a vector of 108 real numbers.

\item Fitness function: The fitness function is defined as
\begin{equation}
E = 1/2\sum_{j=1}^m(d_i-o_i)^2
\end{equation} 

\item Elitism: The tournament selection is used in the elitism process.

\item Crossover: The system will randomly select two parents among the population, then randomly select the number of cross points, and simply exchange the corresponding genes among these two parents to generate a new generation.

\item Mutation: For each individual in the population, the system will randomly
select genes in the chromosome and replace them with randomly generated real
numbers in the given range.
\end{enumerate}

\subsection{Simulations}
There are two phases for applying a fuzzy neural network: training and predicting. In the training phase, we use 150 data entries as training data set. Each
entry consists of three inputs and one expected output. We tune the performance
of the system by adjusting the size of population, the number of generation and
probability of crossover and mutation. Table 1 gives the part of prediction
results with several parameters for output o.

In Table 1, No. of generation = 10000, No. of population = 100, probability
of crossover = 0.7, probability of mutation = 0.3. 
The maximum error of prediction result is 1.64. 
The total prediction error for
150 entries of testing dataset is 19$\%$. By our observation, designing reasonable neutrosophic membership
functions and choosing reasonable training data set which is based on specific
application domain can reduce the prediction error a lot. Here the example is
just for illustration.

\begin{table}
\caption{Prediction Result of Neutrosophic Neural Network}
\begin{center}
\begin{tabular}{|c|c|c|c|c|} \hline
Input x & Input y & Input z & Desired output & Real output o \\ \hline
1 & 0 & 1 & 0 & 0.51 \\ \hline
1 & 2 & 5 & 1 & 1.71 \\ \hline
1 & 4 & 7 & 2 & 2.59 \\ \hline
3 & 2 & 9 & 3 & 3.52 \\ \hline
3 & 6 & 7 & 4 & 3.81 \\ \hline
3 & 10 & 7 & 5 & 4.92 \\ \hline
5 & 8 & 9 & 6 & 5.43 \\ \hline
7 & 10 & 7 & 7 & 5.90 \\ \hline
7 & 10 & 9 & 8 & 6.45 \\ \hline
9 & 10 & 9 & 9 & 7.36 \\ \hline
\end{tabular}
\end{center}
\end{table}

\section{Related Work}

MWSDI (METEOR-S Web Service Discovery Infrastructure) is an infrastructure of 
registries for semantic publication and discovery
of Web services \cite{vssp}. MWSDI supports creating registry federation by grouping
registries that are mapped to the same node in Registries Ontology. MSWDI is
based on the P2P model, so the registries are considered as peers. In our work,
the soft SWS agents also can be regarded as peers. MWSDI uses the Registries
Ontology to maintain a global view of the registries, associated domains and
uses this information during Web service publication and discovery. The limitation of 
MWSDI is that it supports only capability matching of Web services and 
does not consider non-functional properties of Web services. The soft SWS agent
can be viewed as an enhancement over MWSDI as it
provides the service for discovering semantic Web services with the highest
whole QoS.

The MWSDI approach annotates WSDL by associating its input and output types to domain
specific ontologies and uses UDDI structures to store the mapping of input and
output types in WSDL files to domain specific ontologies. It is similar to our
work where we use OWL-S ontology directly to enable the semantic description
of Web services.

SWWS (Semantic Web enabled Web Services) proposes a semantic-oriented service 
Registry which is similar to our
idea \cite{swws}. It has five components: Profile Crawler, UDDI Integration Engine,
Registry API, Ontology Server and Query Interface. The service modelling
ontology is stored in the ontology server. All individual service descriptions 
are stored as instances of the service description ontology and are also
managed by the ontology server. SWWS does not support quality based semantic
Web services discovery.

OASIS/ebXML describes an architecture of service registry \cite{oasis}. The 
registry provides a stable store where information submitted by a submitting
organization is made persistent. Such information is used to facilitate
ebXML based B2B partnerships and transactions. Submitted content may be XML
schema and documents, process descriptions, ebXML Core Components, context
descriptions, UML models, etc. It focuses mainly on the registry information
model and discusses issues like object replication, object relocation and
lifecycle management for forming registry federation. It does not use semantic
Web and semantic Web services technologies.

\section{Conclusions}

In this chapter, we discussed the design of an extensible soft SWS agent and gave one
implementation of Intelligent Inference Engine. The soft SWS agent supports
both keyword based discovery as well as capability based discovery of semantic
Web services. The primary motivation of our work is to solve two challenges
facing current Web services advertising and discovery techniques. One is
how to locate the registry hosting required Web service description and another
is how to find the required Web service with highest QoS in the located registry. 
The soft SWS agent solves both these problems efficiently and effectively.
The soft SWS agent is built upon semantic Web, Web services, and soft computing technologies. 
The soft SWS agent could be used in WWW, P2P, or Grid infrastructures. The soft SWS agent is 
flexible and extensible. With the evolution of soft computing, more and more technologies 
can be integrated into the soft SWS agent. We used specific ontology based semantic Web 
services for bioinformatics and neutrosophic neural network with genetic algorithm as 
our study case. 
The training time is short and training results are satisfactory. 
The soft SWS agent will return the desired semantic Web services based on the entire
QoS of semantic Web services. In the future, we plan to extend the architecture of 
the soft SWS agent to compute the entire QoS workflow of semantic Web services to 
facilitate the composition and monitoring of complex semantic Web services and 
apply it to semantic Web-based bioinformatics applications.

%% file: master.bbl
\providecommand{\bysame}{\leavevmode\hbox to3em{\hrulefill}\thinspace}
\providecommand{\MR}{\relax\ifhmode\unskip\space\fi MR }
\providecommand{\MRhref}[2]{%
  \href{http://www.ams.org/mathscinet-getitem?mr=#1}{#2}
}
\providecommand{\href}[2]{#2}
\begin{thebibliography}{dACM02b}

\bibitem[ABH02]{abh}
A.~Ankolekar, M.~Burstein, and J.~Hobbs, \emph{Daml-s: Web service description
  for the semantic web}, The First International Semantic Web Conference
  (ISWC), 2002.

\bibitem[AG90]{ATA90}
K.~Atanassov and G.~Gargov, \emph{Intuitionistic fuzzy logic}, Compt. Rend.
  Acad. Bulg. Sci. \textbf{43} (1990), 9--12.

\bibitem[AG98]{ATA98}
K.~Atanassov and George Gargov, \emph{Elements of intuitionistic fuzzy logic.
  part i}, Fuzzy Sets and Systems \textbf{95} (1998), 39--52.

\bibitem[AR84]{nr84}
M.~Anvari and G.~F. Rose, \emph{Fuzzy relational databases}, Proceedings of the
  1st International Conference on Fuzzy Information Processing (Kuaui, Hawaii),
  CRC Press, 1984.

\bibitem[Ata86]{ATA86}
K.~Atanassov, \emph{Intuitionistic fuzzy sets}, Fuzzy Sets and Systems
  \textbf{20} (1986), 87--96.

\bibitem[Ata88]{ATA88}
\bysame, \emph{Two variants of intuitionistic fuzzy propositional calculus},
  Preprint IM-MFAIS-5-88, 1988.

\bibitem[Ata89]{ATA89}
\bysame, \emph{More on intuitionistic fuzzy sets}, Fuzzy Sets and Systems
  \textbf{33} (1989), 37--46.

\bibitem[Bag00]{bagai00}
R.~Bagai, \emph{Tuple relational calculus for paraconsistent databases},
  Lecture Notes in Artificial Intelligence, vol. 1952, Springer-Verlag, 2000,
  pp.~409--416.

\bibitem[Bal83]{bld83}
J.~F. Baldwin, \emph{A fuzzy relational inference language for expert systems},
  Proceedings of the 13th IEEE International Symposium on Multivalued Logic
  (Kyoto, Japan), 1983, pp.~416--423.

\bibitem[Bel77a]{BEL77}
N.~D. Belnap, \emph{A useful four-valued logic}, Modern Uses of Many-valued
  Logic (G.~Eppstein and J.~M. Dunn, eds.), Reidel, Dordrecht, 1977, pp.~8--37.

\bibitem[Bel77b]{bln77}
\bysame, \emph{A useful four-valued logic}, Modern Uses of Many-valued Logic
  (G.~Eppstein and J.~M. Dunn, eds.), Reidel, Dordrecht, 1977, pp.~8--37.

\bibitem[Bis83]{bsk83}
J.~Biskup, \emph{A foundation of codd\'s relational maybe--operations}, ACM
  Trans. Database Syst. 8 \textbf{4} (1983), 608--636.

\bibitem[BLHL01]{lhl01}
T.~Berners-Lee, J.~Hendler, and O.~Lassila, \emph{The semantic web}, Scientific
  American \textbf{284} (2001), no.~5, 34--43.

\bibitem[BMS84]{bd84}
M.~L. Brodie, J.~Mylopoulous, and J.~W. Schmidt, \emph{On the development of
  data models}, On Conceptual Modelling (1984), 19--47.

\bibitem[BP82]{bp82}
B.~P. Buckles and F.~E. Petry, \emph{A fuzzy representation for relational
  databases}, Fuzzy Sets Syst \textbf{7} (1982), 213--226.

\bibitem[bpe]{bpel4ws}
\emph{Business process execution language for web service (bpel4ws) 1.1 (may
  2003)}.

\bibitem[BS89]{blr89}
H.~A. Blair and V.~S. Subrahmanian, \emph{Paraconsistent logic programming},
  Theoretical Computer Science \textbf{68} (1989), 135--154.

\bibitem[BS95]{bgs95a}
R.~Bagai and R.~Sunderraman, \emph{A paraconsistent relational data model},
  International Journal of Computer Mathematics \textbf{55} (1995), no.~1--2,
  39--55.

\bibitem[BS96a]{bs96b}
\bysame, \emph{A bottom-up approach to compute the fitting model of general
  deductive databases}, Journal of Intelligent Information Systems \textbf{6}
  (1996), no.~1, 59--75.

\bibitem[BS96b]{bs96a}
\bysame, \emph{Computing the well-founded model of deductive databases}, The
  International Journal of Uncertainty, Fuzziness and Knowledge-based Systems
  \textbf{4} (1996), no.~2, 157--176.

\bibitem[bsm]{bsml}
\emph{The bioinformatic sequence markup language (bsml) 3.1}.

\bibitem[CK78]{ck78}
S.~K. Chang and J.~S. Ke, \emph{Database skeleton and its application to fuzzy
  query translation}, IEEE Trans. Softw. Eng. SE--4 (1978), 31--43.

\bibitem[CNW01]{cnw}
F.~Curbera, W.~Nagy, and S.~Weerawarana, \emph{Web services: Why and how},
  Workshop on Object-Oriented Web Services-OOPSLA 2001 (Tampa, FL), 2001.

\bibitem[Cod70]{cdd70}
E.F. Codd, \emph{A relational model for large shared data banks},
  Communications of the ACM \textbf{13} (1970), no.~6, 377--387.

\bibitem[Cod79]{cdd79}
\bysame, \emph{Extending the database relational model to capture more
  meaning}, ACM Transactions of Database Systems \textbf{4} (1979), no.~4,
  397--434.

\bibitem[Cos77a]{cst77}
N.~C. A.~Da Costa, \emph{On the theory of inconsistent formal systems}, Notre
  Dame Journal of Formal Logic \textbf{15} (1977), 621--630.

\bibitem[Cos77b]{COS77}
N.C.A.D. Costa, \emph{On the theory of inconsistent formal systems}, Notre Dame
  Journal of Formal Logic \textbf{15} (1977), 621--630.

\bibitem[CP87]{cvp87}
R.~Cavallo and M.~Pottarelli, \emph{The theory of probabilistic databases},
  Proceedings of the 13th Very Large Database Conference, 1987, pp.~71--81.

\bibitem[CSZ92]{csz}
D.~Clark, S.~Shenker, and L.~Zhang, \emph{Supporting real-time applications in
  an integrated services packet network: Architecture and mechanism},
  Proceedings of ACM SIGCOMM, 1992.

\bibitem[dACM02a]{ACM02}
S.~de~Amo, W.~Carnielli, and J.~Marcos, \emph{A logical framework for
  integrating inconsistent information in multiple databases}, Proc. PoIKS'02,
  LNCS 2284, 2002, pp.~67--84.

\bibitem[dACM02b]{mcm2002}
\bysame, \emph{A logical framework for integrating inconsistent information in
  multiple databasses}, Proc. PoIKS'02, LNCS 2284, 2002, pp.~67--84.

\bibitem[dama]{damls}
\emph{Daml-s 0.9 draft release}.

\bibitem[damb]{damloil}
\emph{Daml+oil (march 2001)}.

\bibitem[DP04]{dp}
Z.~Ding and Y.~Peng, \emph{A probabilistic extension to ontology language owl},
  Proceedings of the Hawai' International Conference on System Sciences, 2004.

\bibitem[ebx]{ebxml}
\emph{Electronic business xml initiative (ebxml)}.

\bibitem[EN00]{en2000}
Elmasri and Navathe, \emph{Fundamentals of database systems}, third ed.,
  Addison--Wesley, New York, 2000.

\bibitem[FB02]{fb}
D.~Fensel and C.~Bussler, \emph{The web service modelling framework wsmf},
  Electronic Commerce: Research and Application \textbf{1} (2002), 113--137.

\bibitem[FK98]{fk}
S.~Frlund and J.~Koisinen, \emph{Quality-of-service specification in
  distributed object systems}, Distributed System Engineering Journal
  \textbf{5} (1998), no.~4, 179--202.

\bibitem[Gar88]{gar}
David~A. Garvin, \emph{Managing quality: The strategic and competitive edge},
  Free Press, New York, 1988.

\bibitem[GB93]{gb93}
Wen-Lung Gau and Daniel~J. Buehrer, \emph{Vague sets}, IEEE Transactions on
  Systems, Man, and Cybernetics \textbf{23} (1993), no.~2, 610--614.

\bibitem[GCTB]{ctb}
J.~Gonzalez-Castillo, D.~Trastour, and C.~Bartonili, \emph{Description logics
  for matchmaking of services}.

\bibitem[gen]{genebank}
\emph{Genebank database}.

\bibitem[GGPS96]{ggps}
L.~Georgiadis, R.~Guerin, V.~Peris, and K.~Sivarajan, \emph{Efficient network
  qos provisioning based on per node traffic shaping}, IEEE/ACM Transactions on
  Networking \textbf{4} (1996), no.~4, 482--501.

\bibitem[GMSA]{gmsa}
J.~Gardoso, J.~Miller, A.~Sheth, and J.~Arnold, \emph{Modelling quality of
  service for workflows and web service processes}.

\bibitem[Hen01]{hen}
J.~Hendler, \emph{Agents and the semantic web}, IEEE Intelligent Systems
  \textbf{16} (2001), no.~2, 30--37.

\bibitem[HSUW00]{hsuw}
M.~Hiltunen, R.~Schlichting, C.~Ugarte, and G.~Wong, \emph{Survivability
  through customization and adaptability: The cactus approach}, DARPA
  Information Survivability Conference and Exposition (DISCEX 2000), 2000.

\bibitem[KLP97]{klp}
D.~Koller, A.~Levy, and A.~Pfeffer, \emph{P-classic: A tractable probabilistic
  description logic}, Proceedings of AAAI-97, 1997, pp.~390--397.

\bibitem[KM98]{KM98}
N.~Karnik and J.~Mendel, \emph{Introduction to type-2 fuzzy logic systems},
  Proc. 1998 IEEE Fuzz Conf., 1998, pp.~915--920.

\bibitem[KY95]{KY95}
George~J. Klir and Bo~Yuan, \emph{Fuzzy sets and fuzzy logic: Theory and
  applications}, Prentice Hall, Upper Saddle River, New Jersey, 1995.

\bibitem[KZ86]{kz86}
J.~Kacprzyk and A.~Ziolkowski, \emph{Database queries with fuzzy linguistic
  quantifiers}, IEEE Trans. Syst. Man Cybern. SMC--16 \textbf{3} (1986),
  474--479.

\bibitem[LHL03]{lhl}
C.-H. Lee, J.-L. Hong, and Y.-C. Lin, \emph{Type-2 fuzzy neural network systems
  and learning}, International Journal of Computational Cognition \textbf{1}
  (2003), no.~4, 79--90.

\bibitem[Lip79]{lip79}
W.~Lipski, \emph{On semantic issues connected with incomplete information
  databases}, ACM Trans. Database Syst. 4 \textbf{3} (1979), 262--296.

\bibitem[Lip81]{lps:datinc}
\bysame, \emph{On databases with incomplete information}, Journal of the
  Association for Computing Machinery \textbf{28} (1981), 41--70.

\bibitem[LM00]{LM00}
Qilian Liang and Jerry~M. Mendel, \emph{Interval type-2 fuzzy logic systems:
  Theory and design}, IEEE Transactions On Fuzzy Systems \textbf{8} (2000),
  no.~5, 535--550.

\bibitem[LS90]{ls90}
K.C. Liu and R.~Sunderraman, \emph{Indefinite and maybe information in
  relational databases}, ACM Transaction on Database Systems \textbf{15}
  (1990), no.~1, 1--39.

\bibitem[LS91]{ls91}
\bysame, \emph{A generalized relational model for indefinite and maybe
  information}, IEEE Transaction on Knowledge and Data Engineering \textbf{3}
  (1991), no.~1, 65--77.

\bibitem[Mai83]{mai83}
D.~Maier, \emph{The theory of relational databases}, Computer Science Press,
  Rockville, Maryland, 1983.

\bibitem[MDCG03]{mdcg}
E.~Motta, J.~Domingue, L.~Cabral, and M.~Gaspari, \emph{A framework and
  infrastructure for semantic web services}, The Semantic Web-ISWC 2003, 2003.

\bibitem[Men87]{MEN87}
E.~Mendelson, \emph{Introduction to mathematical logic}, Van Nostrand,
  Princeton, NJ, 1987, Third edition.

\bibitem[MJ02]{MJ02}
J.~Mendel and R.~John, \emph{Type-2 fuzzy sets made simple}, IEEE Transactions
  on Fuzzy Systems \textbf{10} (2002), 117--127.

\bibitem[mn]{mn}
\emph{Understanding quality of service for web services}.

\bibitem[NS92]{ngs92}
R.~Ng and V.S. Subrahmanian, \emph{Probabilistic logic programming},
  Information and Computation \textbf{101} (1992), no.~2, 150--201.

\bibitem[oas]{oasis}
\emph{Oasis/ebxml registry services specification v2.5 (june 2003)}.

\bibitem[OCFB03]{ocfb}
B.~Omelayenko, M.~Crubezy, D.~Fensel, and R.~Benjamins, \emph{Upml: The
  language and tool support for making the semantic web alive}, Spinning the
  Semantic Web: Bringing the WWW to its Full Potential, MIT Press, 2003,
  pp.~141--170.

\bibitem[omn]{omnigene}
\emph{Omnigene: standardizing biological data interchange through web services
  technology}.

\bibitem[owla]{owls}
\emph{Owl-s 1.0 release}.

\bibitem[owlb]{owl}
\emph{Web ontology language (owl)}.

\bibitem[Par96]{prs96}
S.~Parsons, \emph{Current approaches to handling imperfect information in data
  and knowledge bases}, IEEE Trans. Knowledge and Data Engineering \textbf{3}
  (1996), 353--372.

\bibitem[PKPS02]{pkps}
M.~Paolucci, T.~Kawamura, T.~Payne, and K.~Sycara, \emph{Semantic matching of
  web services capabilities}, Proceedings of ISWC 2002, 2002.

\bibitem[PPS]{pps}
T.~Payne, M.~Paolucci, and K.~Sycara, \emph{Advertising and matching daml-s
  service descriptions}.

\bibitem[Pra84]{prd84}
H.~Prade, \emph{Lipski\'s approach to incomplete information databases restated
  and generalised in the setting of zadeh\'s possibility theory}, Inf. Syst. 9
  \textbf{1} (1984), 27--42.

\bibitem[PT84]{prt84}
H.~Prade and C.~Testemale, \emph{Generalizing database relational algebra for
  the treatment of incomplete or uncertain information and vague queries},
  Information Sciences \textbf{34} (1984), 115--143.

\bibitem[PT87]{prt87}
\bysame, \emph{Representation of soft constraints and fuzzy attribute values by
  means of possibility distributions in databases}, Analysis of Fuzzy
  Information, Volume II, Artificial Intelligence and Decision Systems (1987),
  213--229.

\bibitem[rdfa]{rdfs}
\emph{Rdf vocabulary description language 1.0: Rdf schema}.

\bibitem[rdfb]{rdf}
\emph{Resource description framework}.

\bibitem[RM88]{rm88}
K.~V. S. V.~N. Raju and A.~K. Majumdar, \emph{Fuzzy functional dependencies and
  lossless join decomposition of fuzzy relational database systems}, ACM Trans.
  on Database Syst. 13 \textbf{2} (1988), 129--166.

\bibitem[Rom95]{rom}
G.~Rommel, \emph{Simplicity wins: how germany's mid-sized industrial companies
  succeed}, Harvard Business School Press, Boston, Mass, 1995.

\bibitem[SB95]{sb95}
R.~Sunderraman and R.~Bagai, \emph{Uncertainty and inconsistency in relational
  databases}, Advances in Data Management (S.~Chaudhuri, A.~Deshpande, and
  R.~Krishnamurthy, eds.), Tata McGraw Hill, 1995, pp.~206--220.

\bibitem[SCMK02]{scmk}
A.~Sheth, J.~Cardoso, J.~Miller, and K.~Kochut, \emph{Qos for service-oriented
  middleware}, Proceedings of COnference on Systemics, Cybernetics and
  Informatics (Orlando, FL), 2002.

\bibitem[SH90]{sh}
G.~Stalk and T.~Hout, \emph{Competing against time: how timebased competition
  is reshaping global markets}, Free Press, New York, 1990.

\bibitem[SKS96]{sks96}
A.~Silberschatz, H.~F. Korth, and S.~Sudarshan, \emph{Database system
  concepts}, third ed., McGraw--Hill, Boston, 1996.

\bibitem[SKWL99]{skwl}
K.~Sycara, M.~Klusch, S.~Widoff, and J.~Lu, \emph{Dynamic service matchmaking
  among agents in open information environmen ts}, SIGMOD Record \textbf{28}
  (1999), no.~1, 47--53.

\bibitem[Sma99]{SMA99}
Florentin Smarandache, \emph{A unifying field in logics. neutrosophy:
  Neutrosophic probability, set and logic}, American Research Press, Rehoboth,
  1999.

\bibitem[Sma03]{SMA03}
\bysame, \emph{A unifying field in logics: Neutrosophic logic. neutrosophy,
  neutrosophic set, neutrosophic probability and statistics}, Xiquan, Phoenix,
  2003, third edition.

\bibitem[soa]{soap}
\emph{Simple object access protocol (soap) 1.2}.

\bibitem[SRT05]{srt}
A.~Sheth, C.~Ramakrishnan, and C.~Thomas, \emph{Semantics for the semantic web:
  The implicit, the formal and the powerful}, International Journal on Semantic
  Web and Information Systems \textbf{1} (2005), no.~1, 1--18.

\bibitem[Str98]{str98}
U.~Straccia, \emph{A fuzzy description logic}, Proceedings of AAAI-98, 1998.

\bibitem[Str04]{str04}
\bysame, \emph{Uncertainty and description logic program: A proposal for
  expressing rules and uncertainty on top of ontologies}, Tech. Report
  2004-TR-14, ISTI-CNR, 2004.

\bibitem[Sub94]{sbr94}
V.~S. Subrahmanian, \emph{Amalgamating knowledge bases}, ACM Transactions on
  Database Systems \textbf{19} (1994), no.~2, 291--331.

\bibitem[SVSM03]{svsm}
K.~Sivashanmugam, K.~Verma, A.~Sheth, and J.~Miller, \emph{Adding semantics to
  web services standards}, International Conference on Web Services (ICWS'03),
  2003, pp.~395--401.

\bibitem[sww]{swws}
\emph{Report on development of web service discovery framework (october 2003)}.

\bibitem[Tur86]{TUR86}
I.~Turksen, \emph{Interval valued fuzzy sets based on normal forms}, Fuzzy Sets
  and Systems \textbf{20} (1986), 191--210.

\bibitem[udd]{uddi}
\emph{Universal description, discovery and integration (uddi) 3.0.1}.

\bibitem[VSSP04]{vssp}
K.~Verma, K.~Sivashanmugam, A.~Sheth, and A.~Patil, \emph{Meteor-s wsdi: A
  scalable p2p infrastructure of registries for semantic publication and
  discovery of web services}, Journal of Information Technology and Management
  (2004), In print.

\bibitem[Won82]{wng82}
E.~Wong, \emph{A statistical approach to incomplete information in database
  systems}, ACM Trans. on Database Systems \textbf{7} (1982), 470--488.

\bibitem[wsd]{wsdl}
\emph{Web services description language (wsdl) 1.1}.

\bibitem[wsf]{wsfl}
\emph{Web services flow language (wsfl) 1.0}.

\bibitem[WZS04]{WZR04}
H.~B. Wang, Y.~Q. Zhang, and R.~Sunderraman, \emph{Soft semantic web services
  agent}, The Proceedings of NAFIPS 2004, 2004, pp.~126--129.

\bibitem[xem]{xembl}
\emph{Xml web services for embl (xembl)}.

\bibitem[xla]{xlang}
\emph{Xlang: Web services for business process design (2001)}.

\bibitem[Zad65a]{zdh65}
L.~A. Zadeh, \emph{Fuzzy sets}, Inf. Control 8 (1965), 338--353.

\bibitem[Zad65b]{ZAD65}
L.A. Zadeh, \emph{Fuzzy sets}, Inform. and Control \textbf{8} (1965), 338--353.

\bibitem[Zad78]{zdh78}
L.~A. Zadeh, \emph{Fuzzy sets as the basis for a theory of possibility}, Fuzzy
  Sets and Systems \textbf{1} (1978), 1--27.

\bibitem[Zad94]{zadeh}
L.~Zadeh, \emph{Fuzzy logic, neural networks, and soft computing},
  Communications of the ACM \textbf{37} (1994), 77--84.

\bibitem[ZBS97]{zbs}
J.~Zinky, D.~Bakken, and R.~Schantz, \emph{Architectural support for quality of
  service for corba objects}, Theory and Practice of Object Systems \textbf{3}
  (1997), no.~1.

\bibitem[ZKLY04]{ZKL04}
Y.-Q. Zhang, A.~Kandel, T.Y. Lin, and Y.Y. Yao, \emph{Computational web
  intelligence: Intelligent technology for web applications, series in machine
  perception and artificial intelligence}, World Scientific, 2004, Volume 58.

\end{thebibliography}
